\documentclass[%
pof,
numerical,
amsmath,
amssymb,
reprint,
]{revtex4-1}
\usepackage{hyperref}
\usepackage{booktabs}
\usepackage{float}
\usepackage{graphicx}
\usepackage{subfigure}
\usepackage[export]{adjustbox}
\usepackage{dcolumn}
\usepackage{bm}
\usepackage{cleveref}
\usepackage[abbreviations]{glossaries-extra}
\usepackage[utf8]{inputenc}
\usepackage[T1]{fontenc}
\usepackage{threeparttable}
\usepackage{mathptmx}
\usepackage{siunitx}
\usepackage{etoolbox}
\usepackage{color}
\usepackage{CJKutf8}
\usepackage{xcolor}
\usepackage{comment}
\usepackage{multirow}
\usepackage{soul}

\soulregister\cite7
\soulregister\ref7
\soulregister\eqref7
\soulregister\pageref7 

\makeatletter
\def\@email#1#2{%
 \endgroup
 \patchcmd{\titleblock@produce}
  {\frontmatter@RRAPformat}
{\frontmatter@RRAPformat{\produce@RRAP{*#1\href{mailto:#2}{#2}}}\frontmatter@RRAPformat}
  {}{}
}%
\makeatother
\begin{document}
\preprint{AIP/123-QED}
\title[]{A fully conservative discrete velocity Boltzmann solver with parallel adaptive mesh refinement for compressible flows}
\author{R.M.~Strässle}
\affiliation{Computational Kinetics Group, Department of Mechanical and Process Engineering, ETH Zürich, 8092 Zürich, Switzerland}%
\author{S.A.~Hosseini}
\affiliation{Computational Kinetics Group, Department of Mechanical and Process Engineering, ETH Zürich, 8092 Zürich, Switzerland}%
\author{I.V.~Karlin}
\affiliation{Computational Kinetics Group, Department of Mechanical and Process Engineering, ETH Zürich, 8092 Zürich, Switzerland}%
\email{rubenst@ethz.ch, shosseini@ethz.ch, ikarlin@ethz.ch}
\date{\today}

\begin{abstract}
This paper presents a parallel and fully conservative adaptive mesh refinement (AMR) implementation of a finite-volume-based kinetic solver for compressible flows.
Time-dependent H-type refinement is combined with a two-population quasi-equilibrium Bhatnagar--Gross--Krook discrete velocity Boltzmann model.
A validation has shown that conservation laws are strictly preserved through the application of refluxing operations at coarse-fine interfaces.
Moreover, the targeted macroscopic moments of Euler and Navier--Stokes--Fourier level flows were accurately recovered with correct and Galilean invariant dispersion rates for a temperature range over three orders of magnitude and dissipation rates of all eigen-modes up to Mach of order $1.8$. 
Results for one- and two-dimensional benchmarks up to Mach numbers of $3.2$ and temperature ratios of $7$, such as the Sod and Lax shock tubes, the Shu--Osher and several Riemann problems, as well as viscous shock-vortex interactions, have demonstrated that the solver precisely captures reference solutions.
Excellent performance in obtaining sensitive quantities was proven, for example in the test case involving nonlinear acoustics, whilst, for the same accuracy and fidelity of the solution, the AMR methodology significantly reduced computational cost and memory footprints.
Over all demonstrated two-dimensional problems, up to a $4$- to $9$-fold reduction was achieved and an upper limit of the AMR overhead of $30$\% was found in a case with very cost-intensive parameter choice.
The proposed solver marks an accurate, efficient and scalable framework for kinetic simulations of compressible flows with moderate supersonic speeds and discontinuities, offering a valuable tool for studying complex problems in fluid dynamics.
\end{abstract}

\maketitle

\section{Introduction}
\label{sec:Introduction}

Understanding compressible and high-speed fluid flows is of paramount importance for science and engineering. 
Development of reliable, accurate and efficient numerical methods for the simulation of compressible flows, especially on large-scale clusters, has been a topic of intense research over the past few decades~\cite{pirozzoli2011numerical, yee2018recent}. 
While different discrete approximations such as finite volumes (FV) and finite differences (FD) to the Navier--Stokes--Fourier (NSF) equations were the main drivers of research in that area, the advent of the lattice Boltzmann method (LBM) in the late 80's opened the door for a new class of numerical methods rooted in the kinetic theory of gases. 

Discrete velocity Boltzmann methods (DVBMs), such as LBM \cite{LBMBookKrueger, succiBook}, solve a discrete velocity version of the Boltzmann equation obtained via a finite-order expansion of the distribution function in terms of a set of orthonormal functions such as Hermite polynomials~\cite{hosseini2023lattice}.
The dynamics of observable properties of a fluid described by the Euler or NSF equations are recovered in the hydrodynamic limit \cite{Chapman}.
In addition to allowing for the possibility to include physics beyond NSF via higher-order expansions, its combination with collision models such as the Bhatnagar-Gross-Krook (BGK) approximation~\cite{bhatnagar1954model} has been shown to provide an efficient alternative to classical solvers \cite{QianLBM, ChenLBM}. 
Simulation of compressible and high-speed flows with DVBMs has witnessed considerable advances in the past decades, illustrated most notably by formulations such as the family of Unified Gas Kinetic and Discrete Unified Gas Kinetic schemes (UGKS, DUGKS)~\cite{guo2021progress}, FD and FV-based models such as those discussed in~\cite{xu2018discrete,mieussens2000discrete} and the LBM~\cite{hosseini2024lattice}.

In the specific context of the LBM, while research was focused on the incompressible regime and stabilization of the isothermal solver through advanced collision models and entropy constrained equilibria construction \cite{Coveney_MRT, Geier_Cascaded, Geier_Cumulant, malaspinas_recursivereg, Karlin_entropicEQ, Karlin_Gibbs, Ali_reviewEntropic} over the past decades, considerable advances are being reported in recent years for compressible flow simulations. These advances have been, in part, motivated by insights from kinetic theory, allowing for, e.g., proper viscous heating in non-unity Prandtl number flows via models such as Shakhov \cite{Shakhov}, Holway \cite{HolwayES} or the quasi-equilibrium approach \cite{gorban1994, QE_ansumali-2007}. The introduction of double-distribution function models, where a second distribution is carrying for example the total or internal non-rotational energy \cite{rykov_model_1976, KarlinTwoPop, ProbingDoubleDist2024}, had considerable impact on the development of efficient compressible LBM-based solvers capturing variable specific heat ratios. The class of double-distribution function LBM solvers relying on standard lattices~\cite{Prasianakis2007, Saadat2019, hosseini2020compressibility, FengCorrection,li2012coupling} and hybrid solvers, modeling the energy balance equation via a FD or FV solver~\cite{FENG_FV_DBM} are practical illustrations of this impact. Models relying on higher-order lattices such as D2Q25 properly recovering the energy balance equation~\cite{ChikatamarlaMultispeed, FrapolliMultispeed} without needing correction terms were also developed in that context. While all these approaches relied on static quadratures and as a result were limited in terms of maximum achievable Mach number, the introduction of dynamic quadratures, starting with shifted lattices \cite{Frapolli2016b,Ali_shiftedStencils, Coreixas2020} and culminating with the particles on demand (PonD) method~\cite{PonD18}, opened the door for hypersonic flow simulations \cite{Reyhanian20, Reyhanian21, sawant2022detonation, Bhaduria23}.

Some comments are in order here; higher-order velocity sets, e.g. the D2Q25, based on construction as the roots of Hermite polynomials \cite{ChikatamarlaMultispeed}, non-integer shifted velocity sets and PonD generally result in non-space-filling lattices, which combined with a Lagrangian discretization in space would require additional interpolation steps to reconstruct populations at lattice sites \cite{BardowMultispeedSSLBM, Wilde_SSLBM_CF, Wilde_CubatureRules}.
This entails a loss of the conservative properties of the scheme. Alternatively, strict conservation can be achieved by construction with a FV discretization \cite{FENG_FV_DBM, XuFV, Ji24} and formulations based on UGKS or DUGKS \cite{UGKS_14, DUGKS_compressiblecase, GuoDUGKS23, Kallikounis22, Kallikounis23}.
However, the larger number of variables and operations associated with higher-order quadratures and adaptive velocity sets combined with double-distribution models render the issue of efficiency and computational cost reduction quite critical.

A commonly employed computational technique is adaptive mesh refinement (AMR), which is designed to enhance the ratio between efficiency and accuracy of simulations on a computational grid by dynamically adjusting the resolution based on some local indicators of the flow.
By adaptively refining the computational grid in regions of interest, AMR can provide high resolution where it is most needed, while reducing computational cost and memory usage in regions where coarser resolution is sufficient, i.e. it ensures efficient usage of computational resources while maintaining or increasing fidelity in critical areas.
In the context of this manuscript, AMR shall refer to the time-dependent H-type grid refinement, i.e. refinement by addition of more grid nodes, as opposed to the R-type, i.e. redistribution or distortion of the grid without changing the number of nodes and their connectivity. 
Note that a P-type refinement, generally described as increasing the order of interpolation or polynomial approximation of the solution on a grid point, was also shown to be a promising avenue that will be revisited in future publications; in the context of DVBMs, this corresponds to making the order of the quadrature adaptive, allowing for the use of lower-order quadratures in lower Mach number regions \cite{Kallikounis_multiscale, PhD_Kallikounis}.
AMR is beneficial for flows exhibiting distinct features, for example,
(1) flows containing localized regions of high complexity such as boundary layers around objects, vortex dominated or turbulent flows,  
(2) sharp interfaces, as in multiphase flows,
and it is especially powerful in
(3) high-speed compressible flows, which contain sharp gradients, shocks and rarefaction waves.
The effectiveness of AMR has been demonstrated for a variety of such problems and several refinement approaches in connection with DVBMs, e.g. LBM \cite{DorschnerRefinementELBM, Guzik, EitelAmor_AMR_LBM, He_AMR_LBM, Schukmann_AMR_LBM}, FD-DVBM \cite{Fakhari_FD_LBM}, and other related models \cite{Huang_AMR_RLBFS}.
However, adopting AMR introduces significant algorithmic changes and presents challenges such as loss of strict conservation properties in conservative schemes
and leveraging the full potential of parallel processing on modern distributed memory machines.

The aim of this paper is to present a parallel AMR realization of an FVM-based kinetic solver, which is fully conservative and thus ideally suited for the simulation of compressible flows.
Throughout the manuscript, emphasis is placed on strict conservation and a correct recovery of the macroscopic moments and dissipation rates for Euler and NSF level dynamics.
The methodology is presented in section~\ref{sec:Methodology}, where the kinetic model is described in section~\ref{sec:kineticModel} and discretized in section~\ref{sec:discretization}, followed by an outline of the AMR methodology and AMR-related implementations in section~\ref{sec:amr}.
The physical capabilities of the solver are validated in section~\ref{sec:validation} and results of several one- and two-dimensional benchmarks of compressible flows are presented in section~\ref{sec:Results}, along with an evaluation of computational efficiency.
Finally, a summary and conclusions are provided in section~\ref{sec:conslusions}.

\section{Methodology\label{sec:Methodology}}

\subsection{Kinetic model description\label{sec:kineticModel}}

This part focuses on the model construction, starting from the Boltzmann equation,
\begin{equation}
    \partial_t f \left (t,\bm{r} , \bm{v}\right ) + \bm{v} \cdot \bm{\nabla} f\left (t,\bm{r} , \bm{v}\right ) = \Omega_f
    , 
    \label{con_Bolt_f}
\end{equation}
where $\Omega_f$ is the collision operator modeled with the BGK approximation \cite{bhatnagar1954model} as $\Omega_f=\left [f^{\rm eq}\left (t,\bm{r} , \bm{v}\right )-f\left (t,\bm{r}, \bm{v} \right ) \right ]/ \tau $.
The distribution function and the local equilibrium distribution function are represented by $f\left (t,\bm{r}, \bm{v} \right )$ and $f^{\rm eq}\left (t,\bm{r} , \bm{v}\right )$, respectively. 
The particle velocity is designated by $\bm{v}$ and $\bm{r}$ marks the position in space.
A single parameter, $\tau$, controls the relaxation rate of the distribution function towards the equilibrium state, known as the local Maxwellian distribution parameterized by the values of local density $\rho$, temperature $T$ and flow velocity $\bm{u}$ as
\begin{equation}
    f^{\rm eq}\left (t,\bm{r} , \bm{v}\right ) = \frac{\rho}{{(2\pi R T)}^{D/2}} \exp{\left[-\frac{{(\bm{v}-\bm{u})}^2}{2RT}\right]}
    ,
    \label{con_feq}
\end{equation}
where $D$ is the dimension of the physical space and $R$ designates the  gas constant. For the sake of convenience, $R=1$ is used for the remainder of this manuscript without loss of generality.
The fluid density and momentum are be obtained as
\begin{align}
    \rho &= \int_{\mathbb{R}^D} f d\bm{v} = \int_{\mathbb{R}^D} f^{eq} d\bm{v}
    , 
    \label{density_1}
\\
    \rho \bm{u} &= \int_{\mathbb{R}^D} \bm{v} f d\bm{v} = \int_{\mathbb{R}^D} \bm{v} f^{eq} d\bm{v}
    . 
    \label{momentum_1}
\end{align}

Following Rykov's original work \cite{rykov_model_1976} to take the effect of internal degrees of freedom of a gas into account and allowing for variable specific heat ratios $\gamma$, Eq. \eqref{con_Bolt_f} is supplemented with a balance equation for a second distribution function $g$ carrying the excess internal energy stemming from non-translational degrees of freedom, i.e., 
\begin{equation}
    \partial_t g \left (t,\bm{r}, \bm{v} \right ) + \bm{v} \cdot  \bm{\nabla} g\left (t,\bm{r} , \bm{v}\right ) = \Omega_g
    ,
    \label{con_Bolt_g}
\end{equation}
with $\Omega_g=\left (g^{\rm eq}\left (t,\bm{r}, \bm{v} \right ) - g\left (t,\bm{r}, \bm{v} \right ) \right )/ \tau$.
The local equilibrium state of the of the second distribution function, $g^{\rm eq}$, is defined as
\begin{equation}
    g^{\rm eq} = 2\left (C_v-\frac{D}{2}\right )T f^{\rm eq}
    ,
    \label{con_geq}
\end{equation}
where $C_v$ indicates the specific heat at constant volume.
As a result, the total energy is obtained as
\begin{multline}
    \rho E = \rho \left (C_v T +\frac{1}{2} \mid \bm{u} \mid ^2 \right )
    = \int_{\mathbb{R}^D} \frac{1}{2} \mid \bm{v} \mid ^2 f  d\bm{v} + \int_{\mathbb{R}^D} \frac{1}{2} g d\bm{v}
    \\
    = \int_{\mathbb{R}^D} \frac{1}{2} \mid \bm{v} \mid ^2 f^{eq}  d\bm{v} + \int_{\mathbb{R}^D} \frac{1}{2} g^{eq} d\bm{v}
    .
    \label{energy_1}
\end{multline}

While allowing for variable specific heat ratios, given that all moments relax at the same rate, the model still is limited to unity Prandtl number defined as
\begin{equation}
    \Pr = \frac{{C}_p \mu}{\kappa}
    ,
\end{equation}
where $C_p$ is the specific heat under constant pressure, $\mu$ the dynamic viscosity and $\kappa$ the thermal conductivity.
The relaxation to the equilibrium is split into two steps to overcome this restriction via the introduction of an intermediary state called the quasi-equilibrium state \cite{gorban1994,QE_ansumali-2007,FrapolliMultispeed,2016Entropic,hosseini2023lattice}. 
The collision terms are rewritten as
\begin{align}
    \label{new_collision_f}
    \Omega_f = \frac{1}{\tau_1}& \left[ f^{\rm eq}\left (t,\bm{r} , \bm{v}\right )-f\left (t,\bm{r} , \bm{v}\right ) \right ] 
    \nonumber \\ &+ \left ( \frac{1}{\tau_1} -\frac{1}{\tau_2}  \right ) \left [ f^{*}\left (t,\bm{r}, \bm{v} \right ) - f^{\rm eq}\left (t,\bm{r} , \bm{v}\right ) \right ]
    , 
\\
    \label{new_collision_g}
    \Omega_g = \frac{1}{\tau_1}& \left[ g^{\rm eq}\left (t,\bm{r}, \bm{v} \right )-g\left (t,\bm{r}, \bm{v} \right ) \right ] 
    \nonumber \\  &+ \left ( \frac{1}{\tau_1} -\frac{1}{\tau_2}  \right ) \left [ g^{*}\left (t,\bm{r}, \bm{v} \right ) - g^{\rm eq}\left (t,\bm{r} , \bm{v}\right ) \right ]
    , 
\end{align}
where $\tau_1$ and $\tau_2$ are the two relaxation parameters related to the viscosity and thermal conductivity, and $f^{*}\left (t,\bm{r}, \bm{v} \right )$ and $g^{*}\left (t,\bm{r}, \bm{v} \right )$ are the quasi-equilibria. 
The quasi-equilibrium state is defined as the minimizer of the Boltzmann entropy function subject to a set of locally conserved fields and quasi-conserved slow fields. 
In order to recover the NSF equations with variable Prandtl number \cite{QE_ansumali-2007}, the quasi-equilibrium distribution functions are required to satisfy the conservation of mass, momentum and total energy, as
\begin{align}
    \int_{\mathbb{R}^D}  f^{*} d\bm{v} &= \int_{\mathbb{R}^D}  f^{\rm eq} d\bm{v}
    , 
    \label{eq_con_fstar1}
\\
    \int_{\mathbb{R}^D} \bm{v}  f^{*} d\bm{v} &= \int_{\mathbb{R}^D} \bm{v} f^{\rm eq} d\bm{v}
    , 
    \label{eq_con_fstar2}
\\
    \int_{\mathbb{R}^D} \left(\mid \bm{v} \mid ^ 2 f^{*} +g^{*}\right) d\bm{v} &=  \int_{\mathbb{R}^D} \left(\mid \bm{v} \mid^2 f^{\rm eq} + g^{\rm eq}\right) d\bm{v}
    . 
    \label{eq_con_gstar}
\end{align}
In addition, the quasi-equilibrium distribution functions are designed to conserve the third-order centered kinetic moments, i.e. the centered heat flux tensor 
$\bm{\bar{Q}}^{*} = \bm{\bar{Q}}$ 
as
\begin{equation}
     \int_{\mathbb{R}^D} \bm{e} \otimes \bm{e} \otimes \bm{e} f^{*} d\bm{v}
     =
     \int_{\mathbb{R}^D} \bm{e} \otimes \bm{e} \otimes \bm{e} f d\bm{v}
     , 
     \label{Q_f}
\end{equation}
and the centered energy flux of the internal degrees of freedom $ \bm{\bar{q}}^{*} = \bm{\bar{q}}$
as 
\begin{equation}
     \int_{\mathbb{R}^D}  \bm{e} g^{*} d\bm{v} 
     = 
     \int_{\mathbb{R}^D} \bm{e} g d\bm{v}
     ,
     \label{q_f}
\end{equation}
where $\bm{e}  = (\bm{v}  - \bm{u})$ and "$\otimes$" denotes the tensor product.
This way, the Prandtl number can be controlled by adjusting $\tau_1$ and $\tau_2$ independently as
\begin{equation}
   \mathrm{Pr} = \frac{\tau_1}{\tau_2}
   .
\end{equation}

Using the rewritten collision terms \eqref{new_collision_f} and \eqref{new_collision_g}, the kinetic equations \eqref{con_Bolt_f} and \eqref{con_Bolt_g} are partial differential equations in time, physical space and phase space (space of particles' velocities).
The presented model recovers the hydrodynamic regimes, i.e. Euler and NSF; a multiscale analysis can be found in appendix~\ref{app:multiscale}.
From the Chapman--Enskog expansion, it follows that the kinematic shear viscosity $\nu$, bulk viscosity $\nu_B$, and the and the thermal diffusivity $\alpha$ in the present model are related to the relaxation coefficient $\tau_1$ and $\tau_2$ as,
\begin{align}
\label{CE:visc.shear}
    \nu &= \frac{\mu}{\rho} = \tau_1 T, 
    \\
\label{CE:visc.bluk}
    \nu_B &= \frac{\mu_B}{\rho} = \left(\frac{2}{D}-\frac{1}{C_v}\right)\tau_1 T,
    \\
 \label{CE:visc.thermal}
   \alpha &= \frac{\kappa}{(C_v+1) \rho} = \tau_2 T.
\end{align}

In the next section, the discretization of these equations is introduced while guaranteeing correct recovery of the hydrodynamic regimes and strictly retaining conservation properties.

\subsection{Model Discretization\label{sec:discretization}}

\subsubsection{Phase space discretization}

The phase space is discretized with a set of $Q$ particles with velocities $\bm{v}_i$, where $i \in \{0, Q-1\}$.
For simplicity, due to the main focus of this work being AMR, the static reference frame was applied as opposed to the adaptive co-moving reference frame used in PonD \cite{PonD18}.
Moments of distribution functions are computed as numerical quadratures, e.g. the conserved density, momentum and energy as
\begin{align}
    \label{eq_quadratur1}
    {\rho} &= \sum_{i=0}^{Q-1} {f}_i = \sum_{i=0}^{Q-1} {f}^{eq}_i 
    \text{,} 
\\
    {\rho \bm{u}} &= \sum_{i=0}^{Q-1} \bm{v}_i {f}_i = \sum_{i=0}^{Q-1} \bm{v}_i {f}^{eq}_i 
    \text{,} 
\\
    \label{eq_quadratur3}
    {\rho E} &= \sum_{i=0}^{Q-1} \frac{1}{2} |\bm{v}_i|^2 {f}_i + \sum_{i=0}^{Q-1} \frac{1}{2} {g}_i
    = \sum_{i=0}^{Q-1} \frac{1}{2} |\bm{v}_i|^2 {f}^{eq}_i + \sum_{i=0}^{Q-1} \frac{1}{2} {g}^{eq}_i
    \text{.} 
\end{align}
All relevant moments of the presented distribution functions are listed in appendix~\ref{app:moments}.

In order to capture the fundamental flow physical properties at Euler Level and the NSF level,
a set of moments of $f$ up to order three and four needs to be correctly recovered.
More precisely, the diagonal components of the full third- and fourth-order tensors, i.e. the contracted third- and fourth-order tensors, are sufficient compared to the full ones, respectively. 
The discrete $f$-equilibria were reconstructed as a Grad expansion \cite{grad1949kinetic},
\begin{equation}
    f_i^{eq} = w_i \sum_{n=0}^{N} \frac{1}{n!} \bm{a}^{eq,(n)} : \bm{\mathcal{H}}^{(n)}
    \text{,}
    \label{eq_gradexpansion}
\end{equation}
where the series is truncated at the expansion order $N$.
The weights of the velocity set are given by $w_i$ and "$:$" denotes the Frobenius inner product. $\bm{\mathcal{H}}^{(n)}$ is the Hermite polynomial tensor of order $n$ and $\bm{a}^{eq,(n)}$ the corresponding coefficient tensor accounting for the required set of equilibrium moments and the reference temperature $T_{ref}$ of the velocity set.
To model the equilibria for Euler level dynamics correctly, an expansion up to $N=3$ is required, whereas order $N=4$ is necessary for NSF.
All relevant Hermite polynomials an the corresponding coefficients are given in appendix~\ref{app:gradexpansion}.
Moreover, forth- and fifth-order quadratures in two spatial dimensions ($D=2$) are necessary to correctly match all moments with the Grad expansion, respectively.
All specifics of the D2Q16 and D2Q25 discrete velocity sets are given in appendix~\ref{app:velocityset}.

Following the discussion on the expansion orders of $f$, according to the non-translational internal energy split, a first- and second-order Grad expansion of the $g$-distribution would be sufficient for Euler and NSF dynamics, respectively.
However, the $g$-equilibria are obtained via their relation to $f^{eq}$ as 
\begin{equation}
    g_i^{\rm eq} = 2\left (C_v-\frac{D}{2}\right )T f_i^{\rm eq}
    .
    \label{disc_geq}
\end{equation}
The requirements on the order of $g_i^{\rm eq}$ are therefore always fulfilled in case the the requirements on $f_i^{\rm eq}$ are also sufficient. 
For simplicity, although second and third-order quadratures would be sufficient, respectively, the $g$-distributions are solved on the same numerical quadratures as the $f$-distributions.

Furthermore, the discrete $f$ and $g$ quasi-equilibria for \mbox{Pr $<1$} are found as corrections to the third- and first-order expansion terms of the Grad expansion for the equilibrium populations, i.e. as
\begin{align}
\label{eq:EponD.discreteQEpops}
    f_i^* &= f_i^{eq} + w_i \frac{1}{3!} \bm{a}^{*,(3)} : \bm{\mathcal{H}}^{(3)}
    \text{,} 
    \\
    g_i^* &= g_i^{eq} + w_i \frac{1}{1!} \bm{a}^{*,(1)} : \bm{\mathcal{H}}^{(1)}
    \text{,}
\end{align}
where $\bm{\mathcal{H}}^{(3)}$ and $\bm{\mathcal{H}}^{(1)}$ are the same full Hermite polynomial tensors as listed in appendix~\ref{app:gradexpansion}.
The coefficients $\bm{a}^{*,(3)}$ and $\bm{a}^{*,(1)}$ are given by ${\bm{\bar{Q}}^{neq}}/{T_{ref}^3}$
and ${\bm{\bar{q}}^{neq}}/{T_{ref}}$, respectively,  
where the centered non-equilibrium heat flux tensor 
and the centered non-equilibrium energy flux tensor of the internal degrees of freedom are computed 
with the centered particles' velocities $\bm{e}_i = (\bm{v}_i - \bm{u})$, as
\begin{align}
     \bar{Q}^{neq} &= \sum_{i=0}^{Q-1} \bm{e}_i \otimes \bm{e}_i \otimes \bm{e}_i (f_i - f_i^{eq})
     \label{Q_fi}
     ,
     \\
     \bar{q}^{neq} &= \sum_{i=0}^{Q-1}  \bm{e}_i (g_i - g_i^{eq})
     \label{q_gi}
     .
\end{align}

\subsubsection{Spatio-temporal discretization}

\paragraph{Time-explicit finite-volume scheme:}

The finite-volume approach is used to discretize the physical space in the phase-space-discretized kinetic equations of \eqref{con_Bolt_f} and \eqref{con_Bolt_g}.
Integration over an infinitesimal control volume $\delta V$ yields the form
\begin{equation}
\label{eq:FVM_form}
    \partial_t \{f_i,g_i\} (\bm{r})
    + \frac{1}{\delta V} \sum_{\sigma \in \Theta}  \{\mathcal{F}_i, \mathcal{G}_i \} (\sigma) 
    = S_{\{f_i,g_i\}} (\bm{r})
    \text{.} 
\end{equation}
$\mathcal{F}_i(\sigma)$ and $\mathcal{G}_i(\sigma)$ are the fluxes trough a discrete surface $\sigma$ of the Hull $\Theta$ of the control volume.
The populations have to be understood as volumetric averages of a cell, defined as
\begin{equation}
    \{f_i,g_i\} (\bm{r}) = \frac{1}{ \delta V}\int_{\delta V} \{f_i,g_i\} (\bm{r}) d\bm{r} 
    \text{,}
\end{equation}
and the source term $S_{\{f_i,g_i\}}$ corresponds to the volumetric average collision term 
\begin{equation}
    S_{\{f_i,g_i\}} (\bm{r}) = \frac{1}{ \delta V}\int_{\delta V} \Omega_{\{f_i,g_i\}} (\bm{r}) d\bm{r} 
    \text{.}
\end{equation}
Hence, also the derived moments are cell averaged values.
For simplicity, time integration is carried out using a first-order accurate Euler-forward scheme
\begin{equation}
    \partial_t \{f_i, g_i\} = \frac{\{f_i, g_i\}\left(t+\delta t\right) - \{f_i, g_i\}\left(t\right)}{\delta t} + \mathcal{O}(\delta t)
    \text{.}
\end{equation}
Finally, if the collision terms are applied, cf. \eqref{new_collision_f} and \eqref{new_collision_g}, the fully discretized hyperbolic partial differential equations read
\begin{widetext}
\begin{align}
\label{eq:finalDiscretePDE}
    f_i (\bm{r}, t+ \delta t) &= f_i(\bm{r},t) - \frac{\delta t}{\delta V} \sum_{\sigma \in \Theta}  \mathcal{F}_i (\sigma, t) 
    + \frac{\delta t}{\tau_1}\Bigl[ f_i^{eq}(\bm{r},t) - {f}_i(\bm{r},t) \Bigr] + \left(\frac{\delta t}{\tau_1} - \frac{\delta t}{\tau_2}\right) \Bigl[  {f}_i^{*}(\bm{r},t)  - {f}_i^{eq}(\bm{r},t) \Bigr] 
    \text{,} 
    \\
    g_i (\bm{r}, t+ \delta t) &= g_i(\bm{r},t) - \frac{\delta t}{\delta V} \sum_{\sigma \in \Theta}  \mathcal{G}_i (\sigma, t) 
    + \frac{\delta t}{\tau_1}\Bigl[ g_i^{eq}(\bm{r},t) - {g}_i(\bm{r},t) \Bigr] + \left(\frac{\delta t}{\tau_1} - \frac{\delta t}{\tau_2}\right) \Bigl[  {g}_i^{*}(\bm{r},t)  - g_i^{eq}(\bm{r},t) \Bigr] 
    \text{.} 
\end{align}
\end{widetext}

\paragraph{Flux reconstruction:}

The reconstruction of fluxes requires interpolation of the populations to the discrete interfaces of the finite volumes, whereas the particles' velocities remain global constants as a consequence of the static reference frame. 
The fluxes become 
\begin{equation}
        \{\mathcal{F}_i, \mathcal{G}_i \} (\sigma) = 
        \bm{v}_i \cdot \bm{n}(\sigma)  
        \{{f}_i, {g}_i \} (\sigma) 
        \delta A(\sigma)
    \text{,}
\end{equation}
where $\bm{n}$ is the surface normal vector and $\delta A$ the infinitesimal area of the discrete surface $\sigma$.
A nearest neighbor deformation (NND) interpolation scheme as proposed by Zhang et al. \cite{NND1, NND2} was employed in this manuscript as a trade off between robustness and accuracy.
On a uniform Cartesian grid, the scheme reads
\begin{align}
    f_{i,x - \delta x/2} = 
    \begin{cases}
    f_{i,x - \delta x} + \frac{1}{2} \phi \left( 
    f_{i,x - \delta x} - f_{i,x - 2 \delta x}
    , 
    f_{i,x} - f_{i,x - \delta x}
    \right)
         \\ \hfill \text{if } v_{i,x - \delta x/2} \geq 0 \text{,} 
    \\
    f_{i,x} - \frac{1}{2} \phi \left( 
    f_{i,x} - f_{i,x - \delta x}
    ,
    f_{i,x + \delta x} - f_{i,x}
    \right)
    \\ \hfill \text{if } v_{i,x - \delta x/2} < 0 \text{,} \\
    \end{cases}
\end{align}
here denoted for $f_i$ on the discrete surface located at \mbox{$\sigma=x-\delta x/2$}.
A simple minmod function was applied for $\phi(a,b)$ to limit the fluxes \cite{limiterRoe1986}.
Depending on the values $a$, $b$ of successive slopes, it can be written as
\begin{equation}
    minmod(a,b) = \begin{cases}
        0 & \text{if } ab \leq 0 \text{,} \\
        \frac{a}{|a|} min(|a|,|b|) & \text{if } ab > 0 \text{.}
    \end{cases}
\end{equation}

The AMR methodology and AMR related implementations are outlined next.

\subsection{Adaptive mesh refinement\label{sec:amr}}

\subsubsection{Refinement overview and employed strategy}

Adopting AMR introduces significant algorithmic changes and presents challenges for strict retainment of a scheme's conservative character as well as efficient parallel processing, the latter being vital in order to fully leverage the potential on modern-day distributed memory machines.
With the restriction to logically rectangular grids, three main approaches can be distinguished based on the refinement algorithm, the grid hierarchy, or the data structure used for the bookkeeping of cell conductivities \cite{OverviewAMRStrategies}:

(1) The first of its kind dynamic grid refinement was the block-structured (SAMR) approach pioneered by Berger and Oliger \cite{Berger_Patchbased1}.
The refinement is based on cells grouped into rectangular patches, hence this approach is sometimes also referred to as patch-based.
The data is organized in levels, with each level containing multiple patches. Hierarchical relations between patches are stored in a graph.
The numerical solution of a hyperbolic system of PDEs can be advanced independent of the level, patch, or position in the domain, with the same time-integration methodology as in the case of a single level uniform grid, and minimal algorithmic changes are needed in order to synchronize information between patches and levels \cite{BeckingsalePhD}. 

(2) In cell-based AMR \cite{KHOKHLOV1998}, each cell is refined individually rather than as a cluster.
A distinct feature is the bookkeeping of cell connectivity using a tree-based data structure, typically a quadtree in two and octree in three dimensions \cite{octtrees}. 
This approach allows for higher local refinement efficiencies, however, the data structure is associated with larger memory requirements and more irregular data referencing overhead as compared to SAMR.

(3) Alternately, the block-based approach \cite{PARAMESH} can be seen as a subclass of the patch-based approach with elements of the cell-based approach.
The refinement is restricted to predefined, rectangular, equally-sized blocks.
This a priori knowledge of possible refinement patches allows for an efficient implementation using a tree-based data structure, where the connectivity information is much simpler than in the cell-based approach and the quadtree/octree is less expensive.
This restriction on the size of blocks rather than individual cells, together with the knowledge of of the data tree in advance, naturally leads to the biggest potential for scalable high-performance AMR implementations \cite{OverviewAMRStrategies}.
On the other hand the refinement efficiency is lower than in both other approaches.

In this work, the block-structured AMR approach was employed.
An AMR library can facilitate the deployment of AMR applications and ensure good parallel efficiency and scalability through optimized data structures and load balancing algorithms.
For this reason, the code developed in this work was built with AMReX \cite{amrex1,amrex2, amrex3} in C++.
The foundations of SAMR can be found in the original works of \cite{Berger_Patchbased1, Berger_Patchbased2, Berger_Patchbased3, Berger_Patchbased4, BergerRigoutsos}.
The detailed theory is well summarized, e.g., in \cite{DeiterdingESAIM, DeiterdingPhD}.
A brief overview of the building blocks and the algorithmic changes associated to SAMR shall be given next together with highlighted implementations.

\subsubsection{Components of the block-structured AMR implementation}
 
\begin{figure*}[t]
\centering
\includegraphics[width=0.99\linewidth]{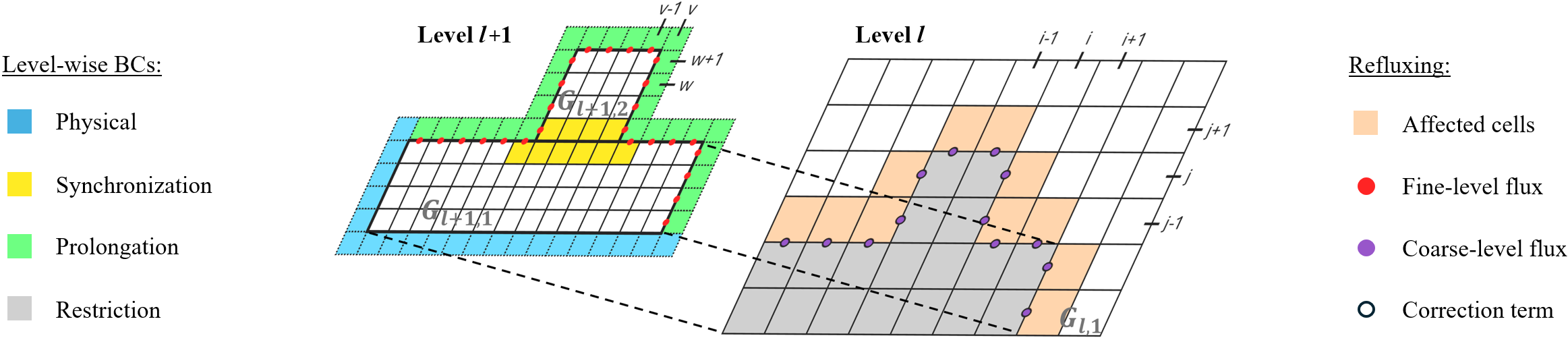}
\caption{Exemplary two-level grid layout with $r_l=2$ for the illustration of level-wise boundary conditions (BCs) and conservative refluxing.}
\label{fig:BCGrid}
\end{figure*}

In the following, without loss of generality, a two-dimensional (2D) uniform Cartesian domain is considered where a scalar time-explicit hyperbolic PDE with first-order in time discretization of the form,
\begin{equation}
\label{eq:generalUpdateScheme}
    Q_{i,j}^{t+\delta t} = Q_{i,j}^{t} - \frac{\delta t}{\delta V} \sum_{\sigma \in S}F_{i,j}^{t}(\sigma) + \delta t S_{i,j}^{t}, 
\end{equation}
is solved with the finite-volume method.
We assume a static computational domain with multiple levels $l$, where the resolution is given by the level-wise refinement factor $r_{l} \geq 2$ with
\begin{equation}
\label{eq:refinementRatio}
    \frac{\delta x_{l}}{\delta x_{l+1}} = \frac{\delta y_{l}}{\delta y_{l+1}} = r_{l} 
    \text{.}
\end{equation}
Each grid is denoted by $G_{l,p}$ with $l \in \{0,L\}$ and patches $p \in \{0,P\}$.
Equation \eqref{eq:generalUpdateScheme} is solved uniformly on every patch.
An exemplary layout of patches on levels $l$, $l+1$ is depicted in Fig.~\ref{fig:BCGrid}.
For simplicity, we denote the cell coordinates of the coarser level with ($i$, $j$) and the finer level with ($v$, $w$), respectively.

\paragraph{Level-wise boundary conditions:}

A halo around $G_{l,p}$ with $h \geq 1$ cells on each patch needs to be filled with boundary conditions (BCs) before eq. \ref{eq:generalUpdateScheme} can be updated.
As depicted for patches $G_{l+1,0}$ and $G_{l+1,1}$ in Fig.~\ref{fig:BCGrid}, there are three types of BCs for level $l+1$.
(1) Physical BCs: Ghost cells extending the physical domain need to be filled; periodic or von Neumann BCs were used in this work.
(2) Synchronization: Ghost cells overlapping a cell of another patch of the same level are populated with the corresponding values of the other patch.
(3) Prolongation:
Ghost cells for which the above does not apply are filled with interpolated values from underlying coarse cells.
The prolongation scheme in this work consisted of a bilinear interpolation 
in space and time, which reads
\begin{multline}
    \label{eq:prolongation}
    Q_{v,w}^{l+1} 
    = (1-a_1)(1-a_2) Q_{i-1,j-1}^{l}
    + a_1(1-a_2) Q_{i,j-1}^{l} \\
    + (1-a_1)a_2 Q_{i-1,j}^{l} 
    + a_1a_2Q_{i,j}^{l}
    \text{,}
\end{multline}
with coefficients $a_1$ and $a_2$, for the example of cell ($v$, $w$) in Fig.~\ref{fig:BCGrid}.

Additionally, another type of BC has to be supplied to level $l$, if a level $l+1$ exists.
(4) Restriction:
Coarse cells which are covered by a patch of fine cells (excluding the halo) need to be filled by an average value of the fine cells.
This is done in order to populate the coarse cells with the most accurate values possible in case the fine-level patch is destroyed in the future, since the fundamental assumption in AMR tells that
a fine-level cell contains a more precise solution than the next coarser level cell.
For the exemplary case of cell ($i-1$, $j-1$) in Fig.~\ref{fig:BCGrid}, the conservative averaging scheme applied in this work reads
\begin{equation}
\label{eq:restriction}
    Q_{i-1,j-1}^{l} 
    = \frac{1}{r_l^2} 
    \sum_{\xi=0}^{r_{l}-1} 
    \sum_{\zeta=0}^{r_{l}-1}
    Q_{v-r_l+\xi, w+r_l+\zeta}^{l+1}
    \text{.}
\end{equation}

\paragraph{Conservative refluxing:}

Application of the above BCs alone can not strictly guarantee global conservation, as there might exist a flux mismatch on the borders of a fine-level patch and the underlying coarse cell interfaces. 
For example in Fig.~\ref{fig:BCGrid}, the sum of both fine-level fluxes at ($v-\frac{1}{2}$, $w$) and ($v-\frac{1}{2}$, $w+1$) indicated in red on level $l+1$ do not necessarily match the flux at ($i-\frac{1}{2}$, $j$) indicated in purple on level $l$.
The correct update for cell ($i$, $j$) reads
\begin{multline}
\label{eq:correctLevelUpdateScheme}
    Q_{i,j}^{t+\delta t_l,l} 
    = Q_{i,j}^{t,l} 
    - \frac{\delta t_l}{\delta x_l} 
    (
    F_{(x),i+\frac{1}{2},j}^{t,l} 
    - 
    \chi
    )
    \\
    - \frac{\delta t_l}{\delta y_l} 
    (
    F_{(y),i,j+\frac{1}{2}}^{t,l} 
    - F_{(y),i,j-\frac{1}{2}}^{t,l}
    ) 
    + \delta t_l S_{i,j}^{t,l} 
    \text{,}
\end{multline}
with 
\begin{equation}
\label{correctFluxTerm}
    \chi = 
    \frac{1}{r_l^2}
    \sum_{\kappa=0}^{r_{l}-1} 
    \sum_{\xi=0}^{r_{l}-1}
    F_{(x),v-\frac{1}{2}, w+\xi}^{t+\kappa \delta t_{l+1},l+1}
    \text{,}
\end{equation}
whereas $\chi$ would read differently, with
\begin{equation}
\label{wrongFluxTerm}
    \chi = F_{(x),i-\frac{1}{2},j}^{t,l}
    \text{,}
\end{equation}
in case the update scheme in eq. \ref{eq:generalUpdateScheme} was uniformly applied to $G_{l,1}$. 
A conservative refluxing routine can correct this mismatch in the form of a correction pass, such that the original scheme (eq. \ref{eq:generalUpdateScheme}) does not have to be modified locally but can still be applied uniformly.
Algorithmically this is done by using flux registers which are initialized with coarse-level fluxes of opposite sign on level $l$, cf. negative eq. \ref{wrongFluxTerm}. 
The level $l+1$ fluxes are then aggregated over space and time, as in eq. \ref{correctFluxTerm}, subtracted from the flux register, and the correction term is applied to level $l$.

\paragraph{Subcycling in time:}

\begin{figure}[t]
\centering
\includegraphics[width=0.85\linewidth]{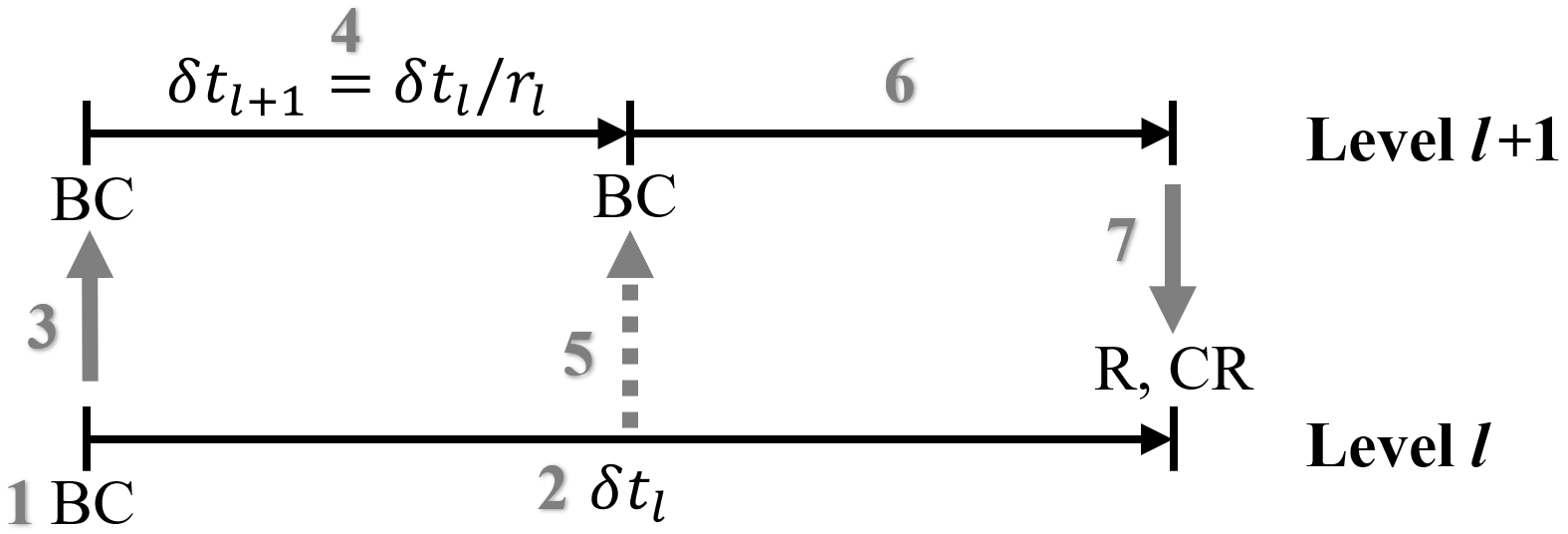}
\caption{Illustration of the updates and level-wise information exchanges in the subcycling approach.
BC refers to the boundary conditions of type physical, synchronization and prolongation, whereas R refers to restriction and CR to conservative refluxing.
The horizontal axis marks time and the vertical axis displays the level $l$ and $l+1$ with an exemplary refinement ratio of $r_l = 2$. 
The order of operations is indicated with gray numbers, interpolations in space with gray arrows, 
and interpolation in space and time with gray dotted arrows, respectively. 
In case a next finer level $l+2$ exists, the same recursive scheme is applied to step number four and six.
}
\label{fig:Subcycling}
\end{figure}

A global time step over all levels is inefficient as the restriction on the Courant--Friedrichs--Lewy (CFL) number is determined by the smallest cell in the domain.
For this reason, and to ensure a constant CFL number across all levels, the subcycling approach for time integration was applied in this work.
This is done by binding the level time steps to the refinement ratio trough acoustic scaling as 
\begin{equation}
    \frac{\delta t_{l}}{\delta t_{l+1}} = r_{l} 
    \text{,}
\end{equation}
together with a recursive time integration routine.
Note that acoustic scaling was adopted in this work due to the greater numerical efficiency compared to diffusive scaling where $\delta t \propto \delta x^2$.
First, the time step size is computed from the most restrictive CFL condition over all cells on all levels in the entire domain and is then extrapolated to the level $l=0$ where the subcycling routine starts. 
The recursive routine,
schematically depicted in Fig.~\ref{fig:Subcycling} for a level $l$ and $l+1$, is realized with the following steps:
\begin{itemize}
    \item (1) The BCs of level $l$ are filled and (2) level $l$ is updated by a time step with $\delta  t_l$.
    \item (3) The BCs of level $l+1$ are filled and (4) level $l+1$ is updated by a time step with $\delta t_{l+1} = \delta  t_l / r_l$.
    \item (5) The BCs of level $l+1$ are filled and (6) level $l+1$ is updated one more time for the case of $r_l=2$ illustrated here, which marks a repetition of steps 3 and 4 until the simulation time $t_{l+1}$ reaches $t_l$.
    \item (7) Restriction is done at $t_{l+1} = t_l$, followed by conservative refluxing, i.e. application of the flux correction term from the flux registers filled by steps 2, 4 and 6.
\end{itemize}
Note that this recursive scheme is also applied to all level $l+1$ steps, continuing as long as a next finer level exists.
Also note for step 5 that the prolongation BCs on level $l+1$ are filled by linear interpolation in time additional to the bilinear interpolation in space, as no values are available on level $l$ at this simulation time.
This is a peculiarity of the formulation as a boundary value problem taken here, as opposed to an initial value problem where typically $r_l$ times the amount of halo layers $h$ would be required. 

\paragraph{Adaptivity and grid generation:}

So far we considered a statically refined multilevel grid.
The adaptive component is given by a regridding routine, which is applied every $n$-th level time step $\delta t_l$ and is responsible for creating, recreating or deleting patches.
It consists of a few steps.
(1) First, refinement sensors are applied to determine cells which require refinement.
In this work, macroscopic variables such as density $\rho$, velocity $\bm{u}$, pressure $p$, Temperature $T$, as well as vorticity $\omega$ and
Gradients thereof, computed as central differences, were used as refinement sensors.
All cells for which refinement sensors exceed their thresholds are subsequently tagged.
(2) A clustering algorithm groups the tagged cells and forms rectangular patches.
Buffer layers around groups are applied with a size such that a local flow state is not advected out of the refined area until the next regridding time.
A refinement efficiency $\eta$ can be imposed which determines the result of more neatly fitted patches for increasing computational cost.
Typical clustering algorithms are \cite{Berger_Patchbased4, BergerRigoutsos}.
(3) The adapted grid layout is integrated into the data hierarchy.
(4) Cells in a newly created patch or in new region of a modified patch have to be filled with initial conditions (ICs). 
This is done by prolongation (e.g. eq. \ref{eq:prolongation}), as in the case for the filling of BCs.

\paragraph{Considerations for a parallel implementation:}
\label{par:parallelAMR}
Some aspects have to be considered for a parallel SAMR implementation on distributed memory machines.
On one hand 
a few changes have to be respected compared to the components discussed in the above paragraphs concerning communications between patches and levels.
Filling of physical BCs and restriction are processor-local operations, whereas prolongation can be made local.
Synchronization and refluxing can require data exchanges.
Further, the outcome of the clustering algorithm generally differs when executed globally versus locally on each processor.
On the other hand
parallel processing efficiency becomes a central challenge, as the grid layout consists of a complex and time-dependent hierarchy, where patches can move, appear and be destroyed, which can require a lot of irregular memory referencing and redistribution of work load.
Different parallelization strategies exist with considerations such as level-wise versus global domain decompositions 
and arguments such as minimization of patch splitting and communication overhead between patches and levels hold.
Sophisticated load balancing algorithms control the distribution of patches between processors.
Usually a Knapsack algorithm or space-filling curves (Morton, Hilbert)
are employed to connect cells as optimization strategy.
The parallel processing approach, e.g. MPI+OMP/GPU, clustering, load balancing strategy and parallel scaling, can be found in the works of AMReX \cite{amrex1, amrex2, amrex3}.

\section{Validation\label{sec:validation}} 

In the following, few aspects of the solver such as strict retainment of conservation properties or correct recovery of dispersion and dissipation rates are validated both with and without AMR, before evaluating more one-dimensional (1D) and 2D Euler and NSF level benchmarks as well as computational cost in section~\ref{sec:Results}.

In all simulations, unless otherwise stated, the parameters for the presented results were set to 
\mbox{$\mu = \num{5e-5}$}, 
\mbox{$\gamma=1.4$}, 
\mbox{Pr $=1$},
the ideal gas law was used with the gas constant set to $R=1$.
The D2Q16 velocity set and a third-order Grad expansion was applied for Euler level problems, whereas the D2Q25 with a fourth-order Grad expansion was used for full NSF flows.
1D tests were run as pseudo-1D with periodic BCs in the pseudo direction, otherwise von Neumann BCs were applied.
Initial conditions were supplied by means of equilibrium populations.
The time step was estimated with
\begin{equation}
    \delta t = \frac{\min({\delta x, \delta y})}{\max(|{\bm{u} \pm c_s}|)} \text{CFL}
    \text{,}
\end{equation}
where the CFL number was uniformly set to $0.01$ in order to also stably simulate higher Ma numbers.
For cases using AMR, $2$ refinement levels with a constant refinement ratio of $r_l = 2$ across all levels were applied.
Two buffer cells around each patch and a regridding time of $2$ were used for all simulations, as well as a grid efficiency of $\eta = 0.7$ in 1D and $\eta = 0.98$ in 2D, respectively.
The resolution of cases without AMR was either matched to the resolution of the base level $l=0$ or to the peak level $l=2$ of an AMR case, in order to make meaningful comparisons.
Physical quantities are reported as non-dimensional values whenever no units are mentioned.

\begin{figure*}[t]
\centering
    \includegraphics[width=.495\linewidth]{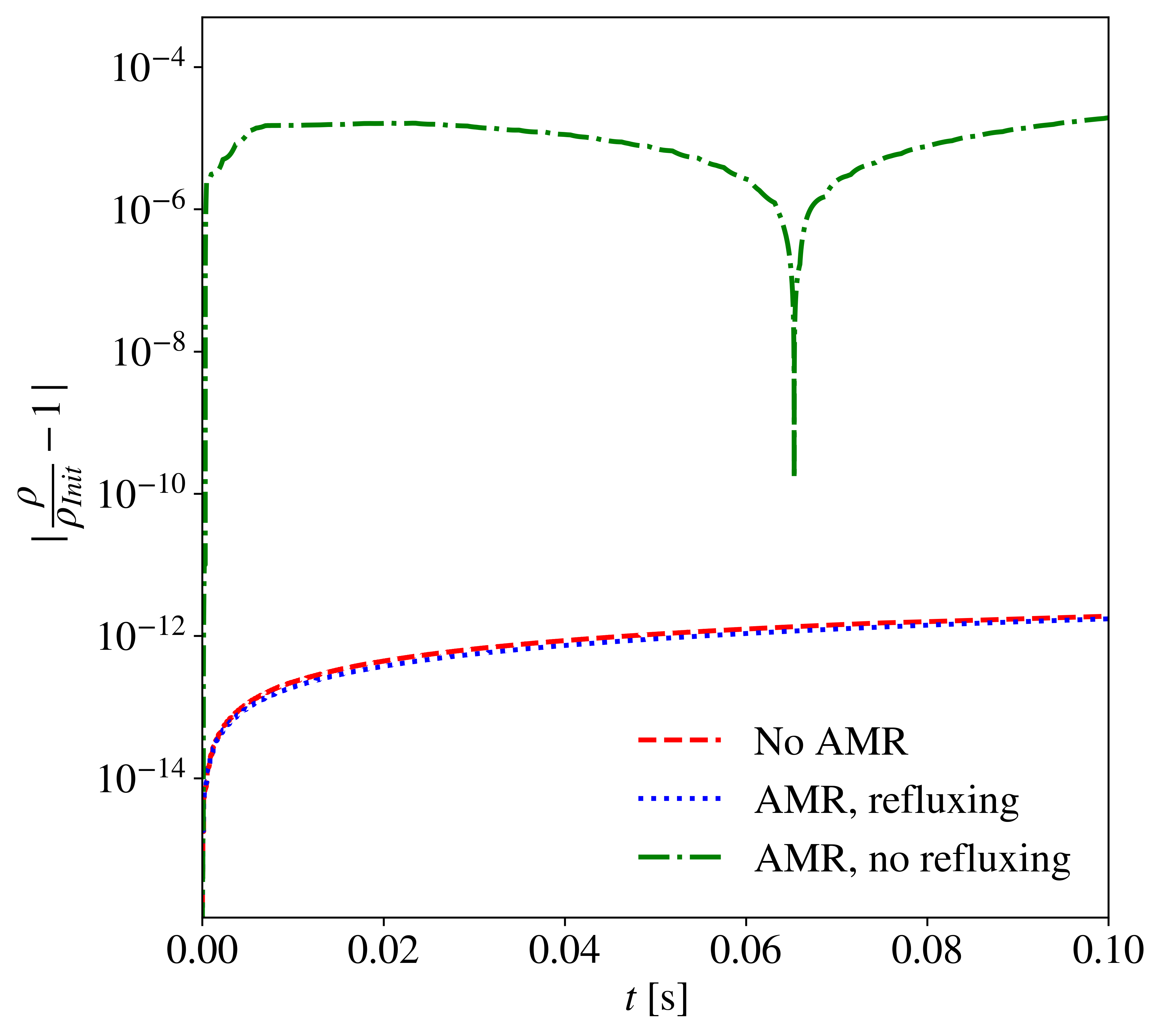}
\hfill
    \includegraphics[width=.495\linewidth]{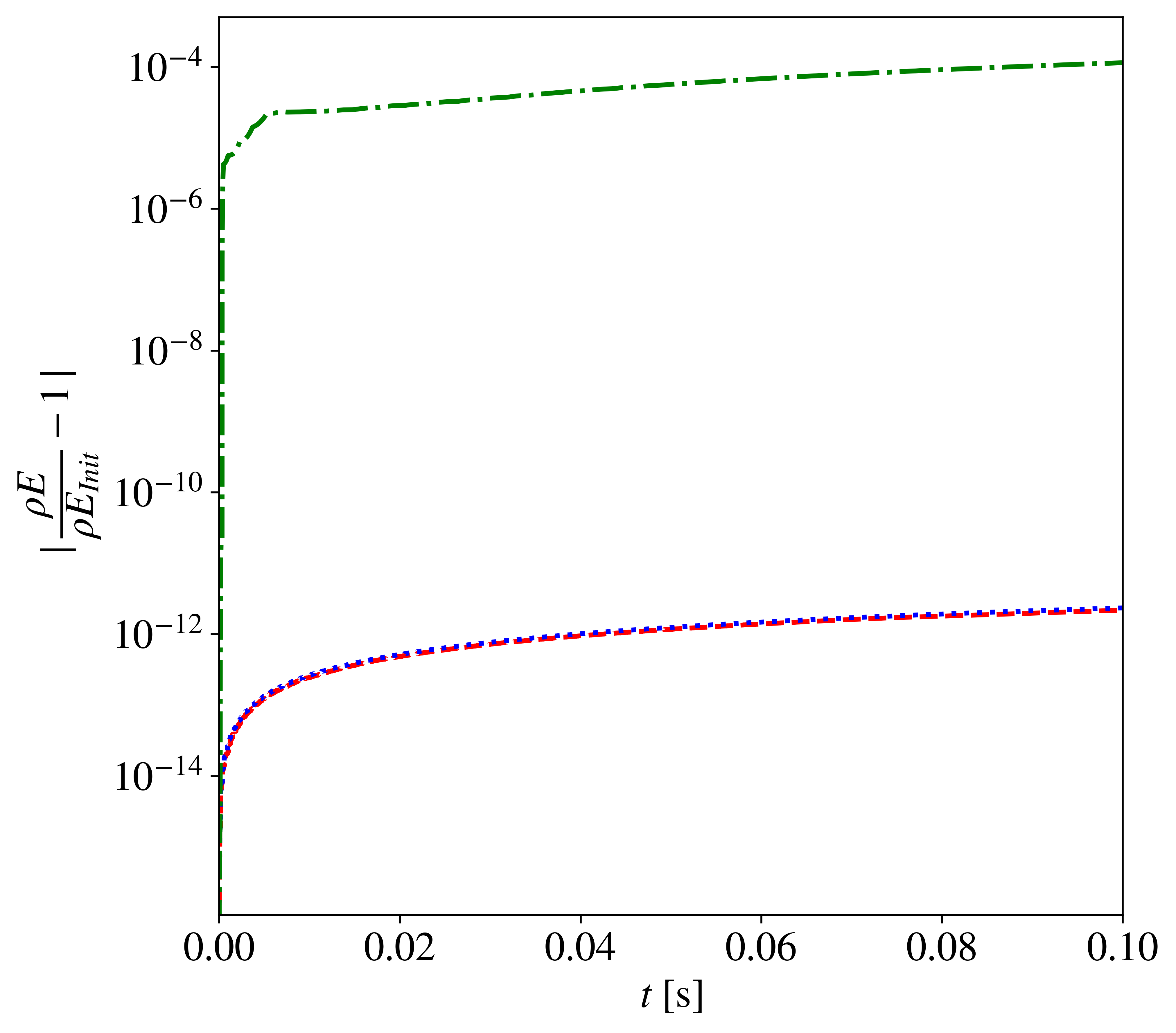}
\caption{Error in conservation of mass (left) and total energy (right) of the system over time.}
\label{fig:ConservationError}
\end{figure*}

\subsection{Conservation properties}

\begin{figure}[b]
\centering
    \includegraphics[width=.99\linewidth]{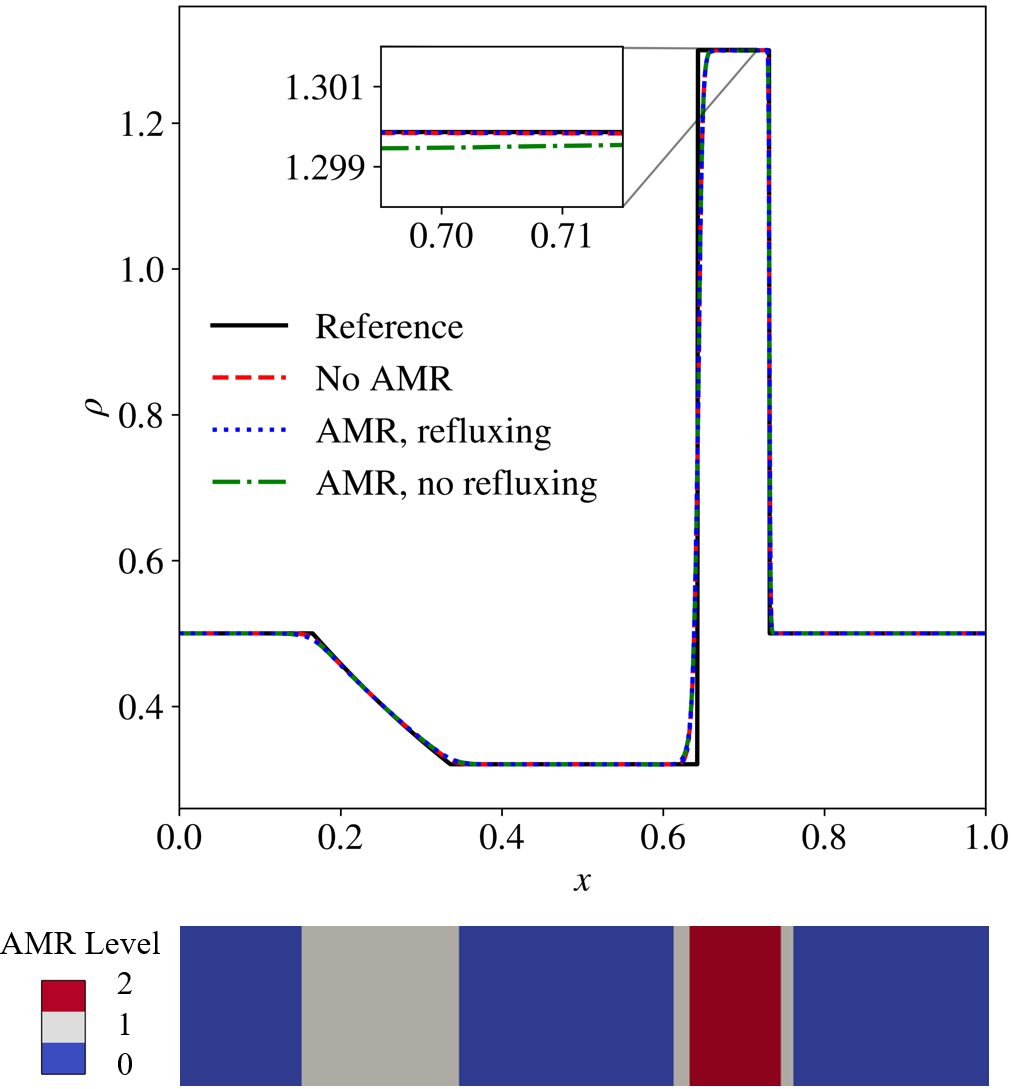}
\caption{
Solution for the density of the shock tube at $t=0.1s$. 
The AMR levels are indicated at the bottom.
}
\label{fig:ConservationSod1}
\end{figure}

First, the conservation properties of the numerical scheme and the AMR methodology are addressed.
This was tested using a 1D shock tube with the initial conditions
\begin{equation}
  (\rho, p, u_x) =
    \begin{cases}
      (0.5, 4, 0), & x \leq 0.5,\\
      (0.5, 0.5, 0), & x > 0.5  
      ,
    \end{cases}       
\end{equation}
in an extended fully periodic domain, $x \in [-0.5,1.5]$. 
Three simulations were run: 
Without AMR on a single level (red line), 
with AMR and the application of conservative refluxing (dashed blue line),
and with AMR, but without the application of conservative refluxing (green dotted line).
Absolute density and the density gradient were used as refinement sensors for both AMR cases in order to target the discontinuities and the plateau with the highest density value.
A base grid level $l=0$ resolution of $\delta x_0 = L_x/256$ was applied to the AMR cases such that the resolution on level $l=2$, i.e. $\delta x_2 = r_0  r_1 \delta x_0 = 4 \delta x_0$, matches the "no AMR" case with $\delta x = L_x/1024$.

The results for the cropped region $x \in [0, 1]$ at $t=0.1$s are depicted in Fig.~\ref{fig:ConservationSod1}, together with a snapshot of the refinement at the bottom.
All simulations capture the reference solution with good agreement,
however, a zoom into the plateau with highest density reveals that the AMR simulation without refluxing does not exactly recover the value of the reference solution, while the other two simulations match perfectly. 
This is a sign of lost mass and non-conservation of the AMR methodology without refluxing.
The same observation can be made in Fig.~\ref{fig:ConservationError}, where the error in conservation of mass and total energy of the system over time is depicted.
It can be seen that the AMR simulation without refluxing produces a significant error, as it is not strictly conservative.
In contrast, AMR with proper conservative refluxing yields the same conservation error as the simulation without AMR, which is in the order of the rounding machine precision ($2^{-53} \approx \num{1.11e-16}$ for double precision) on each cell interface.
Therefore, it is apparent that a correct application of refluxing restores the strict conservation properties of the finite-volume scheme.
Consequently, AMR with refluxing is employed for all results discussed in the remainder of this manuscript.

\subsection{Dispersion and dissipation of hydrodynamic modes}

Next, the ability of the numerical scheme and the AMR methodology to correctly recover the dispersion and dissipation rates of hydrodynamic eigen-modes, i.e. shear, normal and entropic, is addressed.
For this, standard tests were used as, e.g., given in \cite{Ji24, Saadat2019, hosseini2020compressibility, ProbingDoubleDist2024}.
Note that all results reported here correspond to converged simulations in space and time. 
All setups were run in a fully periodic pseudo-1D domain $x \in [0,1]$
and the peak resolution of the AMR cases, i.e. $\delta x_2$ on level $l=2$, were set to match the resolution of the simulations without AMR.

\subsubsection{Dispersion of the normal mode: Speed of sound}

\begin{figure}[b]
\centering
        \includegraphics[width=.99\linewidth]{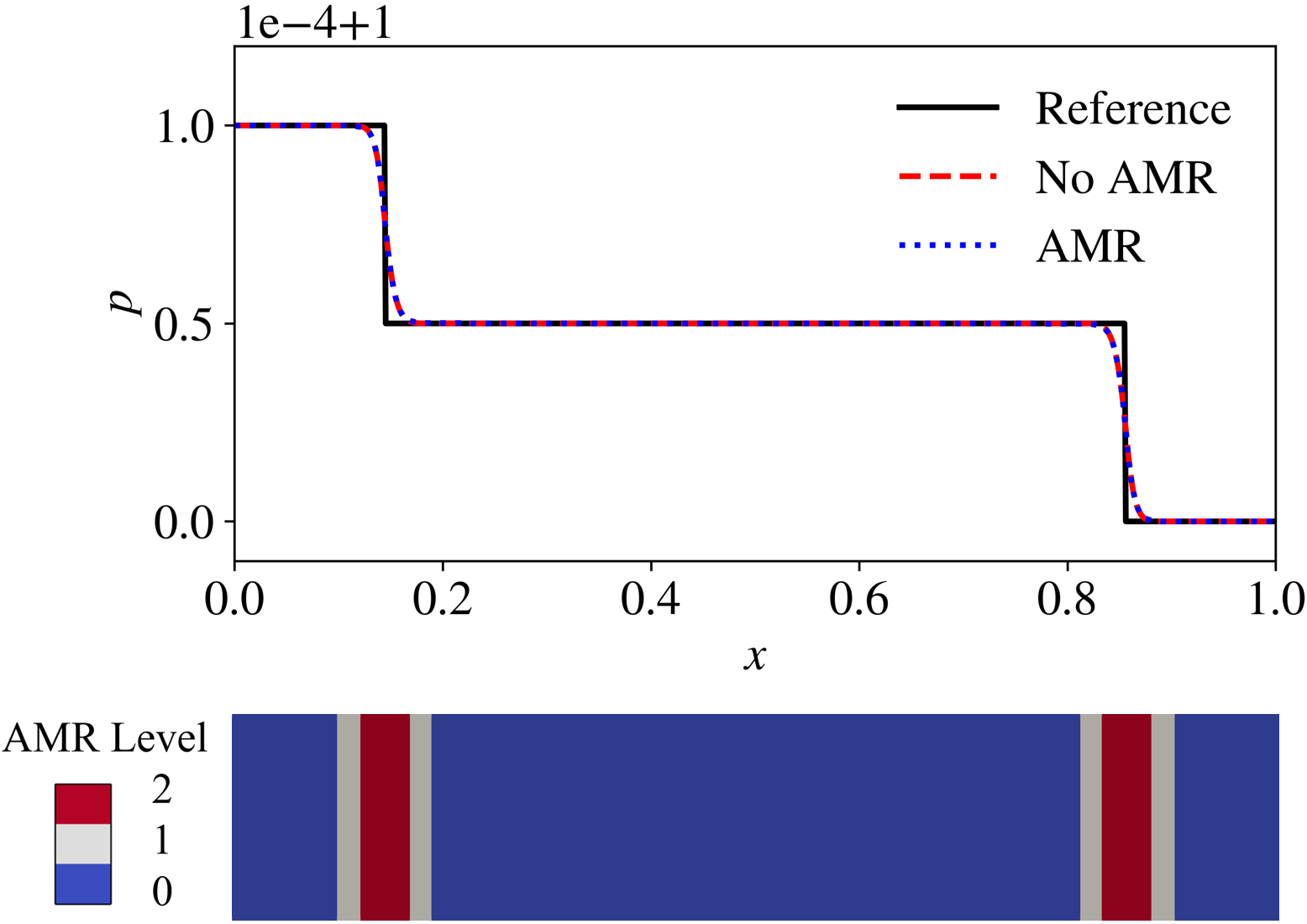} 
\caption{
Refinement based on the gradient of pressure for the dispersion tests: 
Here, $t= 0.3$s, $T_0 = 1$, $\gamma = 1.4$. 
}
\label{fig:SpeedSoundTagging}
\end{figure}

The temperature dependence of the speed of sound was investigated by means of a freely traveling pressure front.
To that end, the domain with a stagnant fluid \(\bm{u}_0 = 0\) was divided into two regions with
\begin{equation}
    p =
    \begin{cases}
        1 + 10^{-4}, & 0 \leq x \leq 0.5,\\
        1, & 0.5 < x \leq 1,
    \end{cases}       
\end{equation} 
and a uniform temperature $T=T_0$ was applied in both regions.
Three different specific heat ratios, namely $\gamma = 5/3$ for monoatomic,  $\gamma = 7/5$ for diatomic, and $\gamma = 8/6$ for triatomic gases were assessed for various temperatures.
The speed of sound was computed by tracking the shock front over time and comparing it with the analytical value of $c_s = \sqrt{\gamma T}$. 
For the AMR cases, refinement was set to target the pressure discontinuities by applying the pressure gradient sensor.
A snapshot of the pressure field and the resulting refinement levels in the case of the AMR simulations can be found in Fig.~\ref{fig:SpeedSoundTagging}.
A resolution of $\delta x = L_x/800$ was applied for the case without AMR and as the peak resolution in AMR ($\delta x_2$ on level $l=2$), i.e. $\delta x_0 = L_x/200$.

\begin{figure}[t]
\centering
    \includegraphics[width=.99\linewidth]{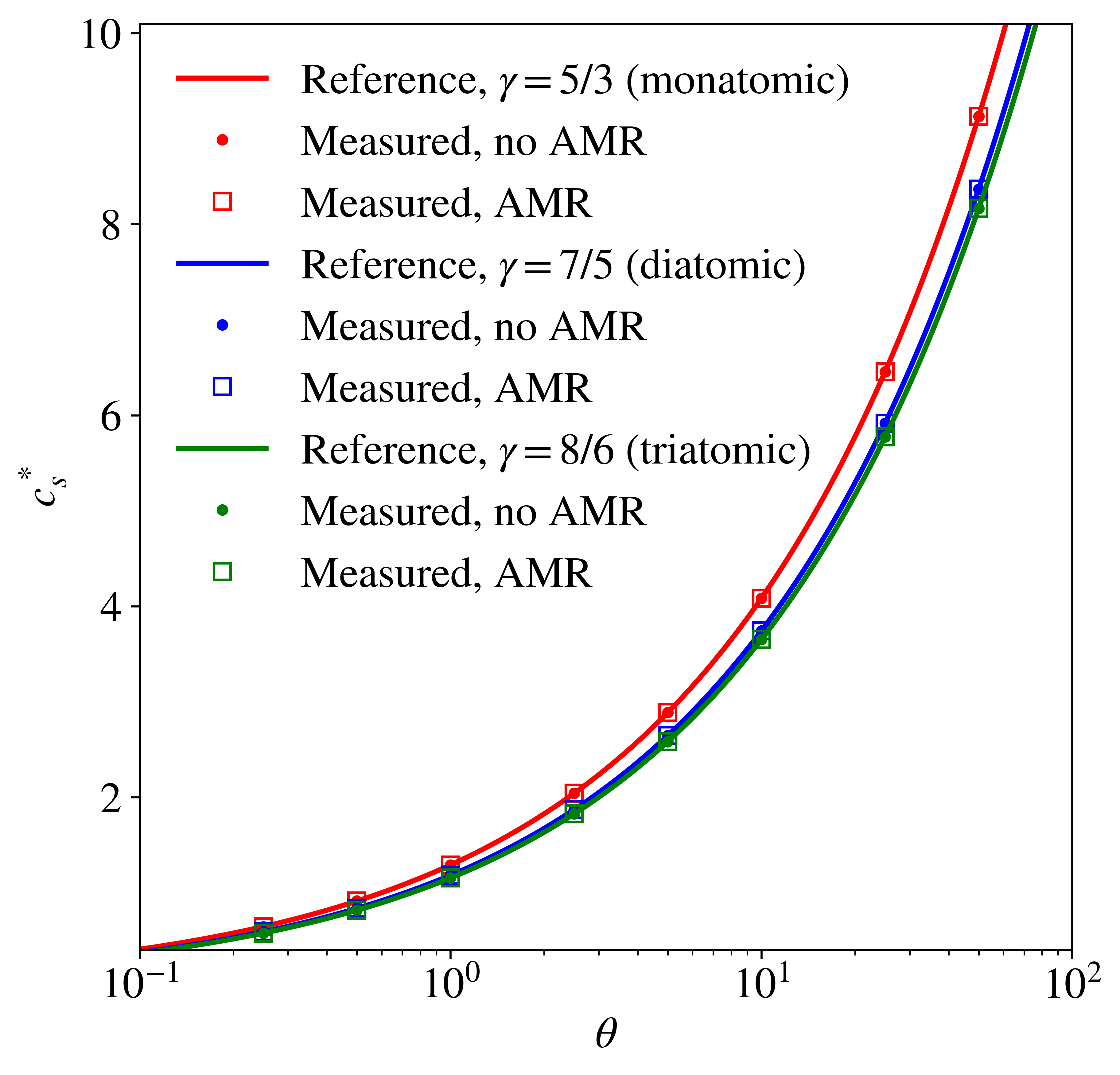}
\caption{
Normalized speed of sound $c_s^* = \sqrt{\gamma \theta}$ at various normalized temperatures $\theta = T/T_{ref}$.
}
\label{fig:SpeedSound}
\end{figure}

Fig.~~\ref{fig:SpeedSound} demonstrates that the present model can correctly capture the speed of sound for different specific heat ratios and a wide temperature range spanning three orders of magnitude. 
Further, it can be seen that the AMR methodology does not distort the outcome.

\subsubsection{Dissipation of the shear mode: Shear viscosity}
\label{subsubsec:dissipationShear}

\begin{figure*}[t]
\centering
    \includegraphics[width=.495\linewidth]{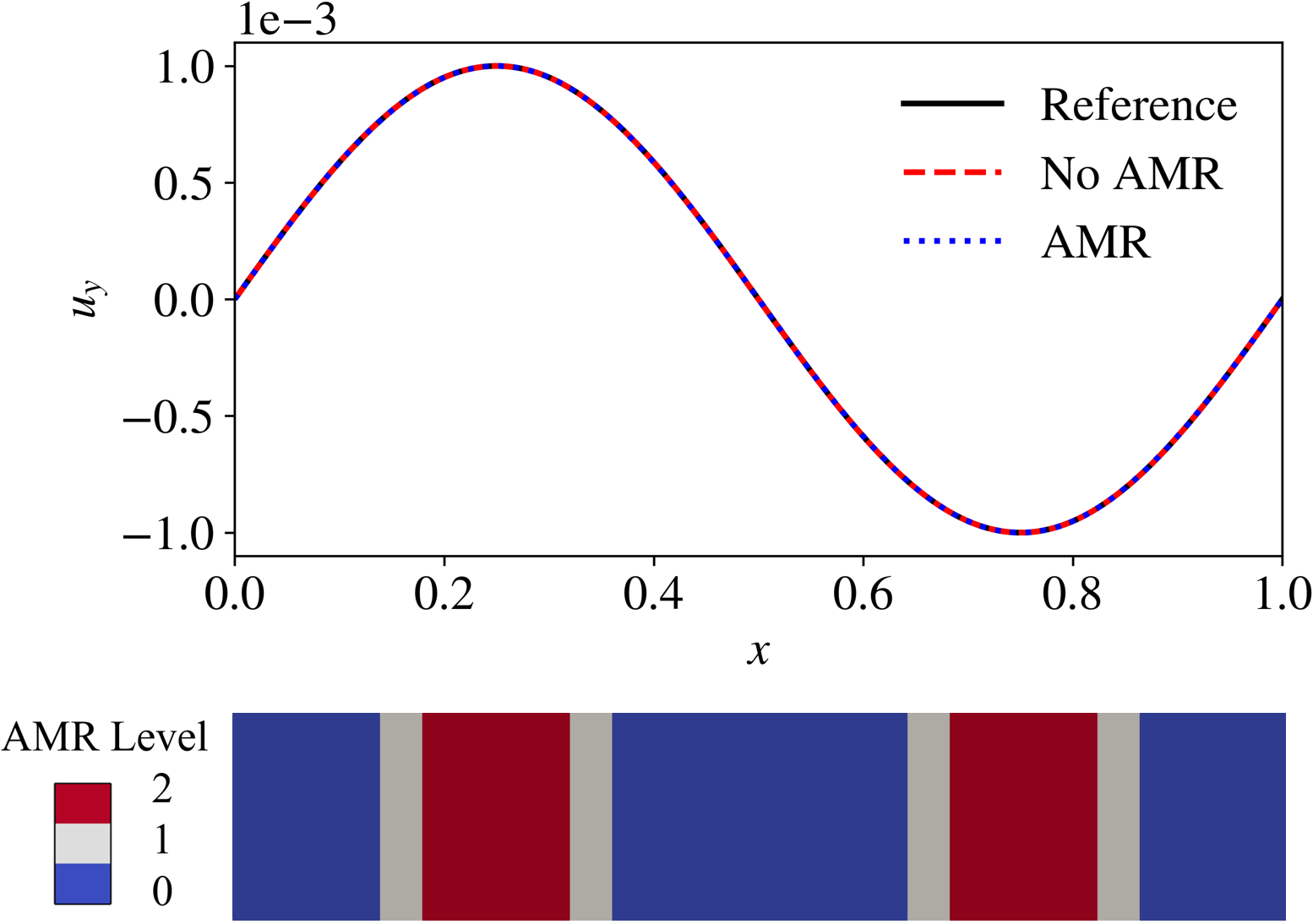}
\hfill
    \includegraphics[width=.495\linewidth]{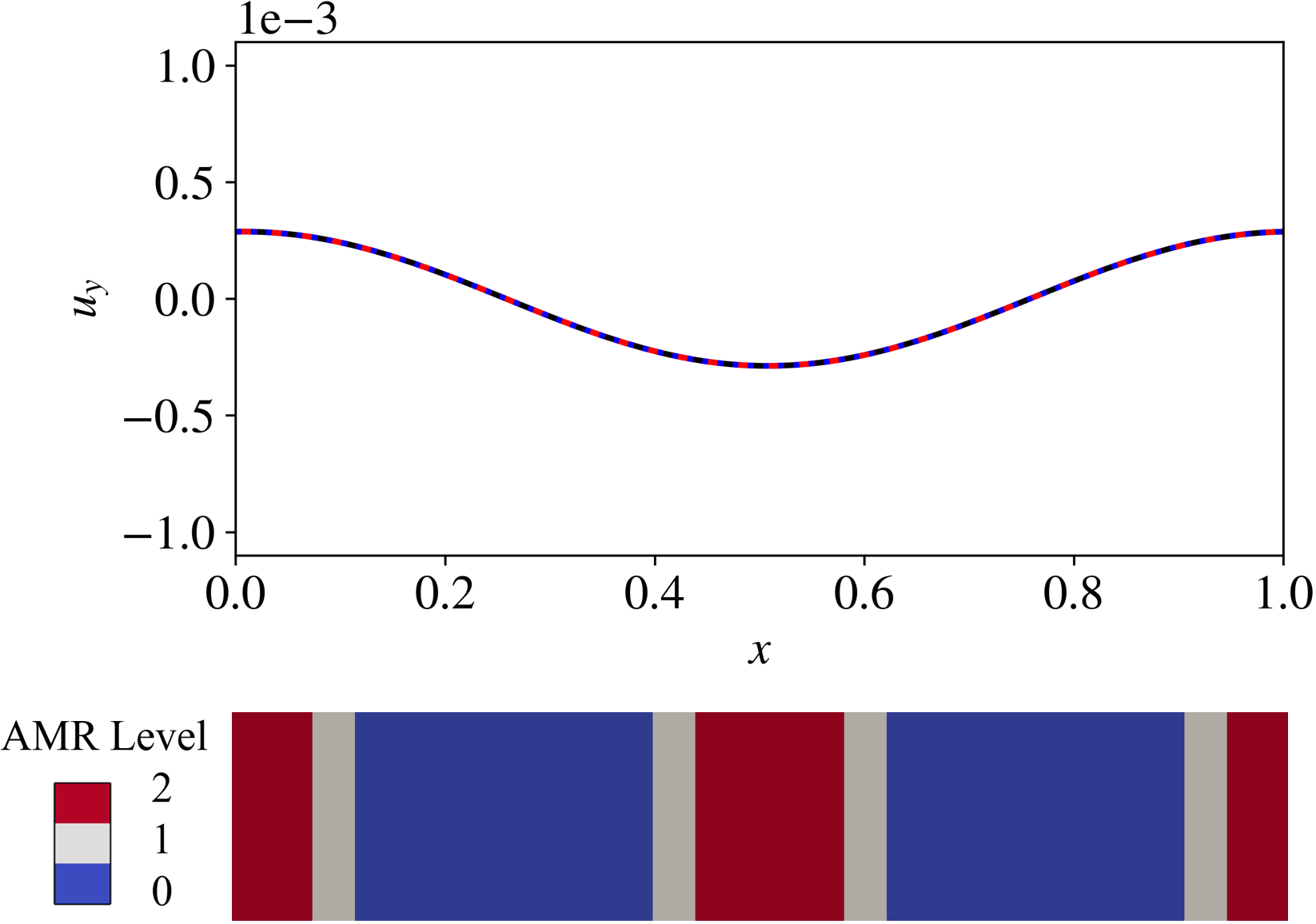}
\caption{Refinement relative to the global maximum for the dissipation test cases: Here, tagging of $|u_y|/|u_y^{max}|$ for the shear viscosity tests at Ma $=0.2$, $t=0.0$s (left) and $t=3.2$s (right).}
\label{fig:PhysicalTestsTagging}
\end{figure*}

\begin{figure*}[t]
\centering
    \includegraphics[width=.495\linewidth]{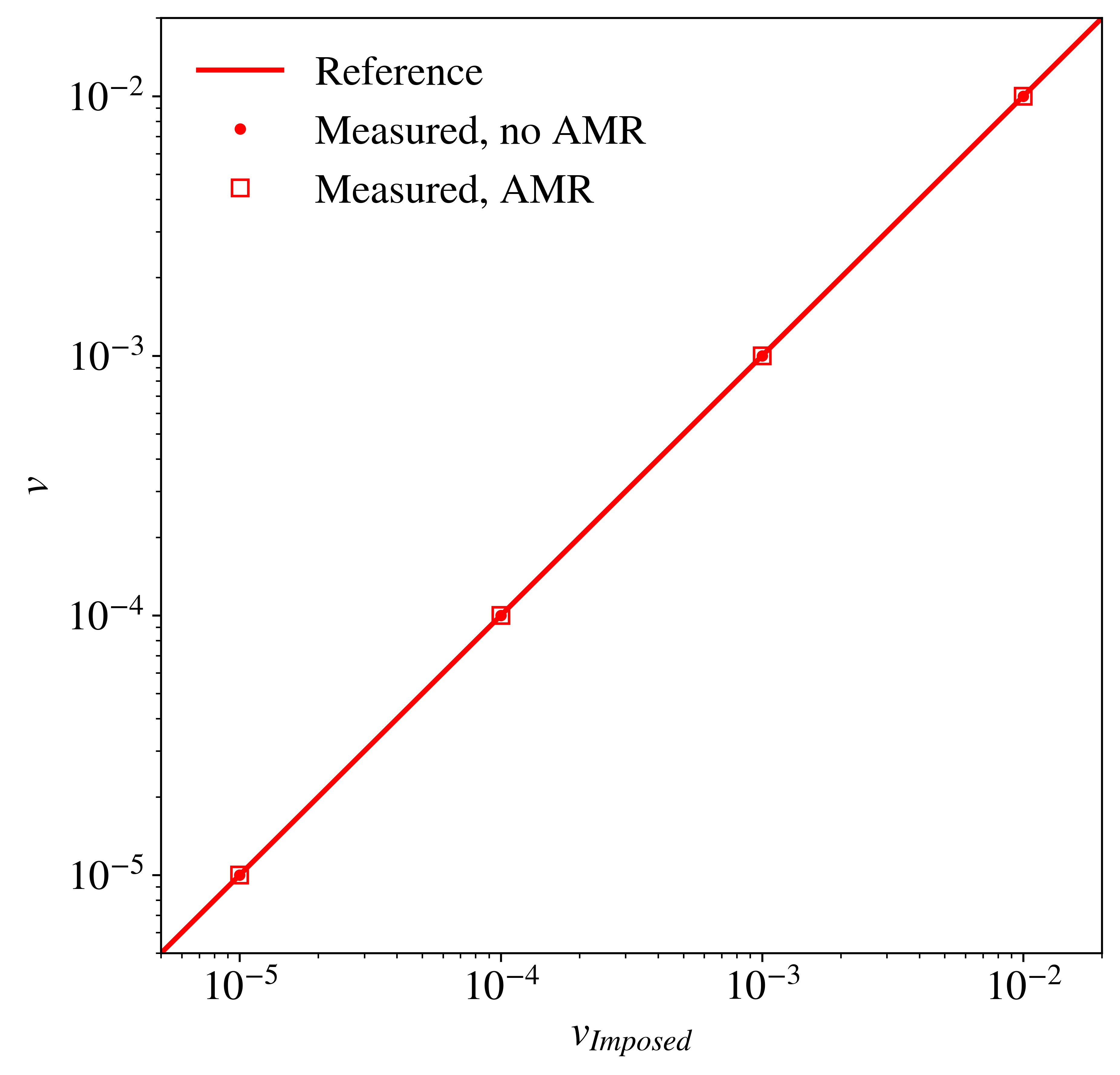}
\hfill
    \includegraphics[width=.495\linewidth]{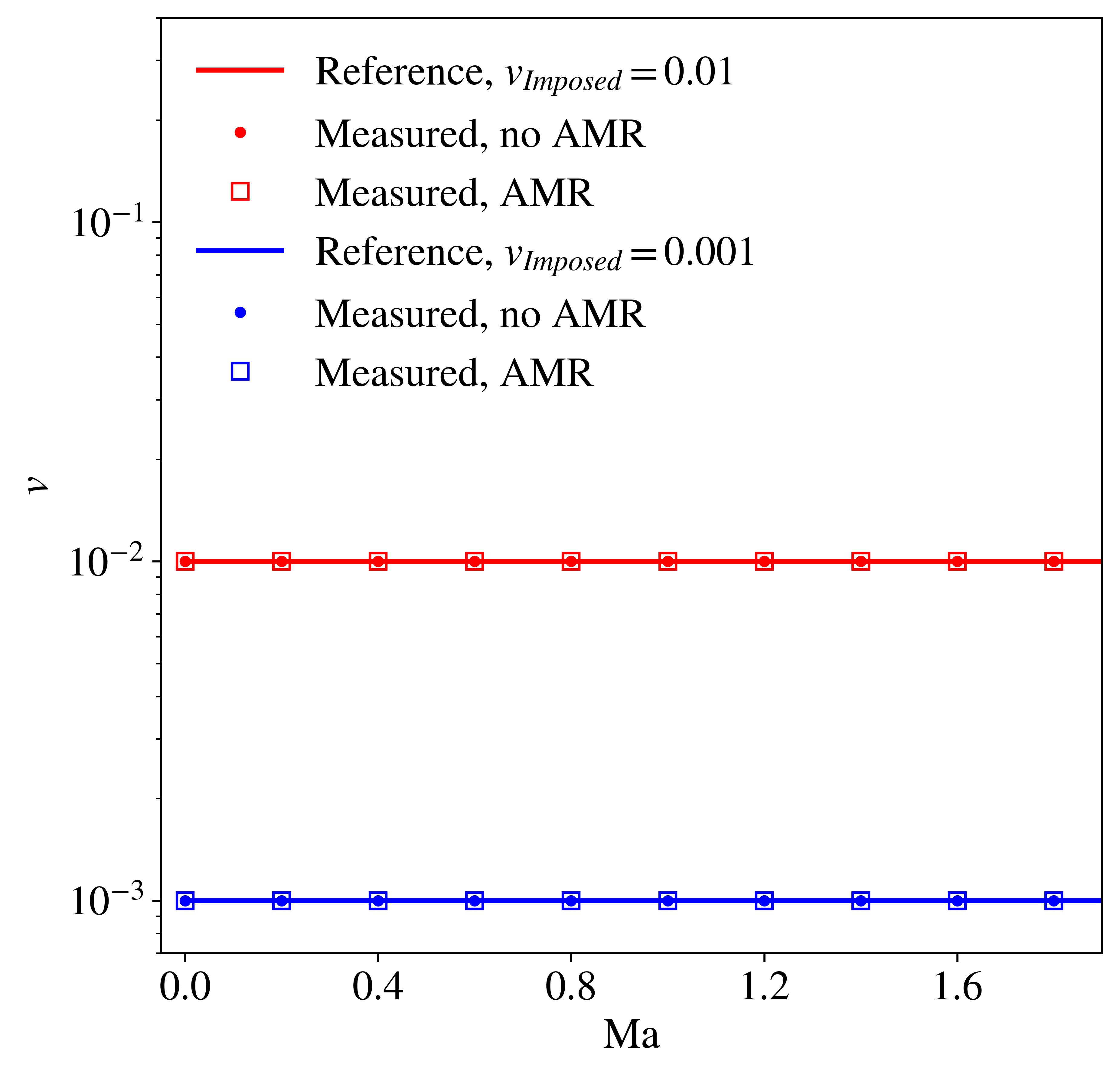}
\caption{Dissipation of the shear mode at Ma $=0$ (left) and at various Ma numbers (right).}
\label{fig:PhysicalTests_Shear}
\end{figure*}

The kinematic shear viscosity $\nu$ was investigated by simulating a plane shear wave with a small sinusoidal perturbation superimposed to the initial velocity field.
The initial conditions read
\begin{equation}
    \rho = \rho_0, \hfill T = T_0, \hfill u_x = u_0, \hfill u_y = A \sin\left(2\pi x/L_x\right)
    ,
\end{equation}
where the initial density and temperature were set to $ \left( \rho _0,T_0 \right) = \left( 1, 1 \right)$. 
The perturbation amplitude was $A=\num{1e-3}$
and $u_0$, which is derived from the Mach number as $u_0 = \mathrm{Ma} \sqrt{\gamma T_0}$, was varied in subsequent simulations.
The evolution of the maximum velocity $u_y^{max}$ in the domain was tracked over time and an exponential function was fitted to it.
The decay rate, i.e. the shear viscosity $\nu$, was then obtained via
\begin{equation}
    u_y^{\rm max}(t) \propto \exp{\left(-\frac{4\pi^2\nu}{L_x^2}t\right)}
    \text{.}
\end{equation}
The refinement criteria for the AMR cases were set to target the peak values of $|u_y|$ by applying 
this sensor relative the maximum value in the domain.
This was done to ensure that $u_y^{\rm max}(t)$ is tagged independent of the actual dissipation rate and simulation time. 
A resolution of $\delta x = L_x/1600$ was applied for the case without AMR and as the peak resolution in AMR ($\delta x_2$ on level $l=2$), i.e. $\delta x_0 = L_x/400$.
A snapshot of the $u_y$-velocity field and the applied tagging relative to the maximum in the domain can be found in Fig.~\ref{fig:PhysicalTestsTagging} for two different simulation times.

\begin{figure*}[t]
\centering
    \includegraphics[width=.495\linewidth]{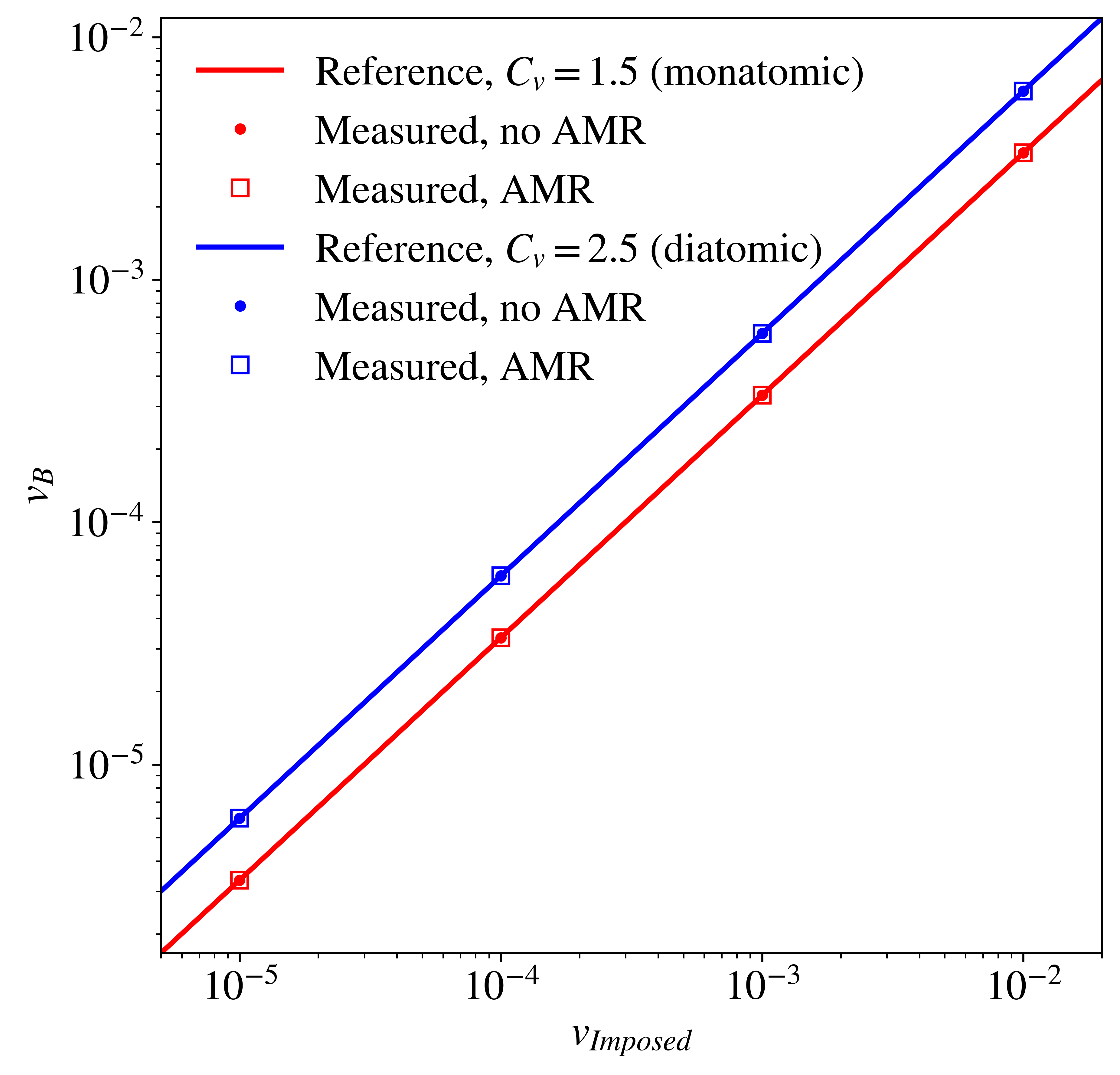}
\hfill
    \includegraphics[width=.495\linewidth]{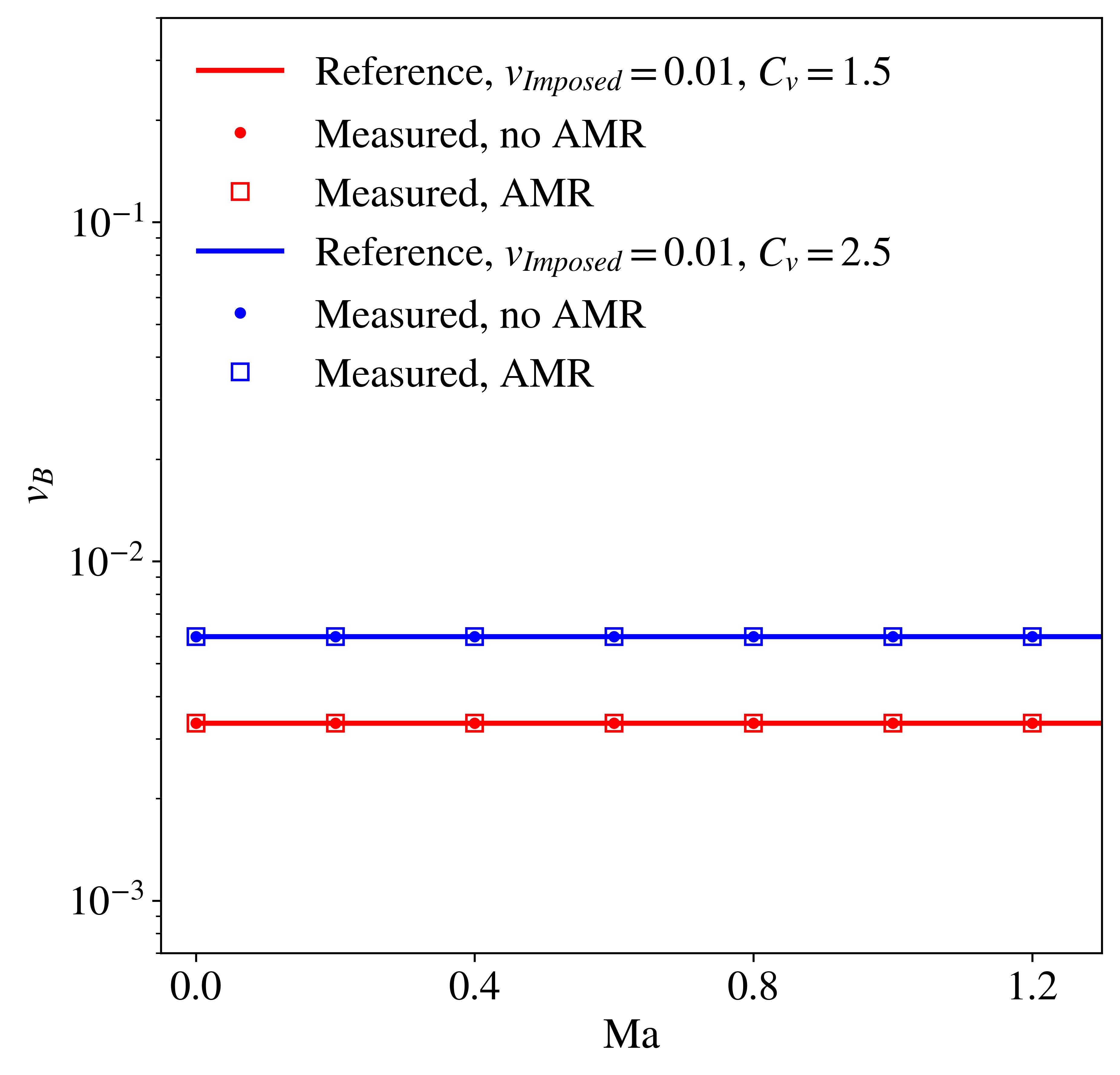}
\caption{Dissipation of the bulk mode for different specific heat capacities at Ma $=0$ (left) and at various Ma numbers (right).}
\label{fig:PhysicalTests_Bulk}
\end{figure*}

The obtained results are depicted in Fig.~\ref{fig:PhysicalTests_Shear}. 
The measured viscosities are in excellent agreement with the imposed values
and recover the expression from the Chapman-Enskog analysis in eq. \ref{CE:visc.shear} for several Ma numbers.
This holds both for the model on a uniform grid as well as with AMR.
After \mbox{Ma $\approx 1.8$}, the model gradually starts to show deviations due to the expected Galilean invariance issues as a consequence of the static reference frame.

\subsubsection{Dissipation of the bulk mode: Bulk viscosity}
\label{subsubsec:dissipationBulk}

The Bulk viscosity $\nu_B$ was investigated via the decay rate of sound waves in the linear regime.
For this purpose, a small perturbation with initial amplitude $A=\num{1e-3}$ was superimposed to the density field.
The flow was initialized as
\begin{equation}
    \rho = \rho_0 + A \sin\left(2\pi x/L_x\right), T = T_0, u_x = u_0, u_y = 0,
\end{equation}
with ($\rho_0$, $T_0$) = ($1$, $1$) and $u_0$ derived from the imposed Mach number.  
The perturbation energy $E'(t) = u_x^2+u_y^2-u_0^2+c_s^2\rho '^2$ of the whole domain, with $\rho' = \rho - \rho_0$, was tracked over time and the exponential function, 
\begin{equation}
    E'(t) \propto \exp{\left(-\frac{4\pi^2\nu_e}{L_x^2}t\right)},
\end{equation}
as defined by \cite{dellar2001bulk}, was fitted to it.
The recovered decay rate is the effective viscosity $\nu_e$, i.e. the combination of shear and bulk viscosities as
\begin{equation}
    \nu_e = \frac{4}{3} \nu + \nu_B
    \text{.}
\end{equation}
The refinement sensor for the AMR cases consisted of monitoring $|\rho'|$ relative the the maximum value $|\rho'^{max}|$ in the domain, similar to the setup for the shear viscosity tests, c.f. Fig.~\ref{fig:PhysicalTestsTagging}.
Also the resolutions were the same as in section~\ref{subsubsec:dissipationShear}.

The obtained results are depicted in Fig.~\ref{fig:PhysicalTests_Bulk} for several imposed viscosities and specific heat capacities at various Ma numbers.
It can be seen that the measured bulk viscosities accurately recover the expression from the Chapman-Enskog analysis in eq. \ref{CE:visc.bluk}, both for the model with and without AMR.
The expected limits of Galilean invariance are gradually encountered after Ma $\approx 1.2$.

 \subsubsection{Dissipation of the entropic mode: Thermal diffusivity}
\label{subsubsec:dissipationThermal}

\begin{figure*}[t]
\centering
    \includegraphics[width=.495\linewidth]{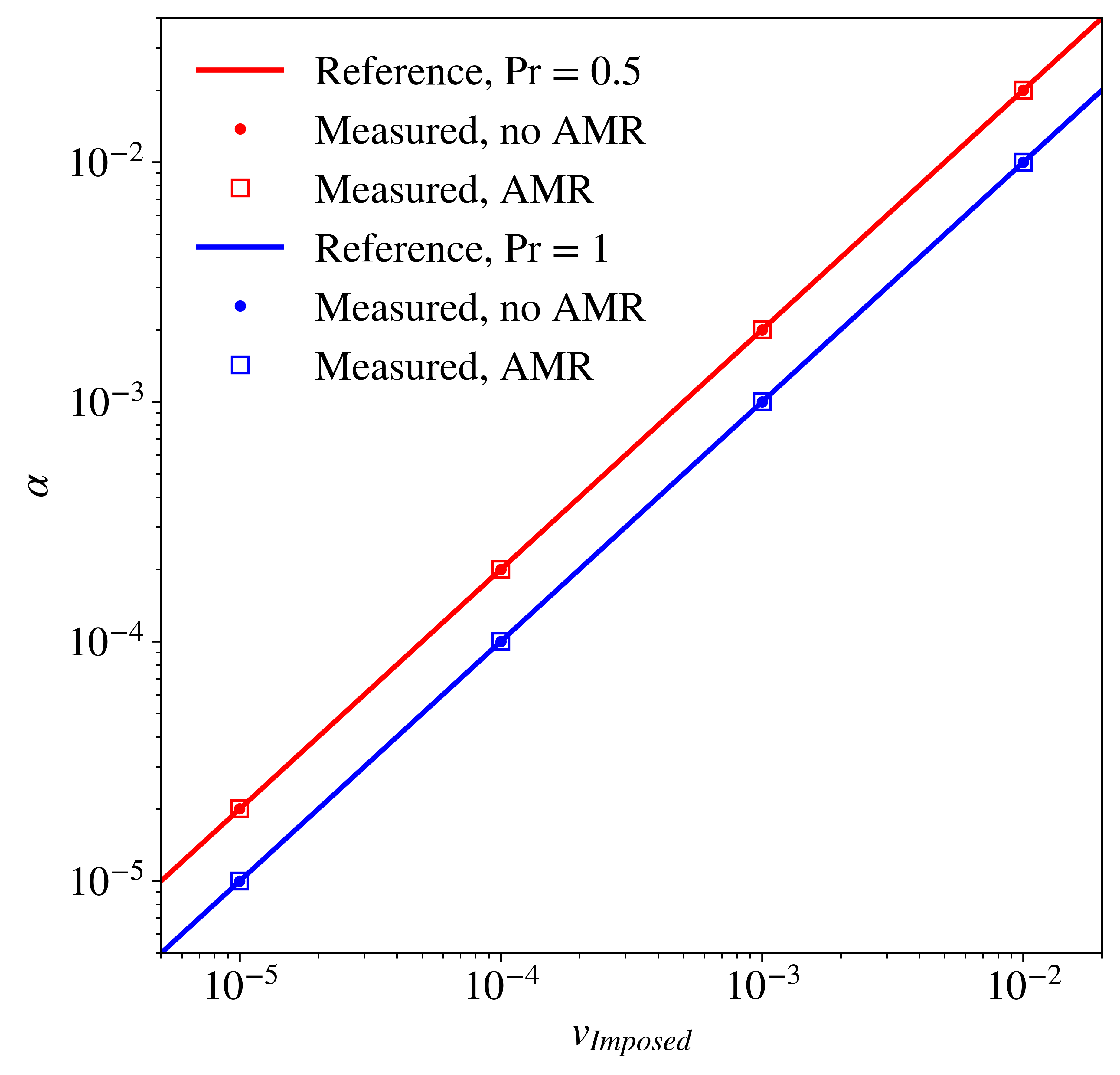}
\hfill
    \includegraphics[width=.495\linewidth]{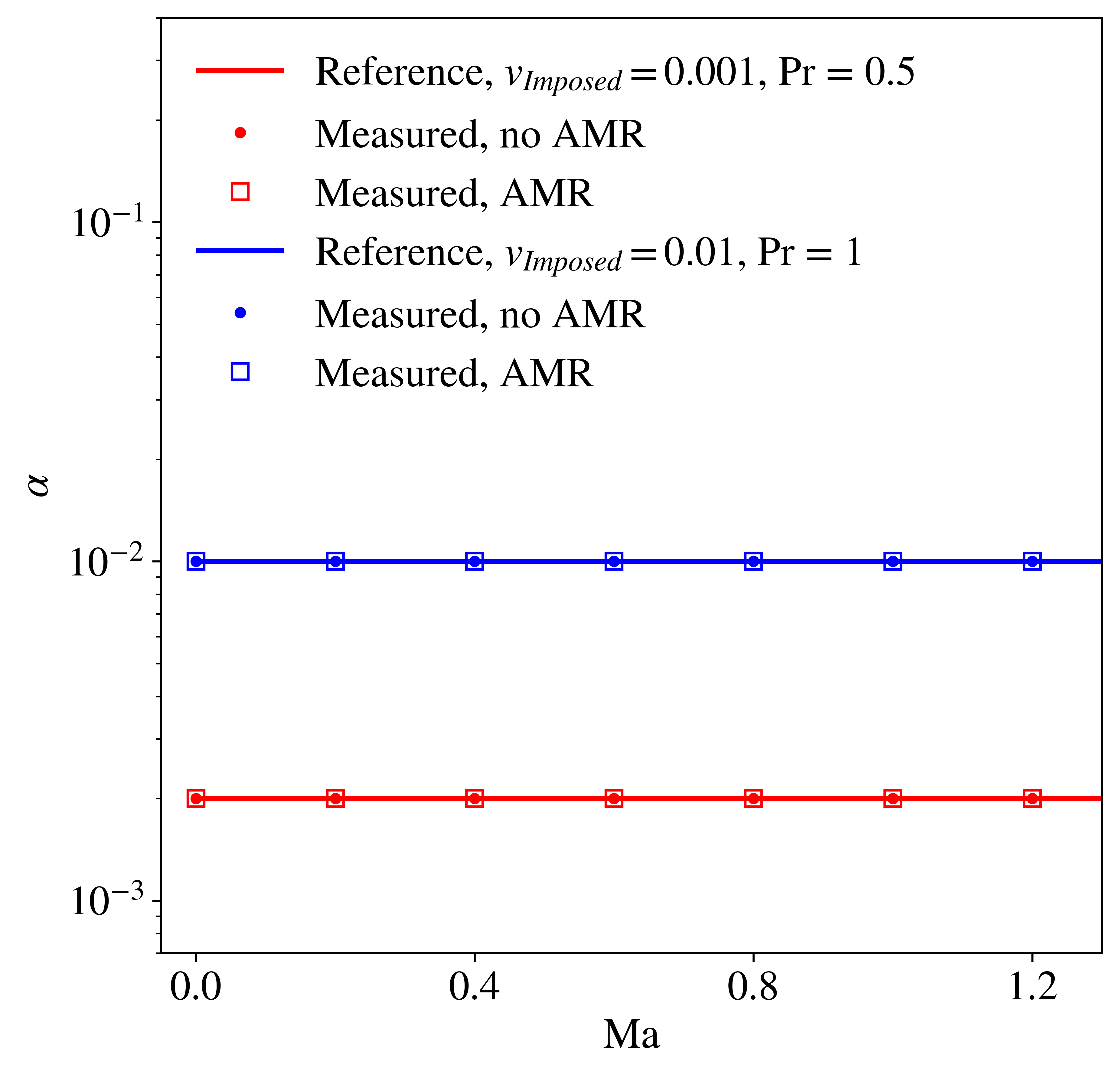}
\caption{Dissipation of the entropic mode for different Prandtl numbers at Ma $=0$ (left) and at various Ma numbers (right).}
\label{fig:PhysicalTests_Thermal}
\end{figure*}

A different type of perturbation was introduced in the ICs of the system to assess the thermal diffusivity $\alpha$. These are
\begin{equation}
    \rho = \rho_0 + A \sin\left(2\pi x/L_x\right), T = \rho_0 T_0/\rho, u_x = u_0, u_y = 0,
\end{equation}
with $ \left( \rho _0,T_0 \right) = \left( 1, 1 \right) $ and a perturbation amplitude of $A=\num{1e-3}$.
The thermal diffusivity was measured by fitting the exponential function,
\begin{equation}
    T'(t) \propto \exp{\left(-\frac{4\pi^2\alpha}{L_x^2}t\right)},
\end{equation}
to the temporal evolution of the maximum temperature difference $T' = T - T_0 $ in the domain.
The refinement sensor was set to monitor $|T'|/|T'^{max}|$ and the same resolutions as for both cases before were applied.

From Fig.~\ref{fig:PhysicalTests_Thermal} it becomes evident that the model also performs well in terms of thermal dissipation rates, including different Prandtl numbers, as the exact expression (eq. \ref{CE:visc.thermal}) is accurately recovered.
Also here, the AMR methodology does not distort the outcome.
Again, the expected limits of Galilean invariance are gradually encountered after Ma $\approx 1.2$.

\section{Results\label{sec:Results}}
After careful validation of the physical properties captured by the solver, a few Euler and NSF level benchmarks of compressible flows are presented next followed by an evaluation of computational efficiency.
The same general considerations and parameters as in section~\ref{sec:validation} were applied in this section. 

\subsection{One-dimensional cases}

\subsubsection{Sod shock tube}

\begin{figure*}[t]
\centering
\begin{minipage}[t]{.5\linewidth}
    \centering
        \includegraphics[width=\linewidth, valign=t]{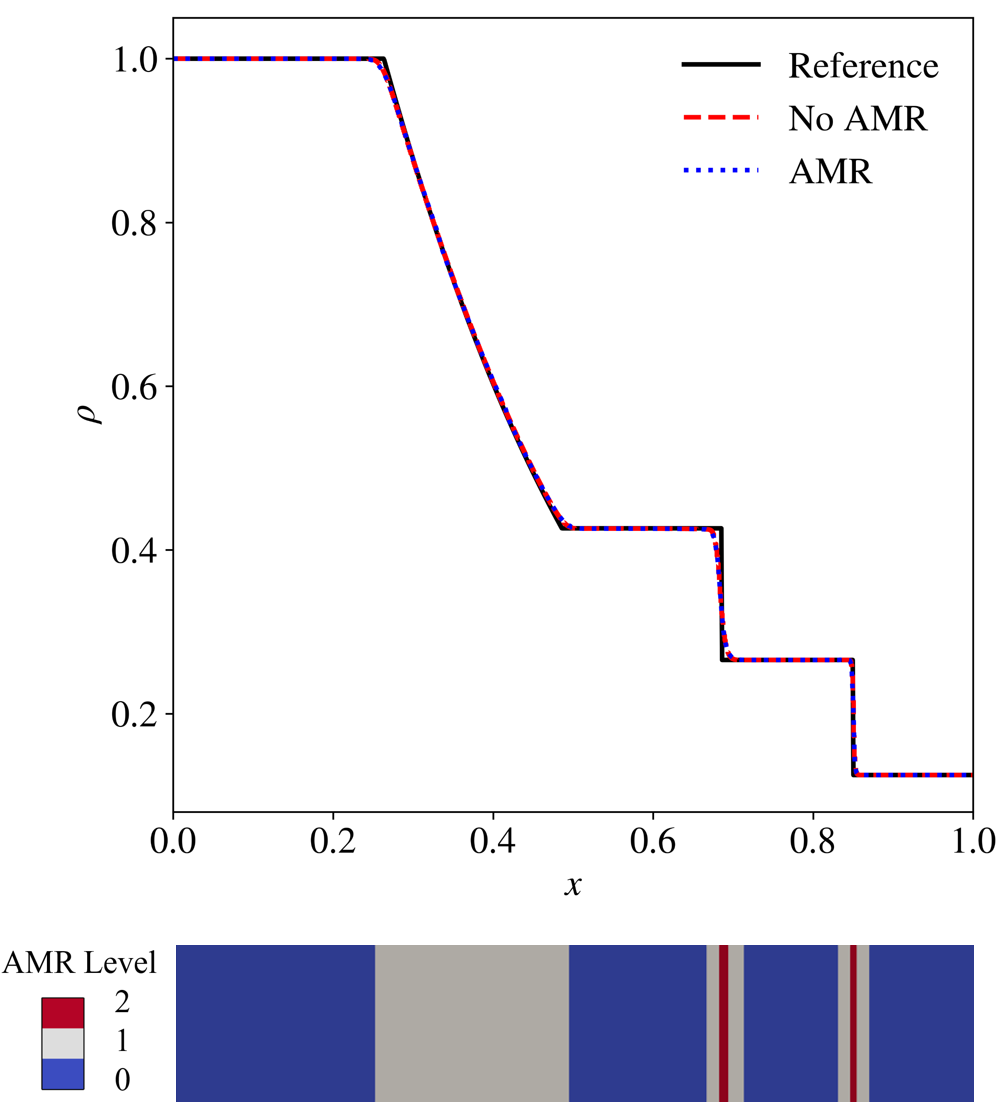}
\\
\hfill
    \begin{minipage}[b]{.90\linewidth}
        \centering
        \vspace{15pt} 
            \caption{Results of the Sod shock tube \cite{Sodtube} at $t=0.2$s for the density (top left), pressure (top right), and velocity (bottom right) compared to the reference solutions. The AMR levels are indicated at the bottom of the results for the density.} 
            \label{fig:SodResults}
    \end{minipage}%
\end{minipage}
\hfill
\begin{minipage}[t]{.48\linewidth}
    \centering
        \includegraphics[width=\linewidth, valign=t]{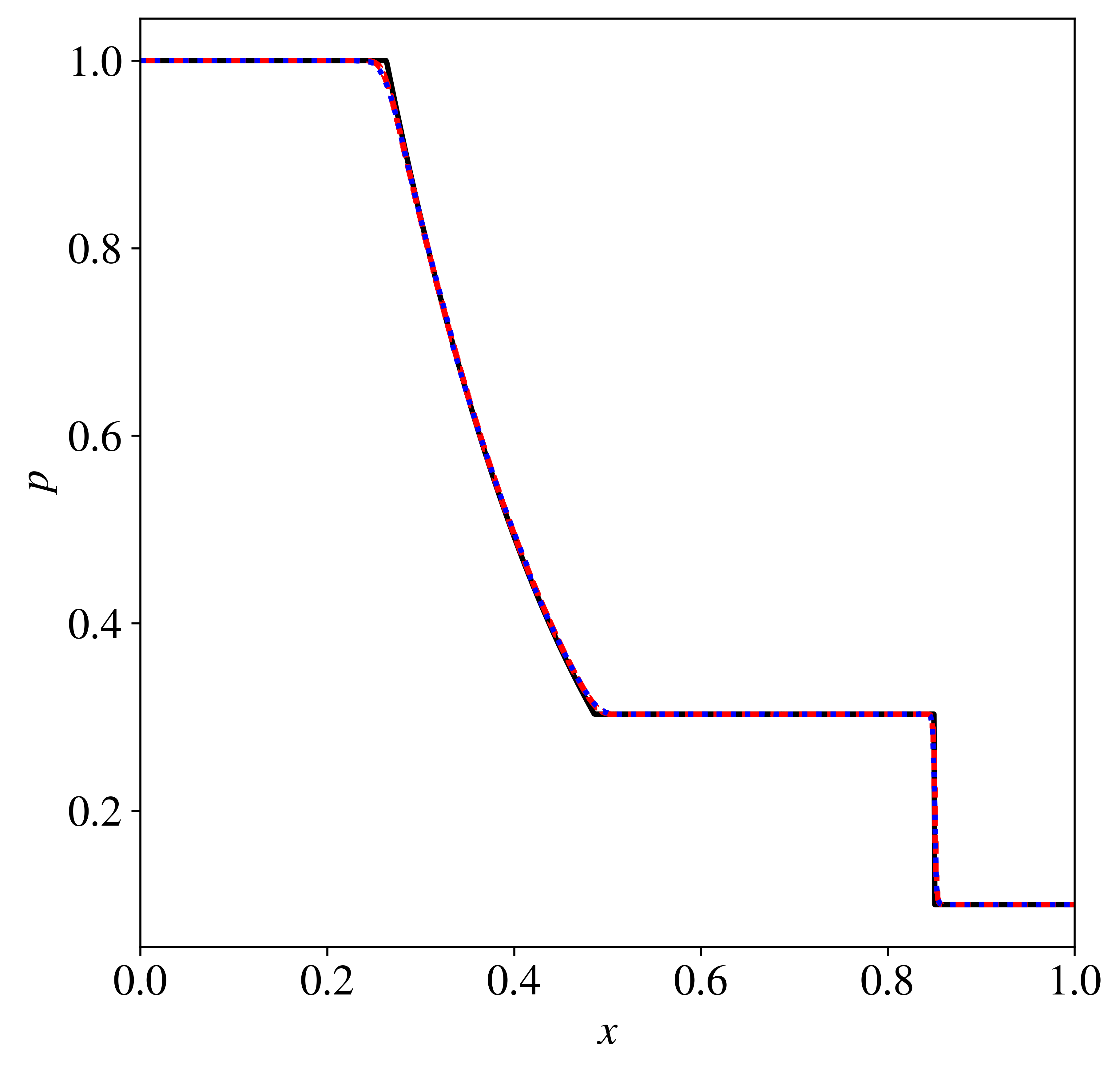}
\\
    \vspace{2pt} 
    \centering
        \includegraphics[width=\linewidth, valign=t]{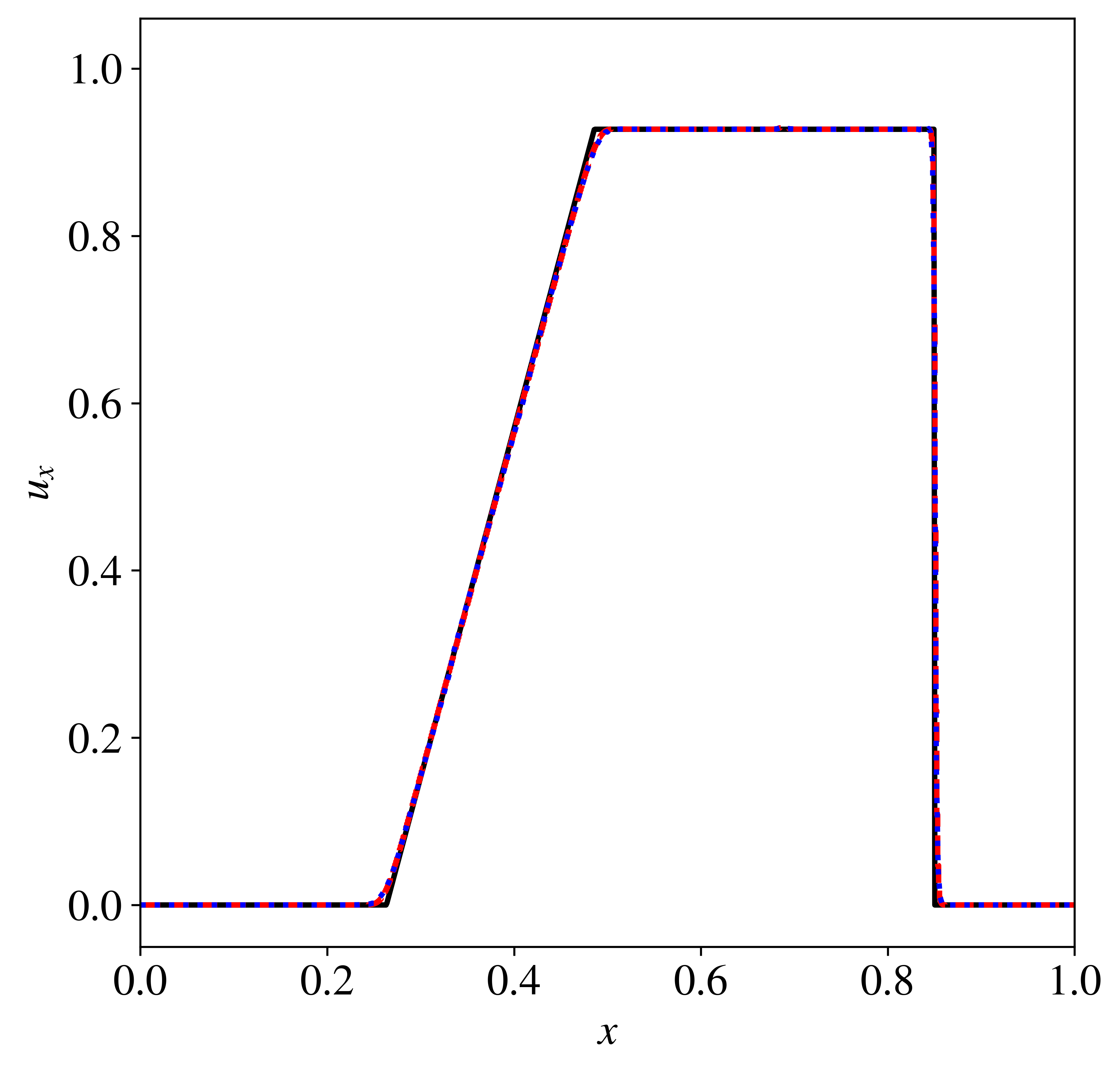}
\end{minipage}
\end{figure*}

The behavior of the model and AMR methodology is evaluated on a few 1D Euler level benchmarks.
For the Sod shock tube \cite{Sodtube}, the flow is initialized as
\begin{equation}
  (\rho, p, u_x) =
    \begin{cases}
      (1, 1, 0), & 0 \leq x \leq 0.5,\\
      (0.125, 0.1, 0), & 0.5 < x \leq 1,
    \end{cases}       
\end{equation}
in a domain $x \in [0,1]$.
The density gradient was used as refinement sensor to target the shock, contact discontinuity and rarefraction wave. 
A resolution of $\delta x = L_x/1024$ was applied for the case without AMR and as the peak resolution in AMR ($\delta x_2$ on level $l=2$), i.e. $\delta x_0 = L_x/256$.

From Fig.~\ref{fig:SodResults}, it can be seen that the correct solution is obtained both with and without AMR, while computational effort is saved in the lower resolved regions of AMR and overall no accuracy is lost.

\subsubsection{Lax shock tube}

\begin{figure*}[t]
\centering
\begin{minipage}[t]{.5\linewidth}
    \centering
        \includegraphics[width=\linewidth, valign=t]{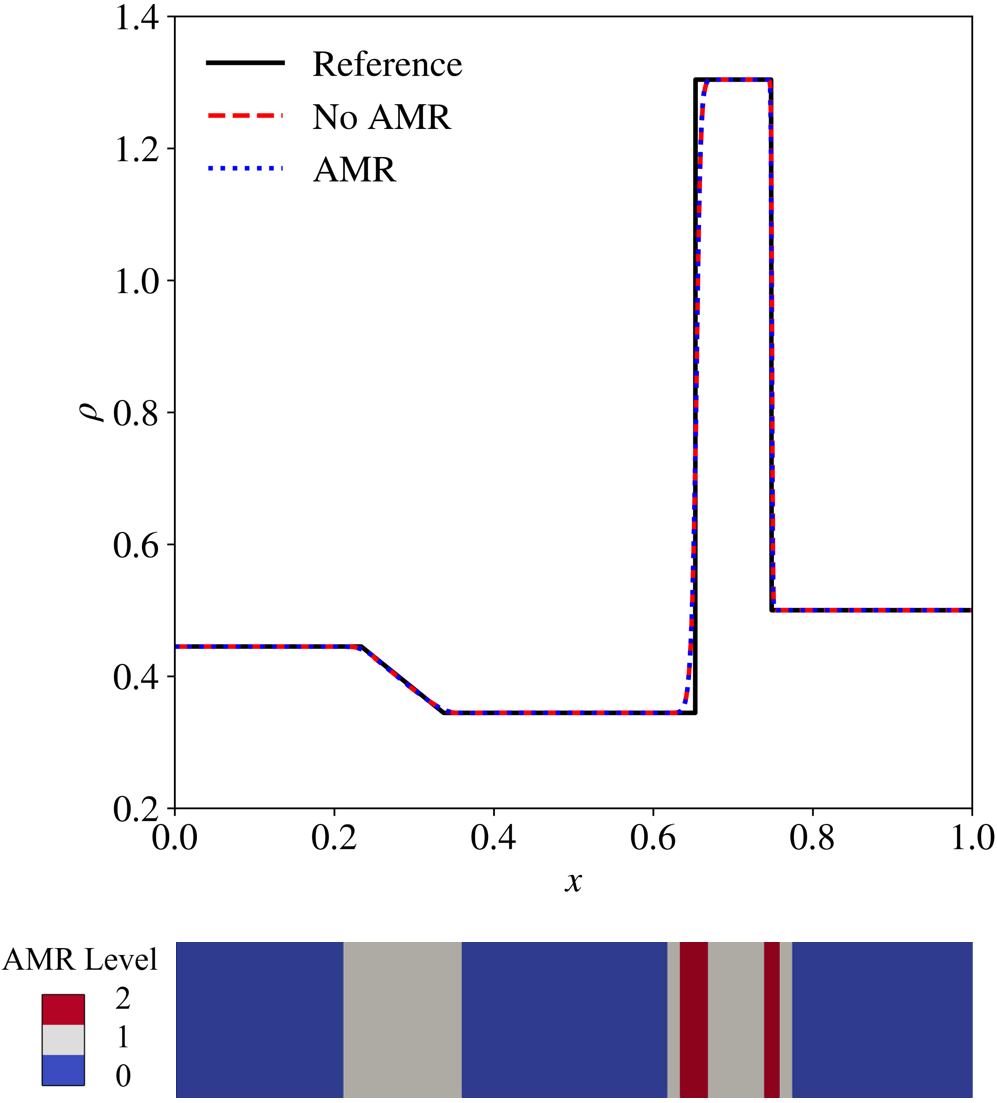}
\\
\hfill
    \begin{minipage}[b]{.90\linewidth}
        \centering
        \vspace{15pt} 
        \caption{Results of the Lax shock tube \cite{Laxtube} at $t=0.1$s for the density (top left), pressure (top right), and velocity (bottom right) compared to the reference solutions. The AMR levels are indicated at the bottom of the results for the density.} 
            \label{fig:LaxResults}
    \end{minipage}%
\end{minipage}
\hfill
\begin{minipage}[t]{.48\linewidth}
    \centering
        \includegraphics[width=\linewidth, valign=t]{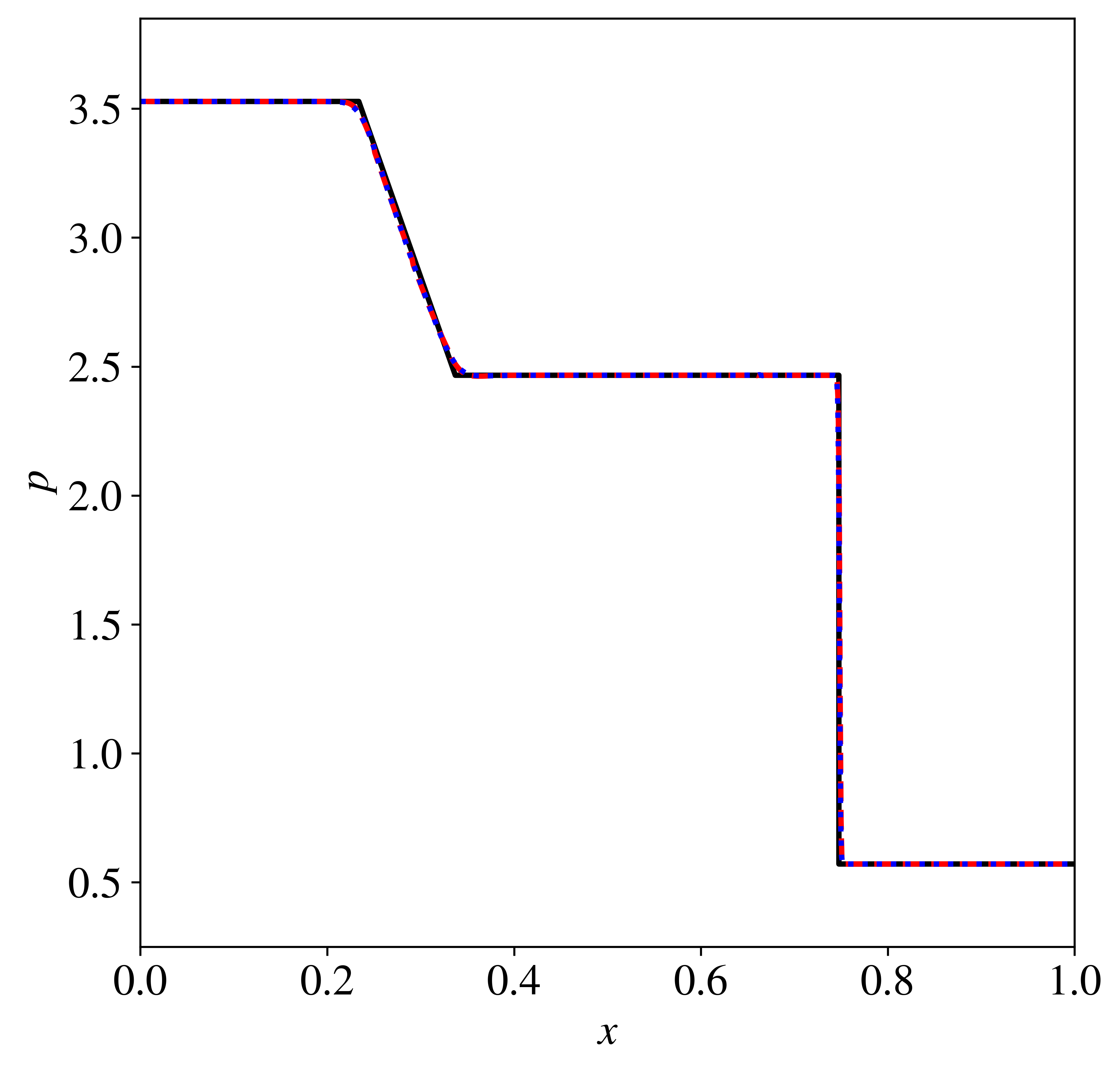}
\\
    \vspace{2pt}
    \centering
        \includegraphics[width=\linewidth, valign=t]{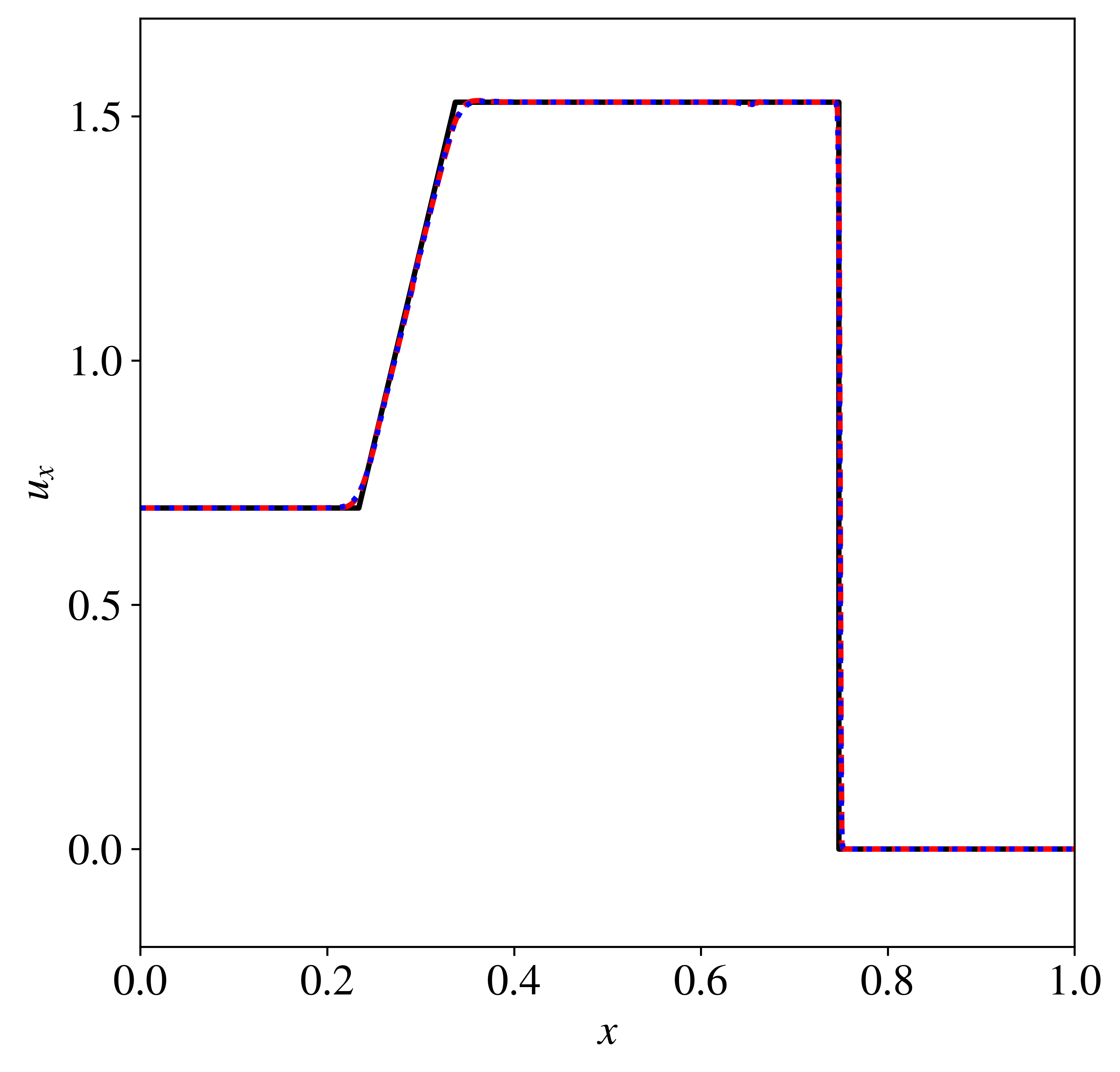}
\end{minipage}
\end{figure*}

The initial condition of the Lax shock tube \cite{Laxtube} reads,
\begin{equation}
  (\rho, p, u_x) =
    \begin{cases}
      (0.445, 3.528, 0.698), & 0 \leq x \leq 0.5,\\
      (0.5, 0.571, 0), & 0.5 < x \leq 1,
    \end{cases}       
\end{equation}
in a domain $x \in [0,1]$.
The temperature gradient was used as refinement sensor. 
A resolution of $\delta x = L_x/1024$ was applied for the case without AMR and as the peak resolution in AMR ($\delta x_2$ on level $l=2$), i.e. $\delta x_0 = L_x/256$.

The solution is accurately recovered for both cases, as can be seen in Fig.~\ref{fig:LaxResults}.

\subsubsection{Shu--Osher problem}

\begin{figure*}[t]
\centering
\begin{minipage}[t]{.49\linewidth}
    \centering
        \includegraphics[width=\linewidth, valign=t]{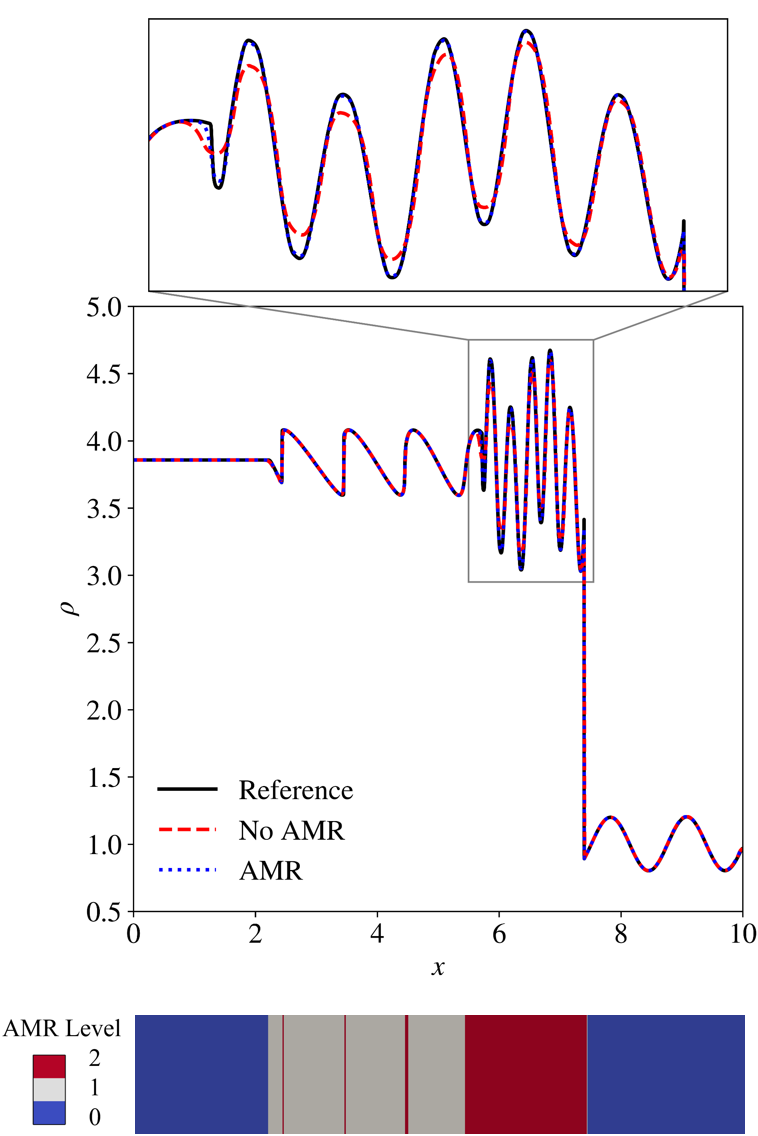}
\\
\hfill
    \begin{minipage}[b]{.90\linewidth}
        \centering
        \vspace{15pt} 
        \caption{Results of the Shu--Osher problem \cite{ShuOsher} at $t=1.8$s for the density (left), including a zoom into the most sensitive region of the density wave on the post-shock side (top left), pressure (top right), and velocity (bottom right) compared to the reference solutions. The AMR levels are indicated at the bottom of the results for the density.} 
            \label{fig:ShuOsherResults}
    \end{minipage}%
\end{minipage}
\hfill
\begin{minipage}[t]{.48\linewidth}
    \centering
        \includegraphics[width=0.99\linewidth, valign=t]{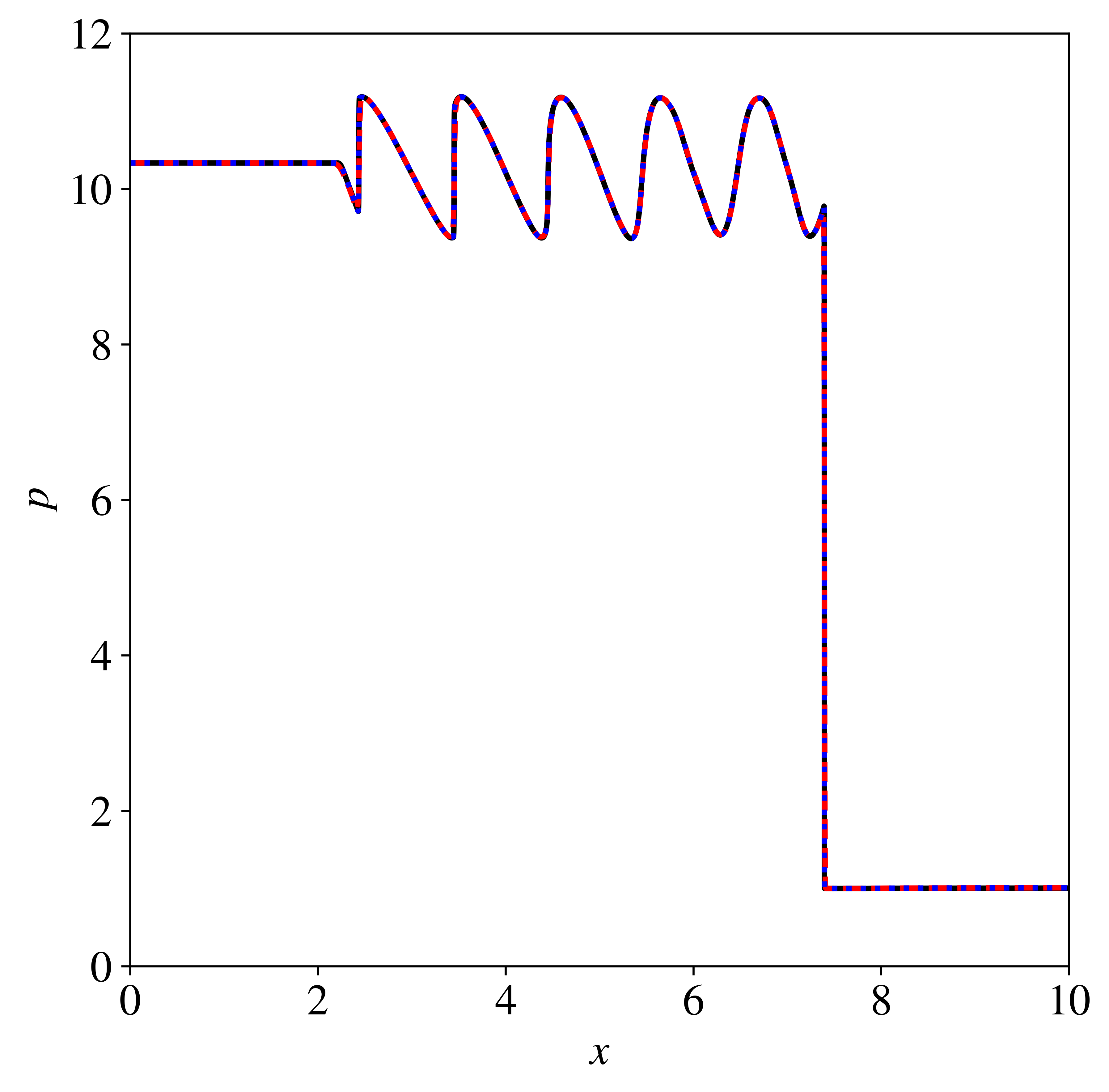}
\\
    \centering
        \includegraphics[width=\linewidth, valign=t]{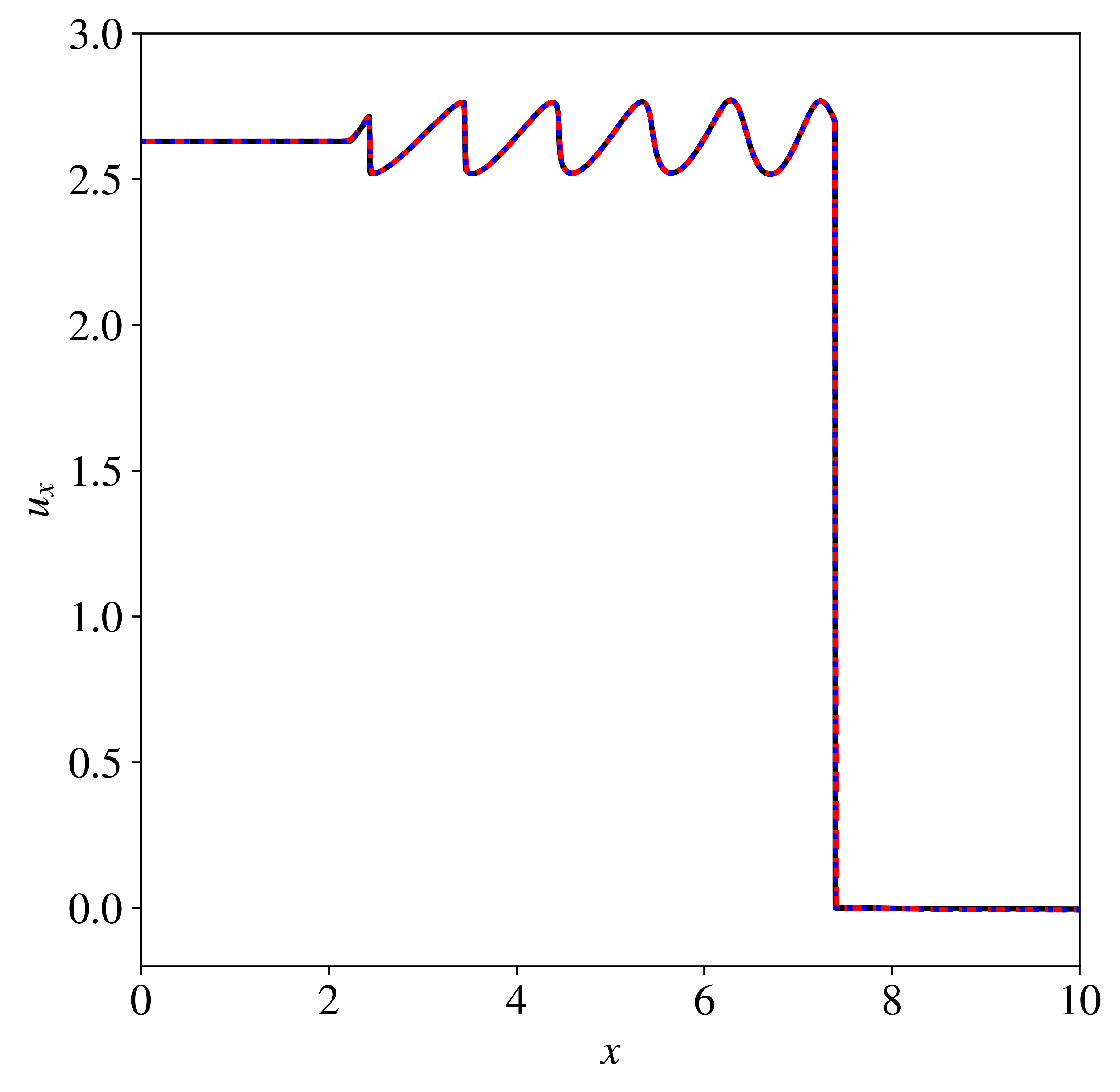}
\end{minipage}
\end{figure*}

The Shu-Osher problem \cite{ShuOsher} consists of a shock–density wave interaction initialized as
\begin{equation}
  (\rho, p, u_x) =
    \begin{cases}
      (3.857,10.333,2.629), & 0 \leq x \leq 1,\\
      (1+\frac{1}{5} \sin (5(x-5)), 1, 0), & 1 < x \leq 10,
    \end{cases}       
\end{equation}
in a domain $x \in [0, L_x]$ with $L_x = 10$.
The density and pressure gradients were used as refinement sensors to target the shock as well as strong gradients and discontinuities that emerge during the simulation in the post-shock region from the interaction with the density waves. 
A resolution of $\delta x = L_x/2560$ was applied for the case without AMR and this time as the base resolution in AMR ($\delta x_0$ on level $l=0$) in order to get an increased resolution of $\delta x_2 = L_x/10240$ and more accurate solution for the challenging density wave pattern in the high Ma number post-shock region of the problem.

Fig.~\ref{fig:ShuOsherResults} depicts the results.
It can be seen that the correct solution is obtained both with and without AMR.
A zoom into the most sensitive region of the density wave on the post-shock side reveals that the simulation employing AMR captures the reference solution more accurately whilst only a fraction of the total domain is covered with grids of higher resolution. 

\subsection{Two-dimensional cases}
 
\subsubsection{Riemann configurations}

\begin{figure*}[t]
\centering
    \includegraphics[width=.33\linewidth]{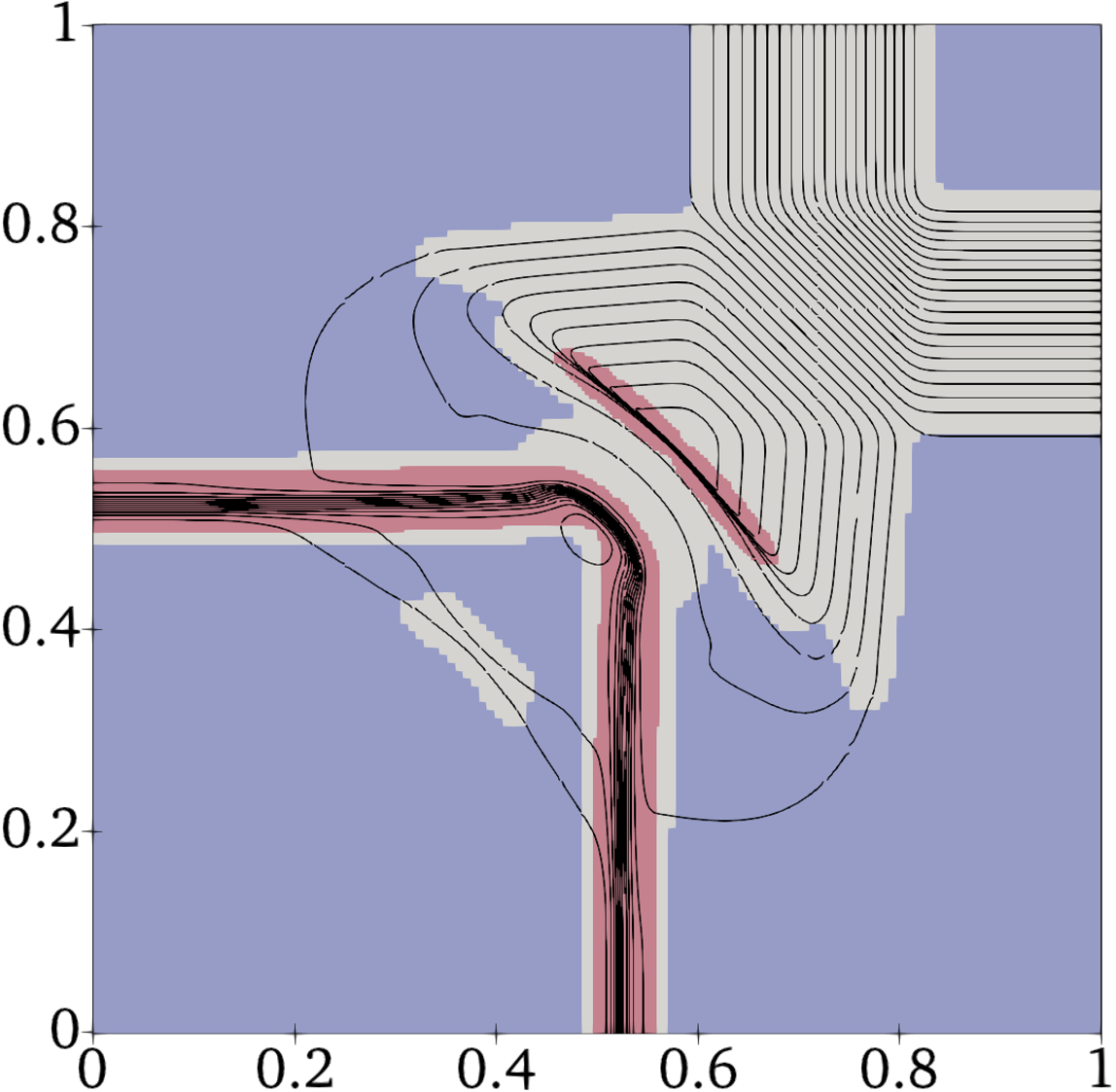}
\hfill
    \includegraphics[width=.33\linewidth]{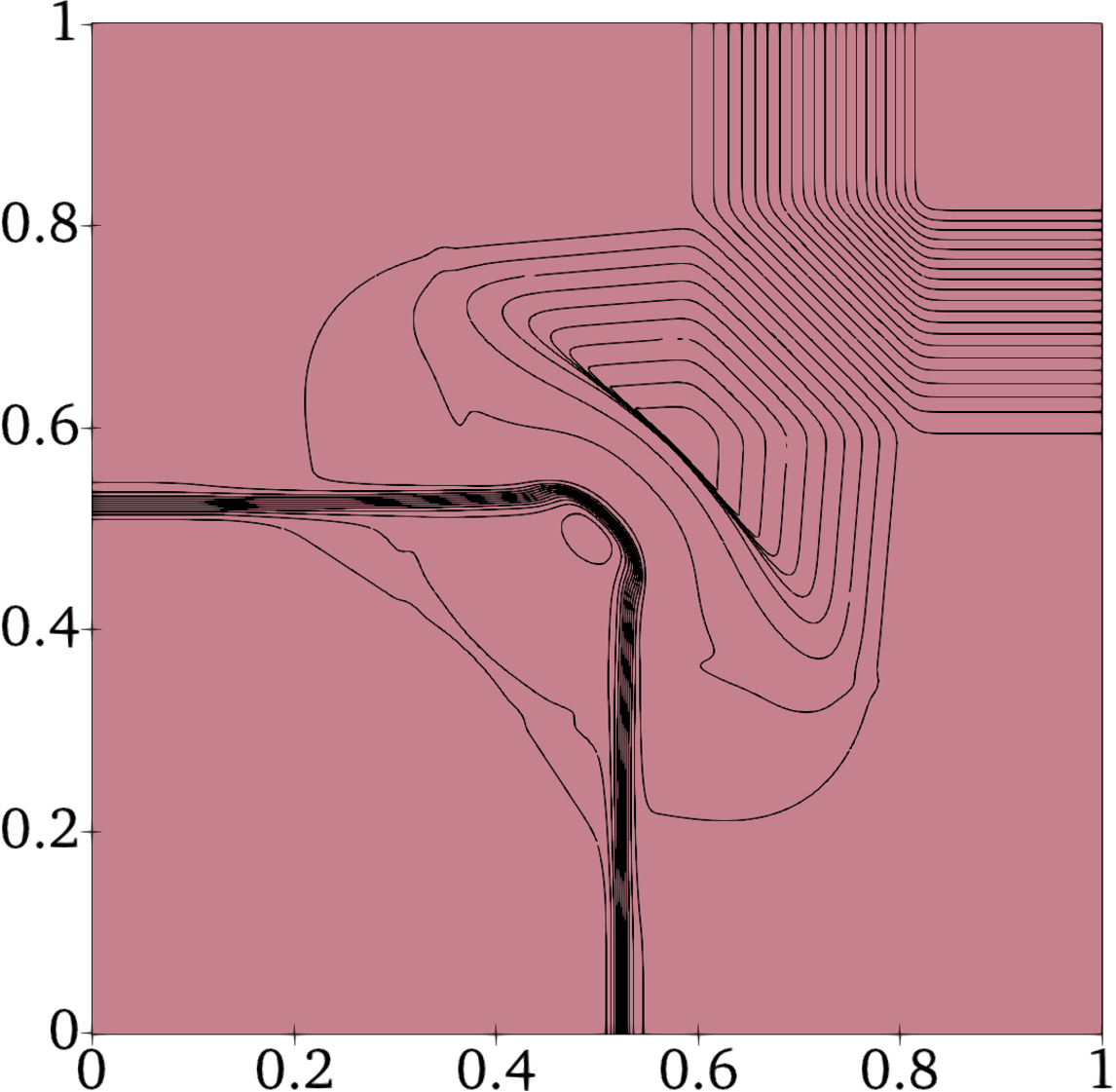}
\hfill
    \includegraphics[width=.33\linewidth]{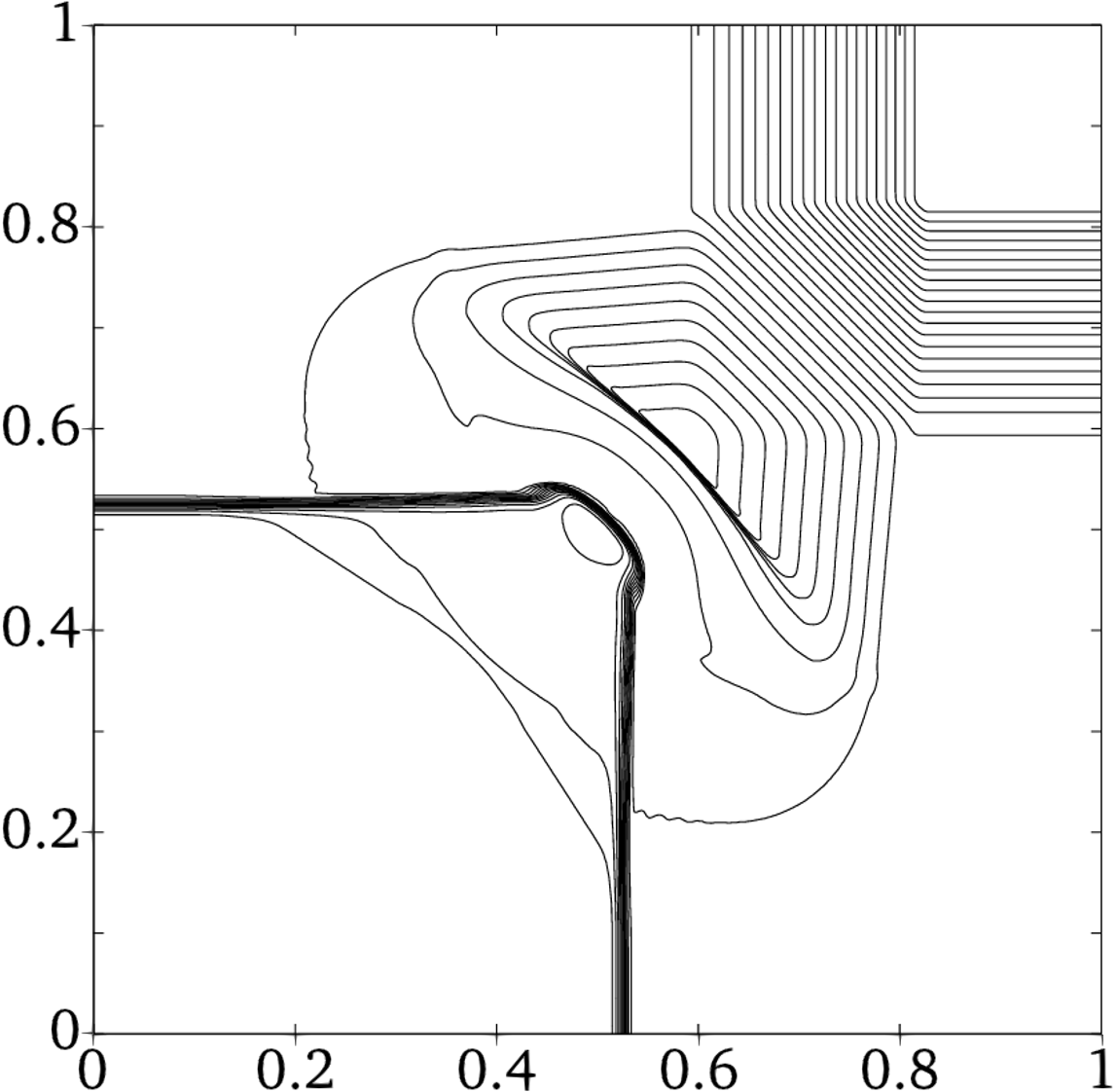}
\\ \vspace{0.25cm}
    \includegraphics[width=.33\linewidth]{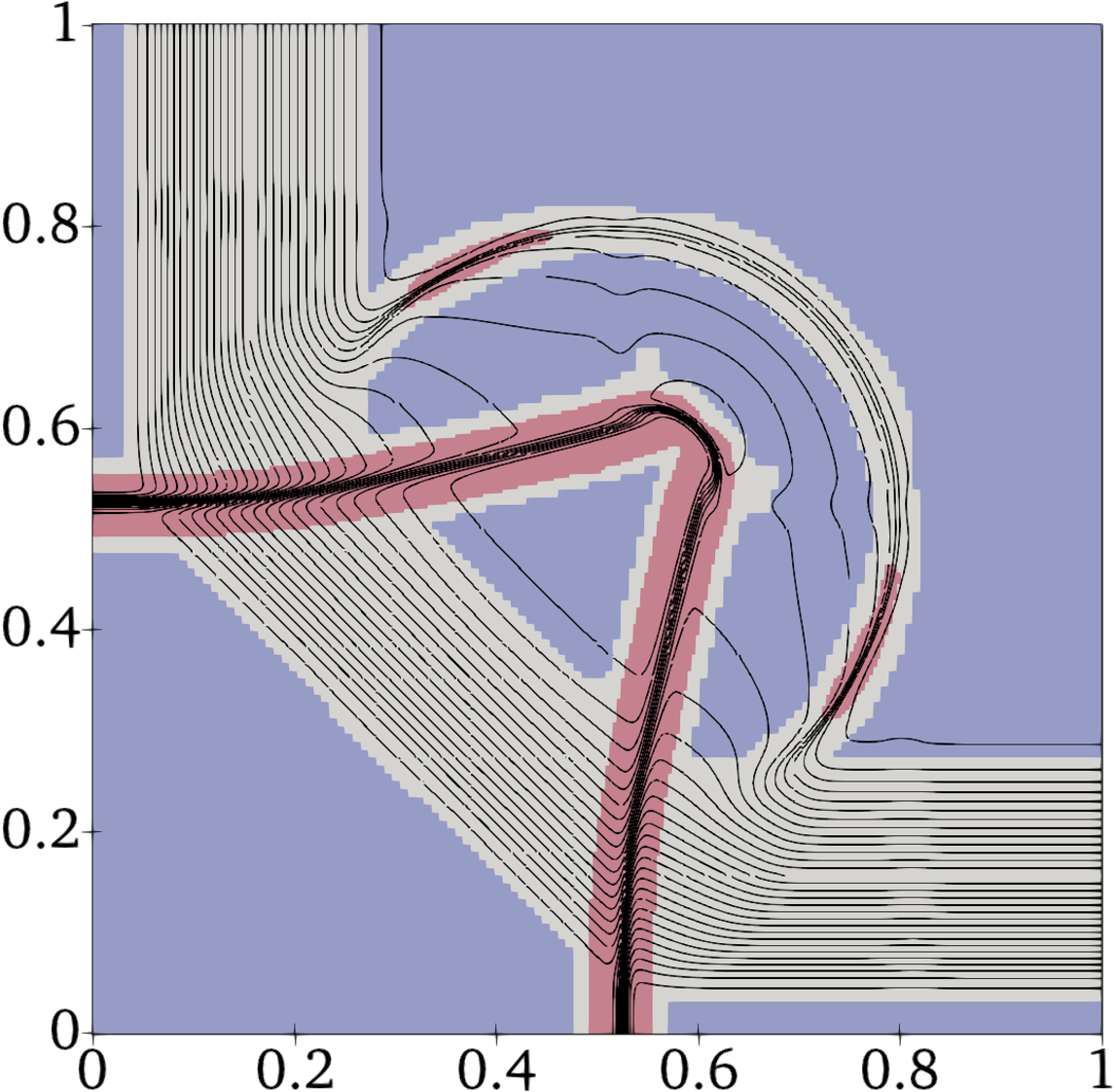}
\hfill
    \includegraphics[width=.33\linewidth]{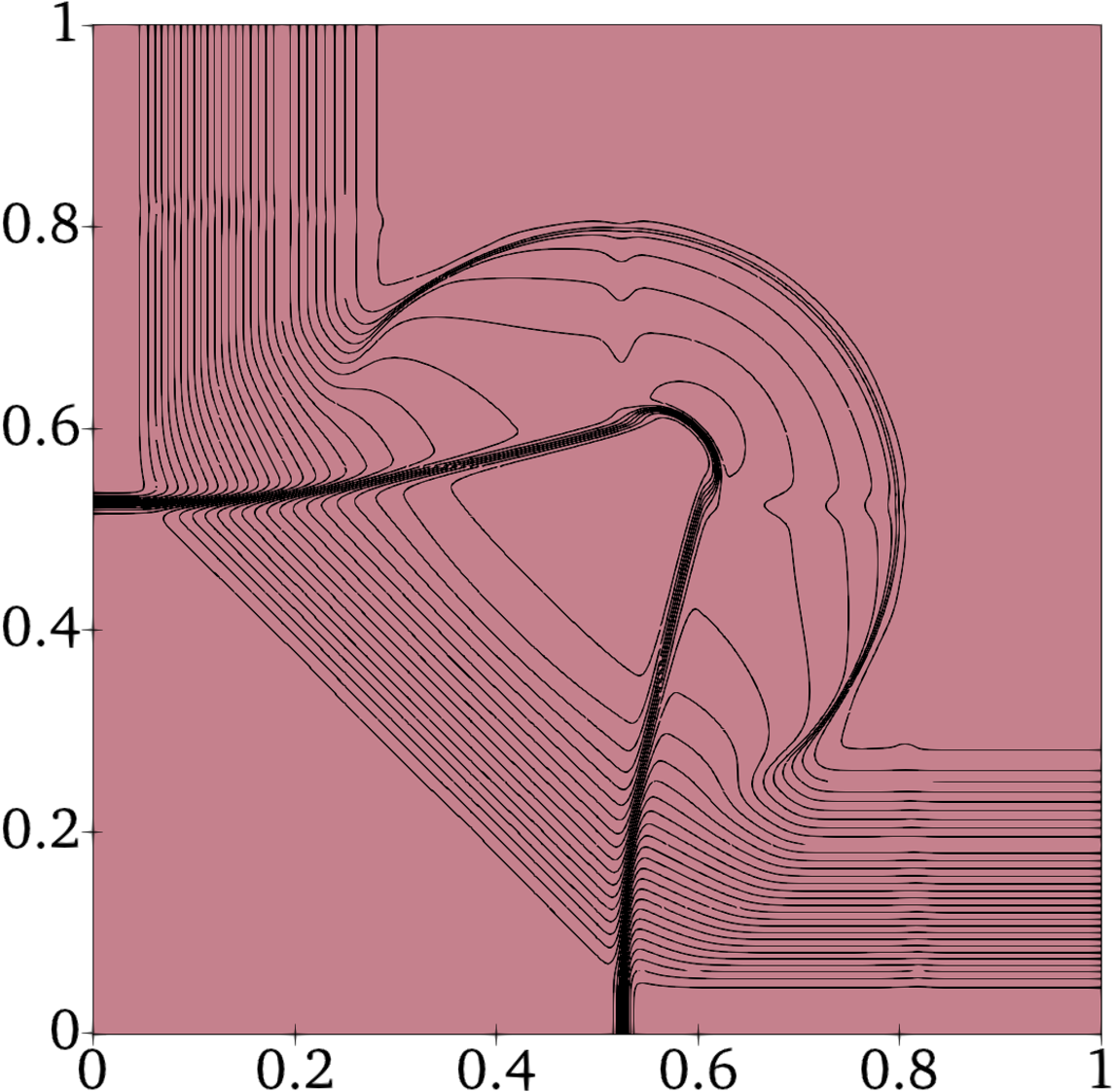}
\hfill
    \includegraphics[width=.33\linewidth]{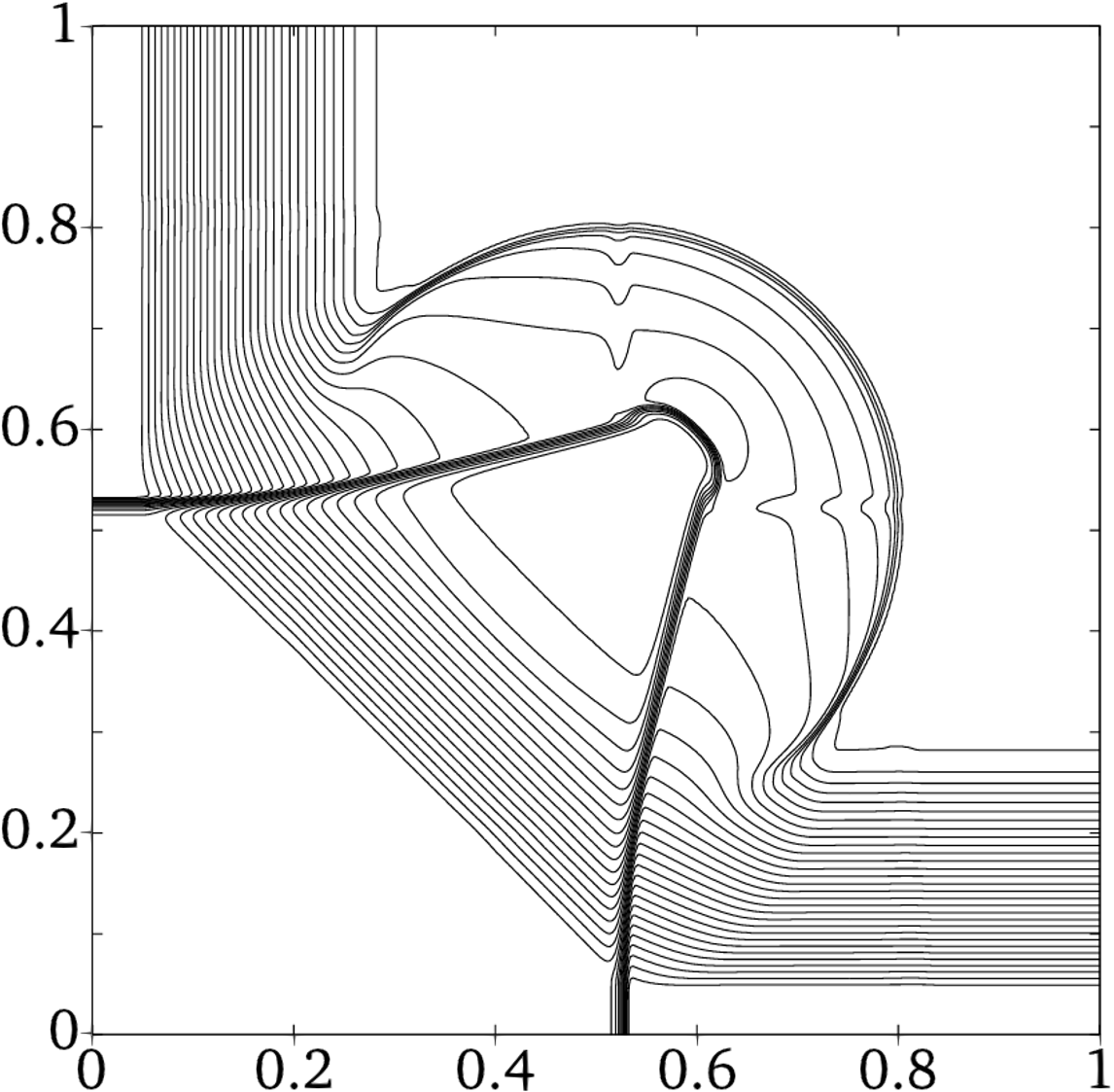}
\\ \vspace{0.25cm}
    \includegraphics[width=.33\linewidth]{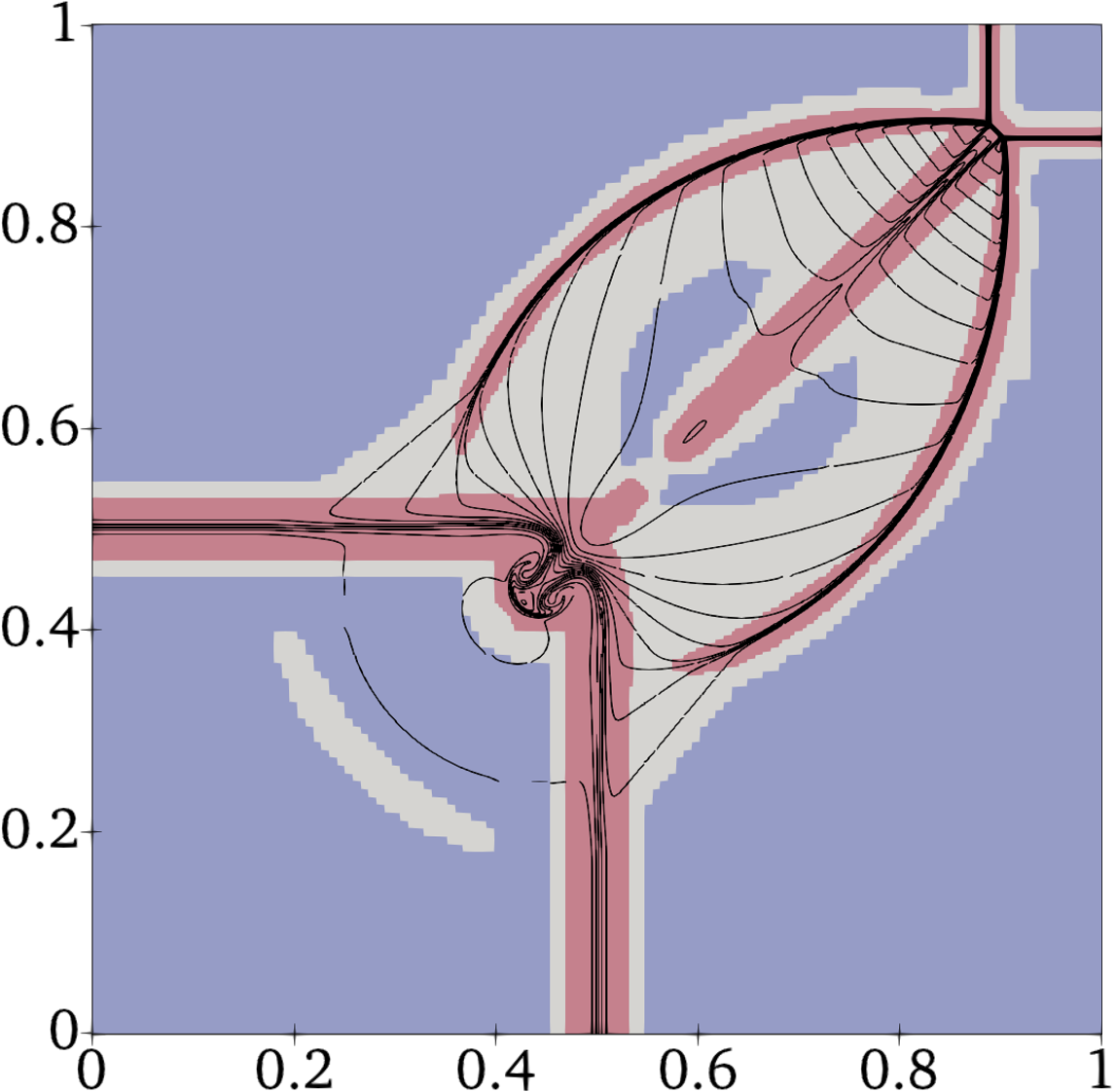}
\hfill
    \includegraphics[width=.33\linewidth]{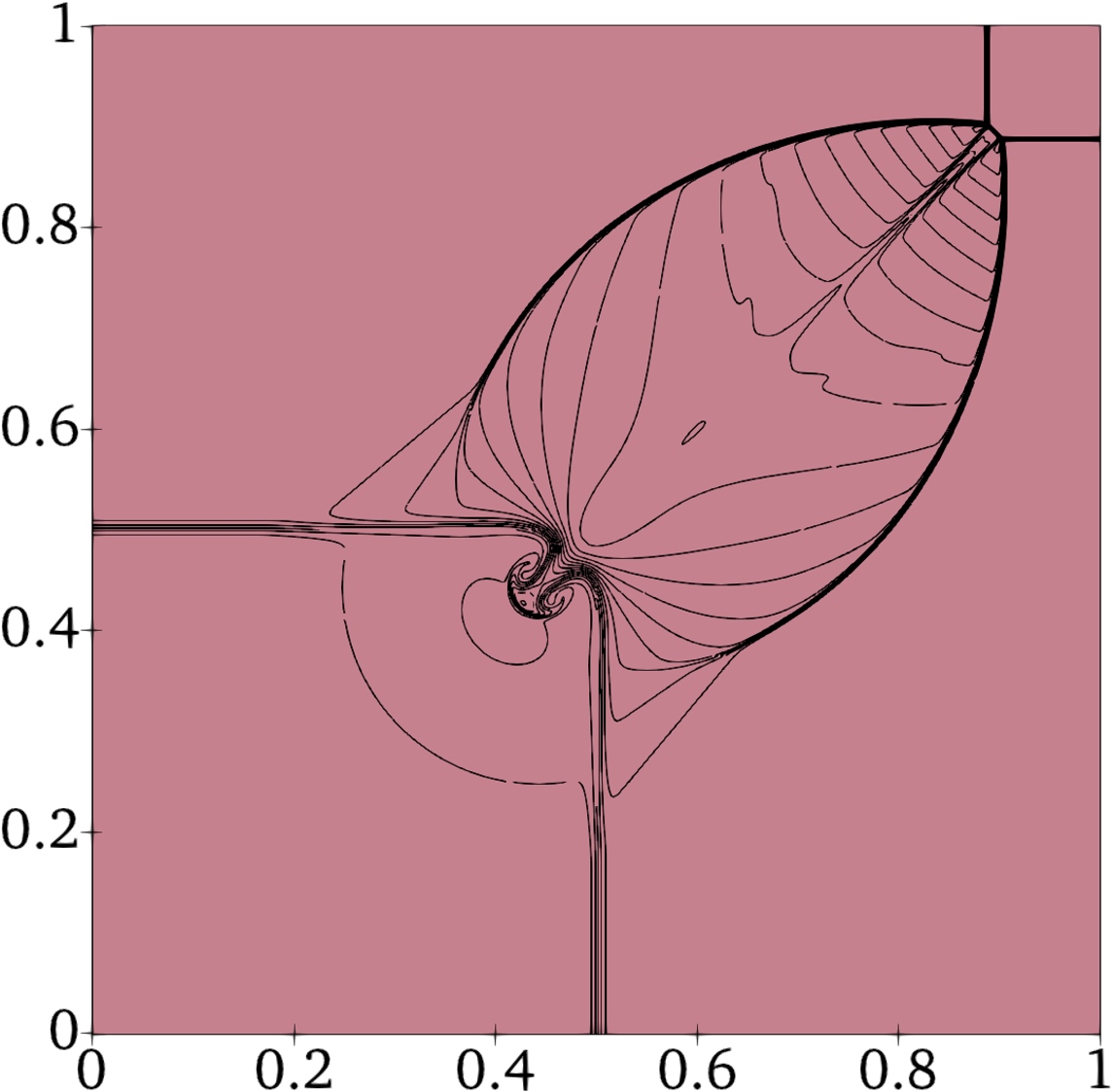}
\hfill
    \includegraphics[width=.33\linewidth]{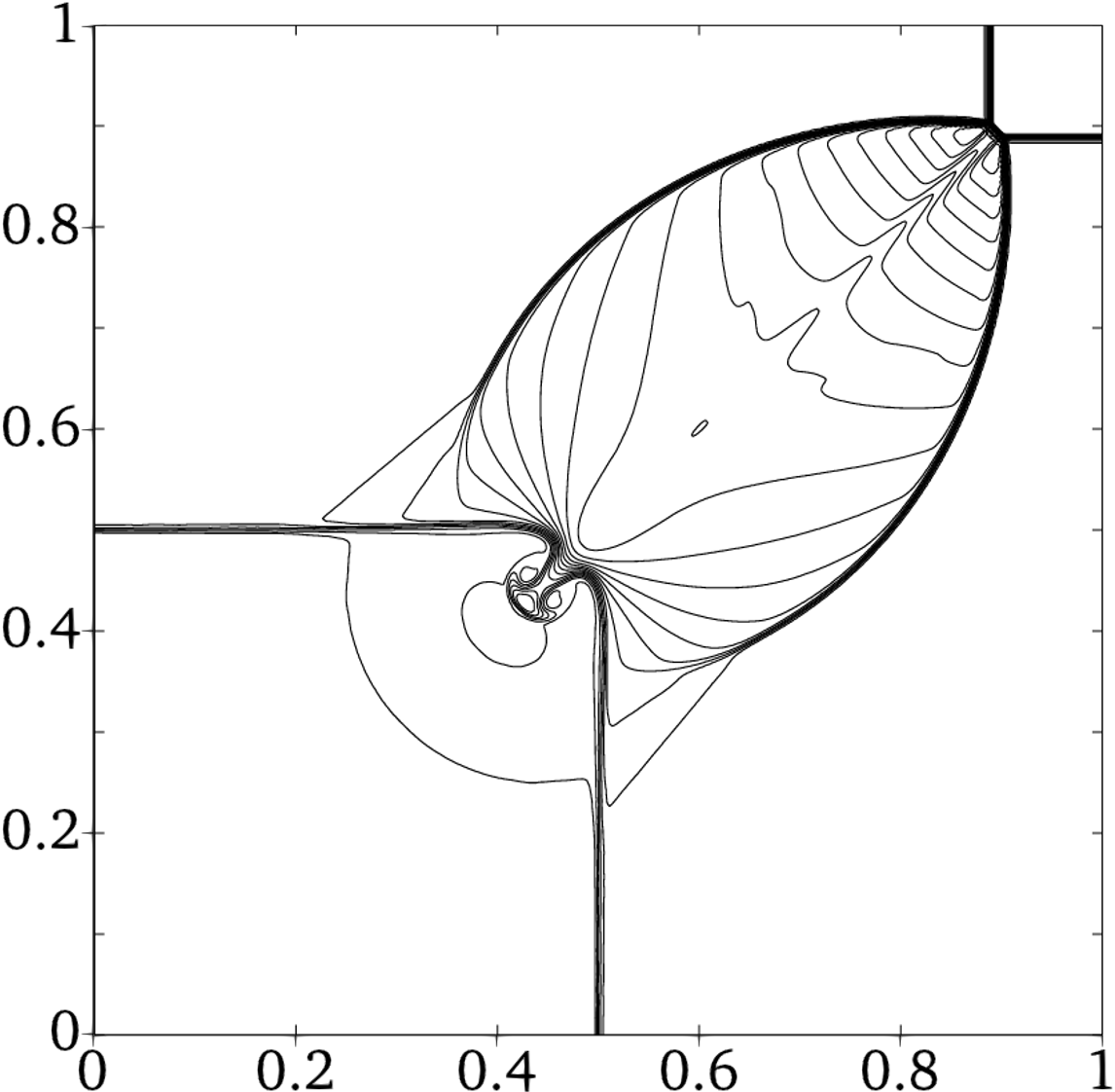}
\caption{
Solution of the Riemann configurations $\#7$, $\#8$ and $\#12$ at $t=0.25$s depicted in order from top to bottom with $30$ equidistant density contours.
The columns from left to right depict the solutions with AMR, without AMR, and the reference solutions from Lax \& Liu \cite{LaxLiu1998}, respectively. 
The spatial resolution of the present solver is indicated as colored background (blue: $\delta x = \delta y = L_x/256 = L_y/256$, gray: $\delta x = \delta y = L_x/512 = L_y/512$, red: $\delta x = \delta y = L_x/1024 = L_y/1024$).
}
\label{fig:Riemann}
\end{figure*}

\begin{figure*}[t]
\centering
\includegraphics[width=.33\linewidth]{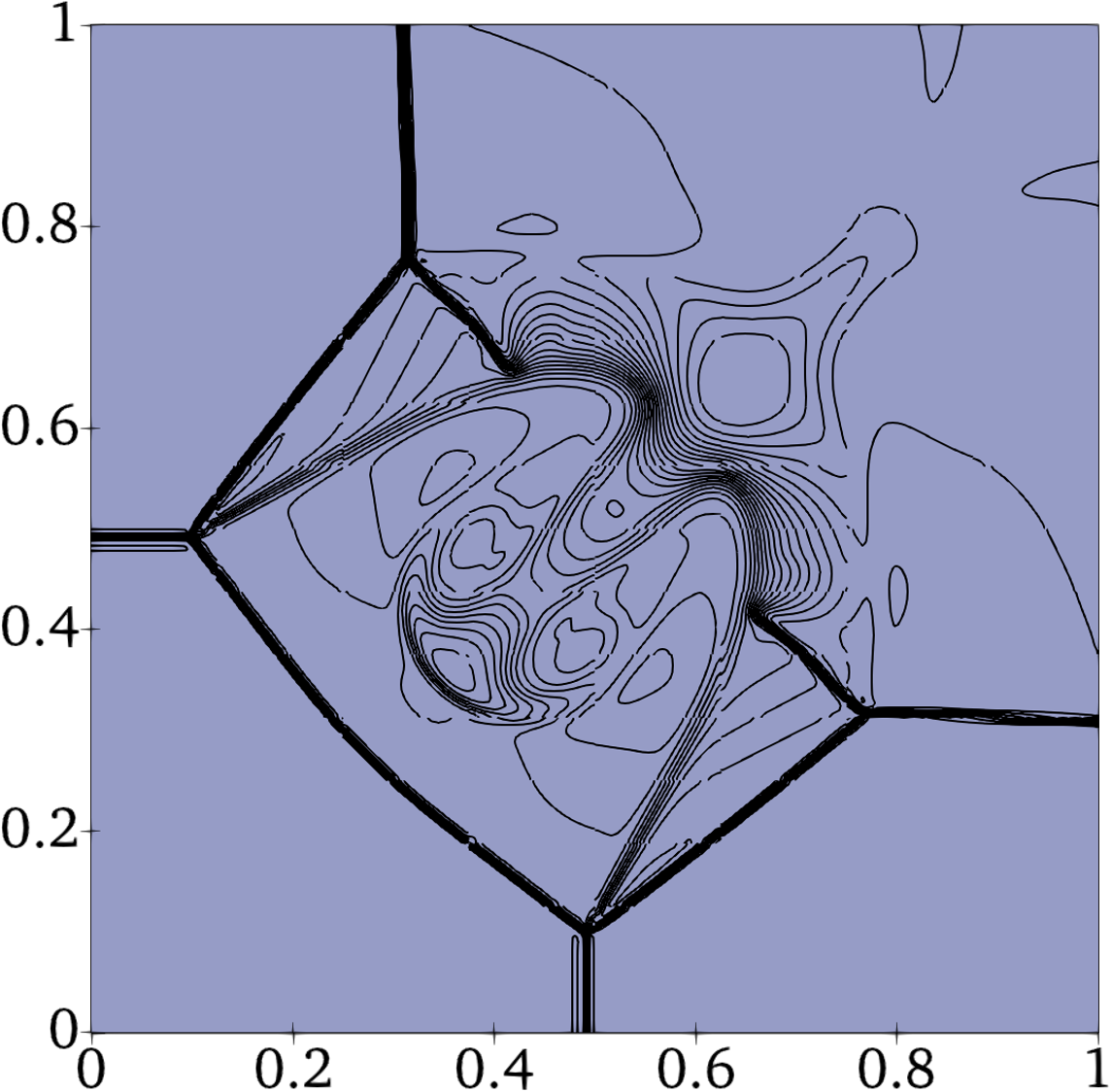}
\hfill
\includegraphics[width=.33\linewidth]{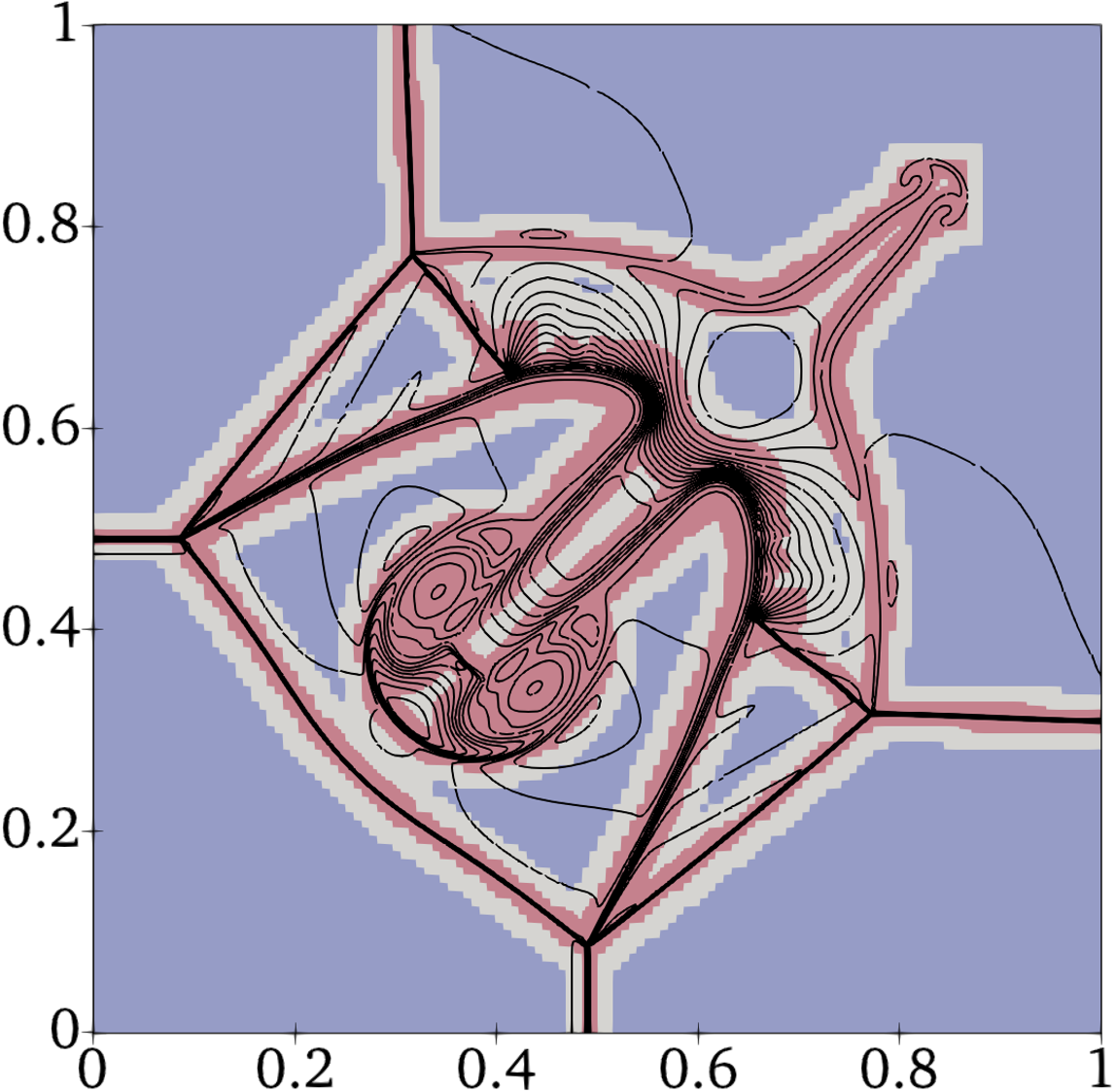}
\hfill
\includegraphics[width=.33\linewidth]{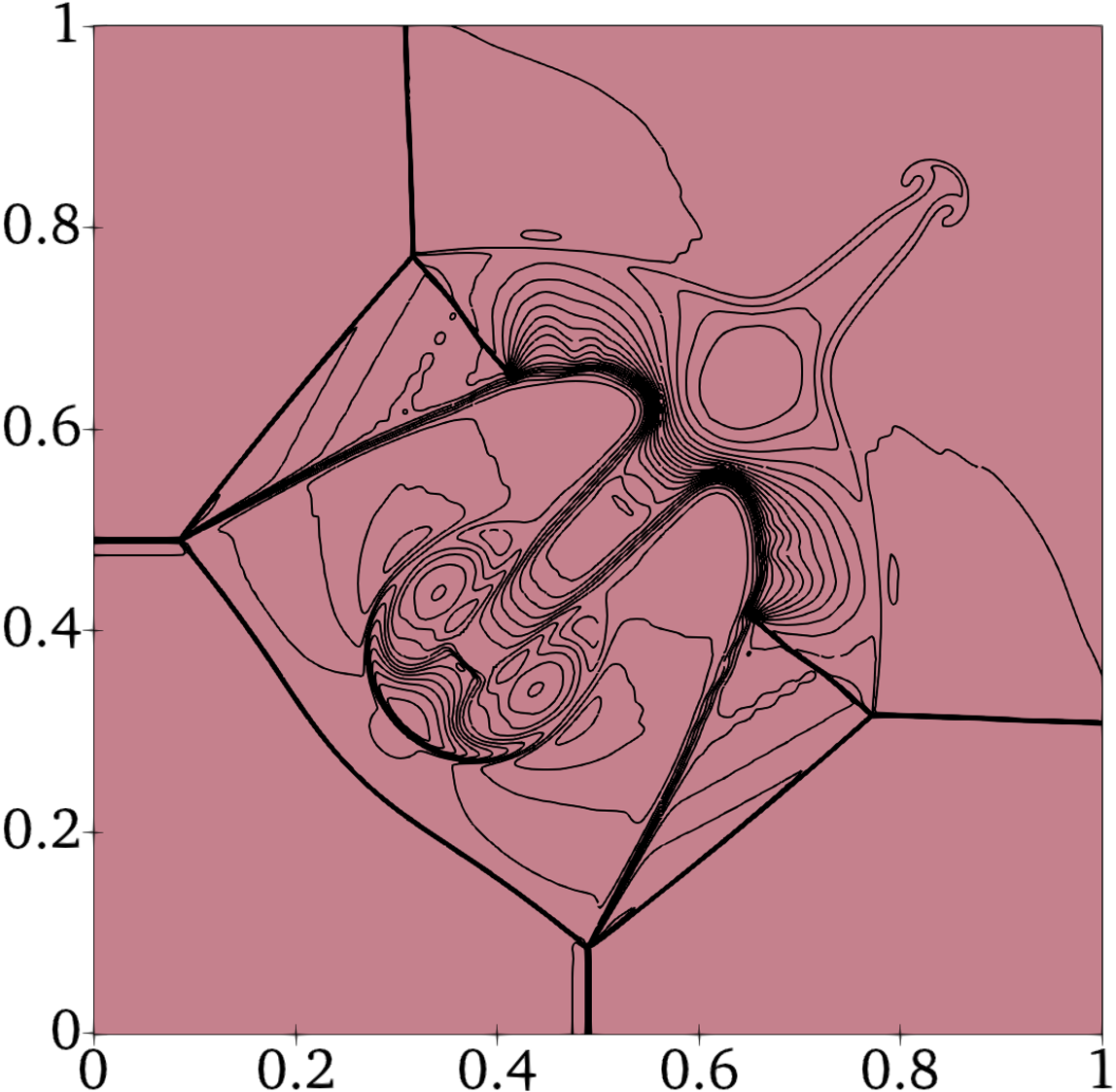}
\caption{
Solutions of the Riemann configuration $\#3^*$ at $t=0.85s$ depicted with $30$ equidistant density contours.
Left and right show the solutions on uniform grids, whereas the AMR solution is depicted in the middle.
The spatial resolution is indicated as colored background (blue: $\delta x = \delta y = L_x/256 = L_y/256$, gray: $\delta x = \delta y = L_x/512 = L_y/512$, red: $\delta x = \delta y = L_x/1024 = L_y/1024$).
}
\label{fig:LR3}
\end{figure*}

\begin{table*}[t]
\footnotesize	
\centering
\begin{tabular}{|c|c|c|c|c|c|c|}
    \hline
   Configuration & 
   Quadrant 1 (NE) &  Quadrant 2 (NW) &  Quadrant 3 (SW) & Quadrant 4 (SE) & Quadrant borders & 
   $t_{end}$  
   \\    \hline
    $ $ [$\#$] & ($\rho$, p, $u_x$, $u_y$)  &  ($\rho$, p, $u_x$, $u_y$)  &  ($\rho$, p, $u_x$, $u_y$) &  ($\rho$, p, $u_x$, $u_y$) & ($x$, $y$) & [$s$] \\
    \hline
7  & {(1, 1, 0.1, 0.1)}        & {(0.5197, 0.4, -0.6259, 0.1)}  & {(0.8, 0.4, 0.1, 0.1)}               & {(0.5197, 0.4, 0.1, -0.6259)}  & (0.5, 0.5) & 0.25 \\ 
8  & {(0.5197, 0.4, 0.1, 0.1)} & {(1, 1, -0.6259, 0.1)}         & {(0.8, 1, 0.1, 0.1)}                 & {(1, 1, 0.1, -0.6259)}         & (0.5, 0.5) & 0.25 \\ 
12 & {(0.5313, 0.4, 0, 0)}     & {(1, 1, 0.7276, 0)}            & {(0.8, 1, 0, 0)}                     & {(1, 1, 0, 0.7276)}            & (0.5, 0.5) & 0.25 \\ 
3* & {(1.5, 1.5, 0, 0)}        & {(0.5323, 0.3, 1.206, 0)}      & {(0.138, 0.029, 1.206, 1.206)}       & {(0.5323, 0.3, 0, 1.206)}   & (0.85, 0.85)   & 0.85  \\ 
    \hline
\end{tabular}
\caption{
Setup for the 2D Riemann configurations. 
}
\label{tab:2DRiemannIC}
\end{table*}

The Riemann configurations are further classical benchmarks to investigate the behavior of the solver for 2D Euler level solution fields involving complex interactions between shocks, contact discontinuities, rarefraction waves and vortexes. 
A total of $19$ configurations have been thoroughly studied in the CFD and mathematics literature \cite{LaxLiu1998, KurganovR2D, Schulz-Rinne, ZhangZhengRiemann, TungChang1, TungChang2, RiemannBookZheng, RiemannBookSheng}.
Here we look into more detailed behavior of the solver for three of these configurations.
The fields were initialized with different values for $\rho$, $p$, $u_x$ and $u_y$ in a domain $x \in [0,1]$ and $y \in [0,1]$ separated into four quadrants, i.e. north-east (NE), north-west (NW), south-west (SW) and south-east (SE).
Additionally a modified configuration \#3, here termed \#$3^*$, was initialized with the quadrant borders at ($x$, $y$) $=(0.85, 0.85)$ instead of ($x$, $y$) $=(0.5, 0.5)$, as done by many researchers, e.g. \cite{LR3case}, in order to evaluate the capabilities of a numerical scheme in a long-term evolution of the solution field, generating more complex wave and vortex interactions.
All setups are given in table~\ref{tab:2DRiemannIC}, including the simulation time $t_{end}$.
The sensor for mesh refinement consisted of a combination of vorticity as well as gradients of density and pressure.
The refinement efficiency was elevated to $\eta = 0.98$ in order to accommodate more neatly fitting meshes for these 2D cases.
A resolution of $\delta x = \delta y = L_x/1024 = L_y/1024$ was applied for the case without AMR and as the peak resolution with AMR ($\delta x_2 = \delta y_2$ on level $l=2$), i.e. $\delta x_0 = \delta y_0 = L_x/256 = L_y/256$.

The resulting density contours for configurations $\#7$, $\#8$ and $\#12$ can be seen in Fig.~\ref{fig:Riemann}, where the resolution is indicated by color in the background. 
A comparison between the column in the middle (no AMR results) and on the right (reference solutions) reveals that the model on a uniform mesh accurately captures the relevant features in the reference solutions.
More precisely, 
in configuration $\#7$,
the contact discontinuities on the interfaces of the SW quadrant result in negative slip lines at the correct positions. 
The rarefraction waves on the interfaces of the NE quadrant propagate towards the top right corner, while building up a small shock behind them.
In configuration $\#8$, the correct flow field is recovered with great detail as compared to the reference solution. 
This time, the rarefraction waves of the interfaces from the NE quadrant propagate away from the top right corner, forming a semi circular shock and curving the negative slip lines stemming from the contact discontinuities on the SW quadrant interfaces considerably. 
Also the ripples reported in the literature, e.g. \cite{LaxLiu1998, KurganovR2D}, are recovered where the initial interfaces of the NE quadrant were located.
Similarly in configuration $\#12$, 
the distinct interaction patterns in the NE quadrant, emerging from the shocks traveling to the top right corner, as well as the position of the vortexes formed in the SW quadrant, due to the interaction with the positive slip lines, are correctly captured.
The solution fields are perfectly symmetric, which further validates the model.
Consulting the left column in Fig.~\ref{fig:Riemann} (AMR results) proves that the AMR worked as intended by refining where it is most crucial for an accurate result, i.e. more resolution was provided in regions of moderate gradients and vorticity, e.g. rarefraction waves and various interaction patterns, and the most accurate resolution occurred near strong gradients and discontinuous solutions, e.g. shocks and contact discontinuities.
Note that the AMR solution (left column) is only expected to match the solution in the middle column at the locations where the resolutions match, i.e. those with an underlying red color.
As a consequence of the expected hyperviscosities associated with any underresolved numerical scheme, higher frequency features are damped more strongly in more coarsely resolved simulations than in high resolution simulations.
This behavior is also intrinsic to AMR and can be found in the presented results for features which have a coarse resolution, e.g. the more strongly damped ripples in configuration $\#8$.
However, the relevant parts of the solution which were targeted with the sensors are captured with great accuracy, both compared to the simulations with uniform high resolution and the reference solutions.

\begin{figure}[t]
\centering
\includegraphics[width=.99\linewidth]{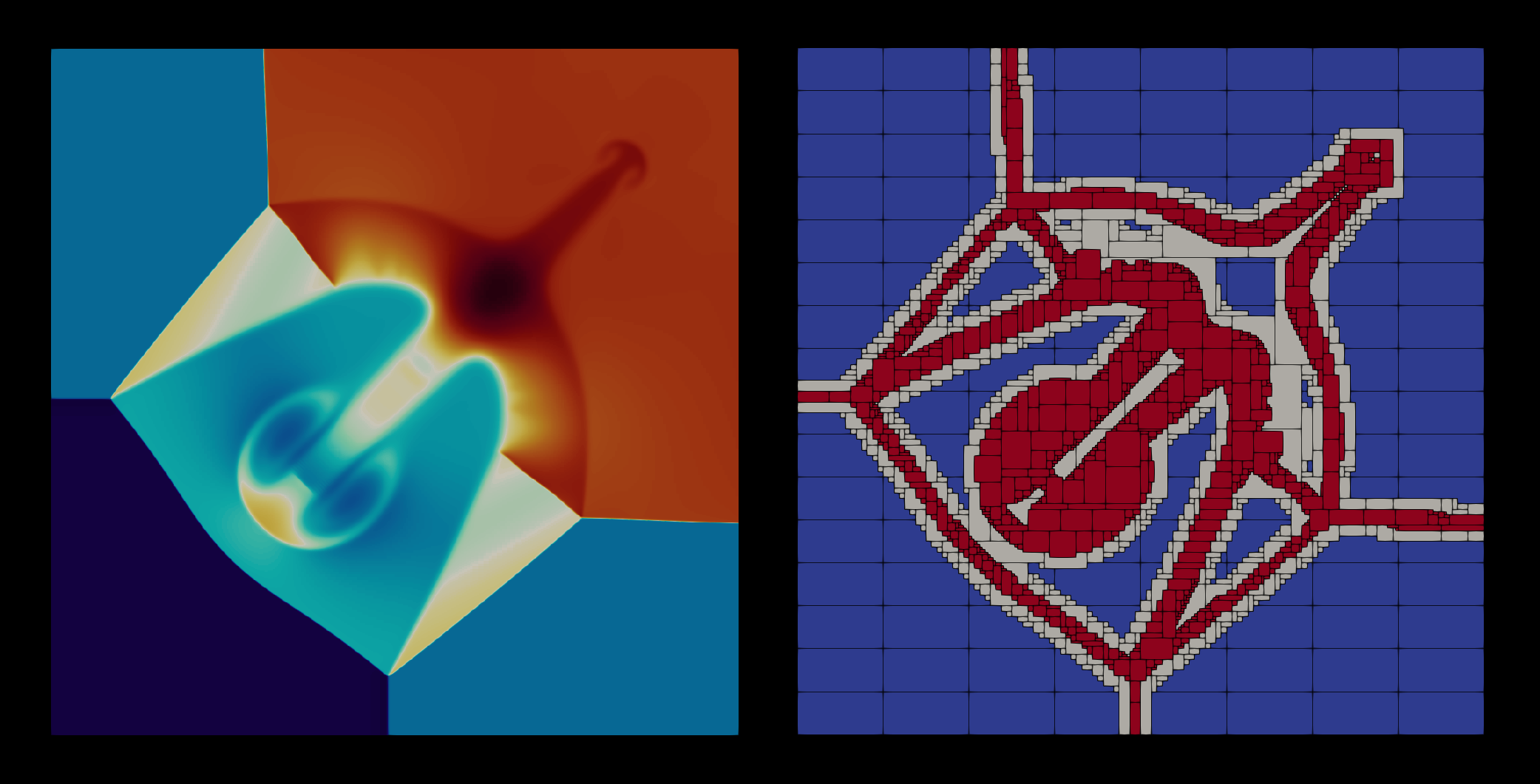}
\caption{
A video (Multimedia available online) of the AMR simulation in Fig.~\ref{fig:LR3} displaying the temporal evolution of the density and AMR levels including individual AMR patches indicated by black lines (color bars are given in the video and the coloring of resolution follows Fig.~\ref{fig:LR3}).
}
\label{fig:LR3_video}
\end{figure}

Fig.~\ref{fig:LR3} depicts the solutions of the density for configuration $\#3^*$, for a single level simulation with coarse and fine resolution as well as an AMR simulation. 
In this configuration, triple shock structures, interactions patterns with slip lines and a central vortex pair within them is formed. 
The kinetic solver accurately recovers the solution fields reported in the literature for the high resolution simulation in terms of the density contours.
However, the the solution in the case of a uniformly coarse resolution significantly differs.
Predominantly noticeable is the mismatch in the position of the inner vortex pair and the curvature of the shock in the SW quadrant as well as the manifestation of the vortex pair in the NE quadrant. 
As expected, the AMR solution correctly depicts the mentioned flow features targeted with the refinement sensors.
It provides a solution with an accuracy indifferent to the uniform high resolution case despite being covered with high resolution grids only on a relatively small fraction of the domain.
Hence, these results demonstrate the potential of AMR for Euler level flows, where features requiring good resolution predominantly occupy a rather small part of the computational domain.
A video of the AMR simulation presented in Fig.~\ref{fig:LR3} is given as an integral multimedia in Fig.~\ref{fig:LR3_video} (Multimedia available online).

\begin{figure*}[t]
\centering
    \includegraphics[width=.325\linewidth]{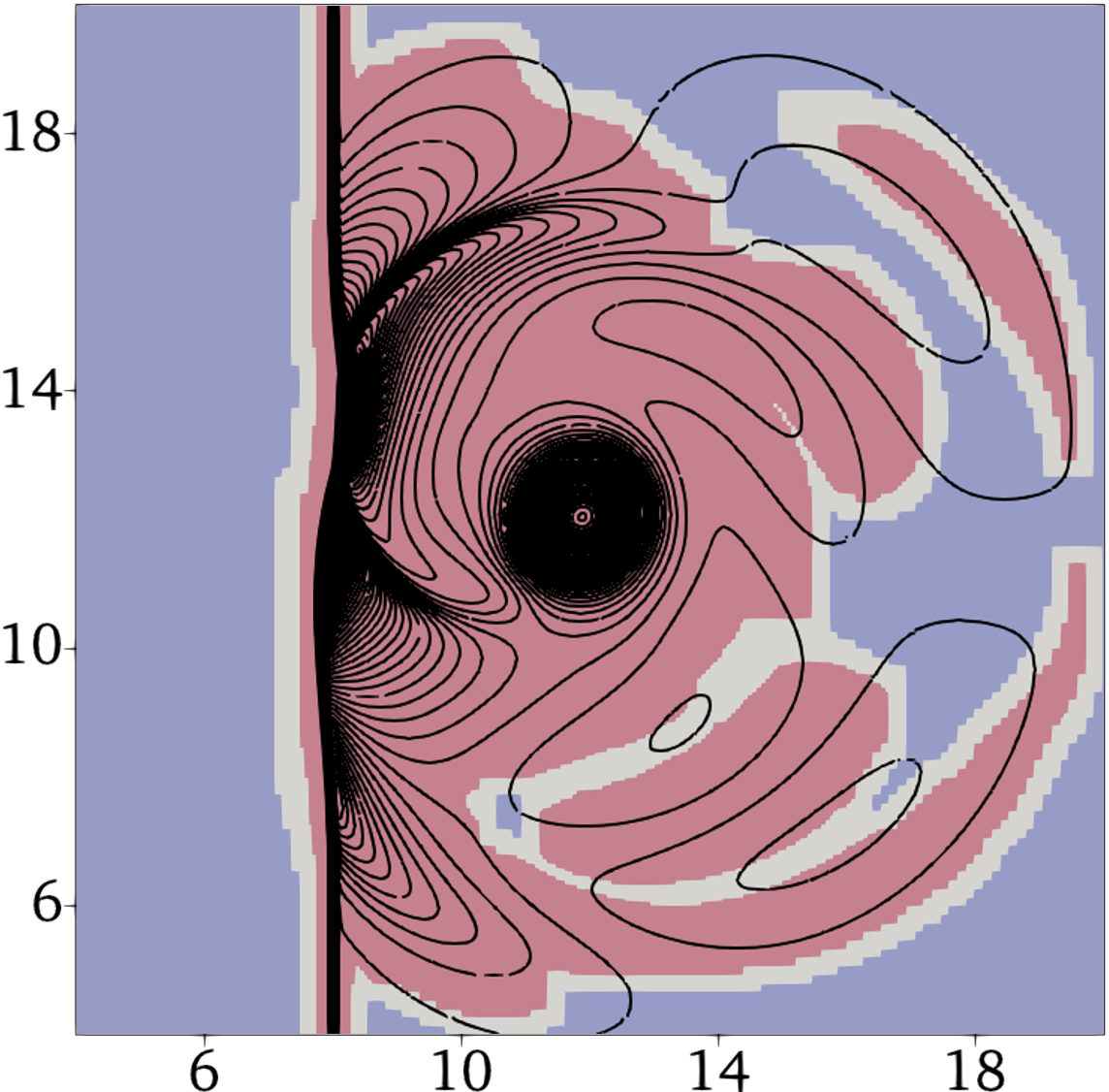}
\hfill
    \includegraphics[width=.325\linewidth]{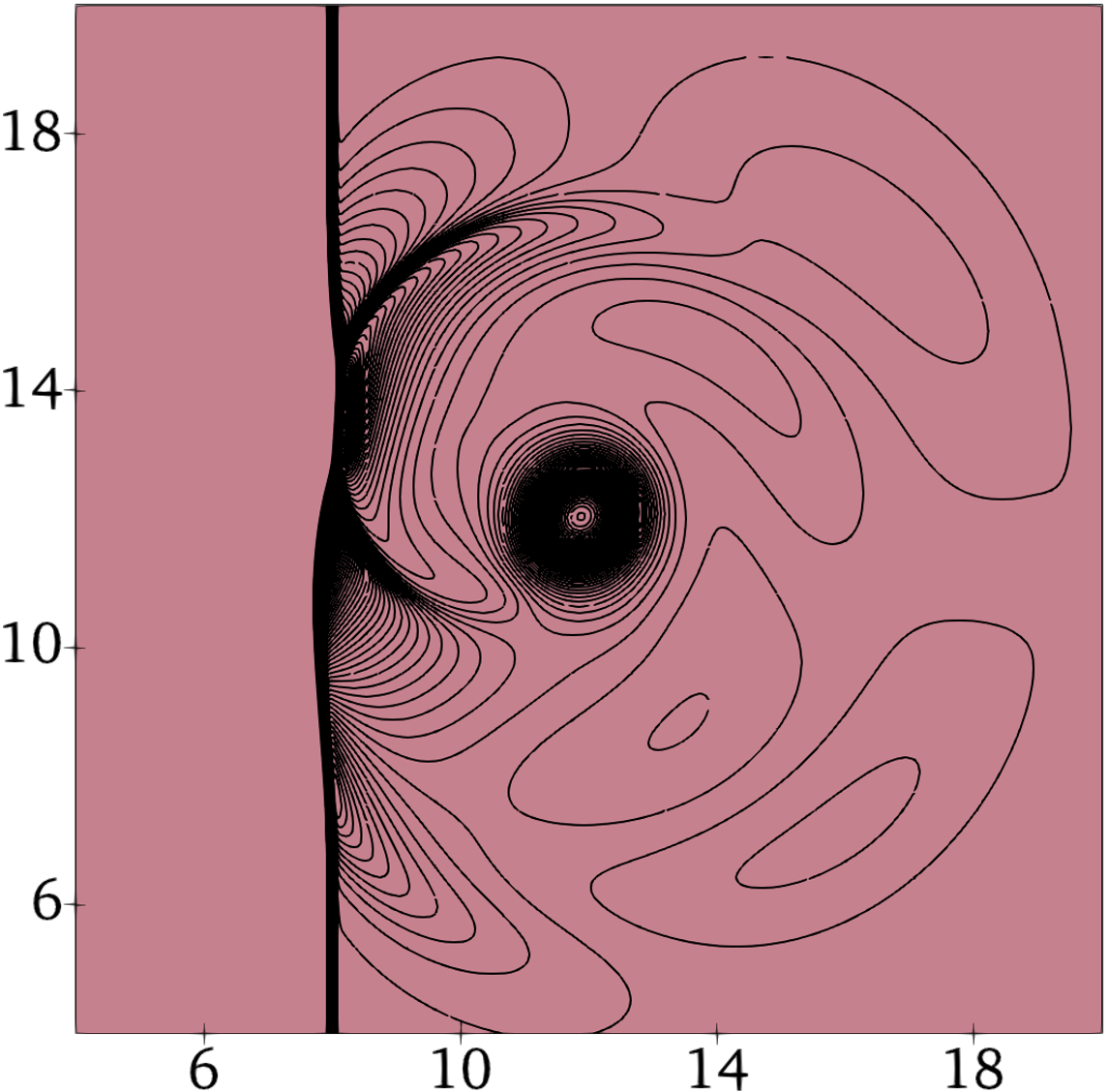}
\hfill
    \includegraphics[width=.325\linewidth]{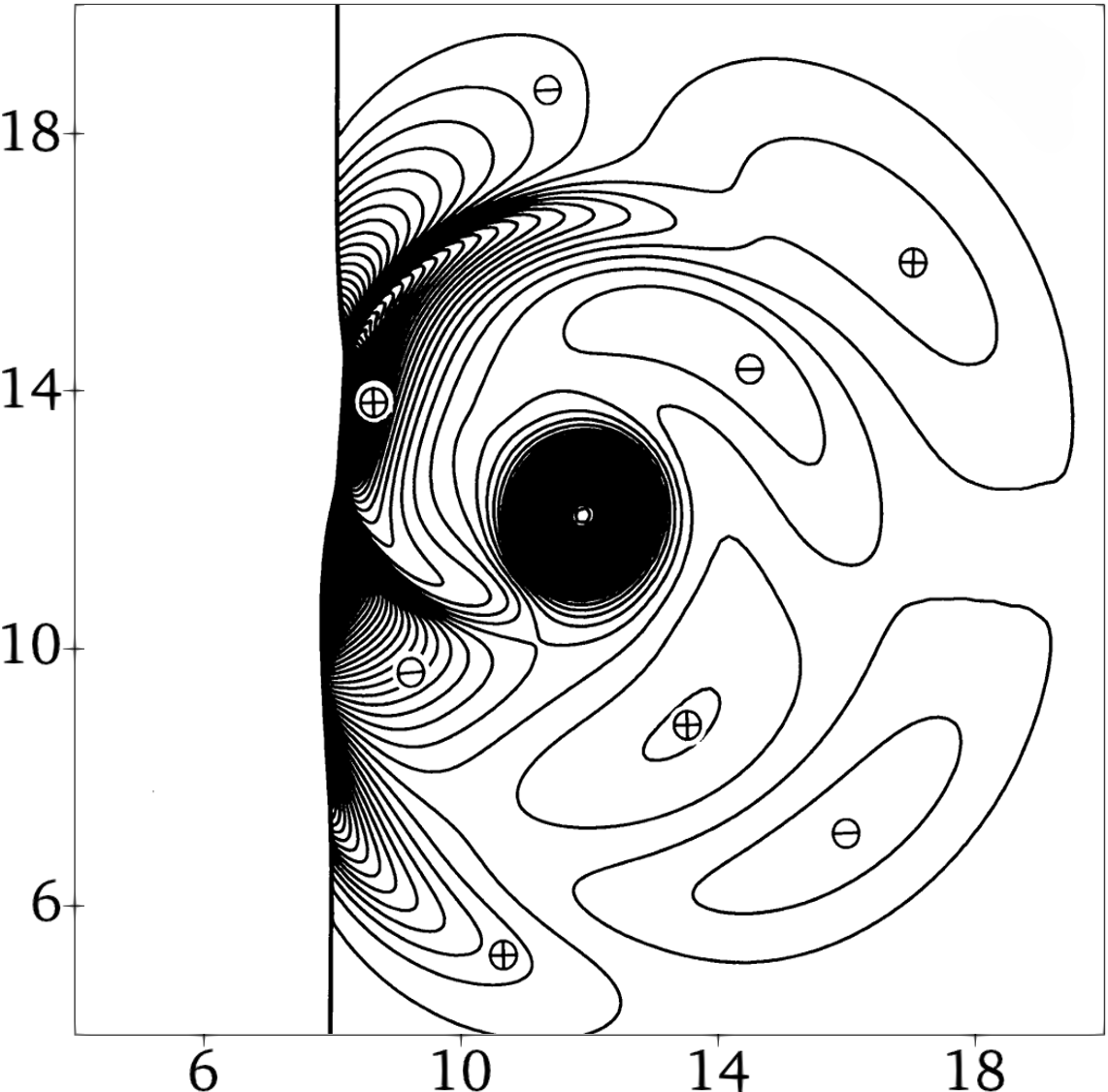}
\caption{
Solutions of the shock-vortex interaction 
case C (\text{Ma}$_s = 1.2$ and \text{Ma}$_v = 0.25$)
at $t^*=6$ depicted with $140$ equidistant contours of the sound pressure $p_{Sound}$. 
The solution with AMR is depicted on the left, while the solution on a uniform grid is shown in the middle and the reference solution from Inoue \& Hattori \cite{inoue1999sound} on the right 
(The signs $\oplus$ and $\ominus$ in the reference denote regions of positive and negative sound pressure, respectively).
The spatial resolution of the present solver is indicated as colored background (blue: $\delta x = \delta y = L_x/420 = L_y/360$, gray: $\delta x = \delta y = L_x/820 = L_y/760$, red: $\delta x = \delta y = L_x/1680 = L_y/1440$).
}
\label{fig:ShockVortexLevelsA}
\end{figure*}

\subsubsection{Shock–vortex interaction}

Next, a NSF level flow, namely a shock–vortex interaction, is presented in which proper recovery of the dissipation rates together with high resolution is crucial.
The setup of Inoue \& Hattori \cite{inoue1999sound} as adopted in \cite{saadat2021extended} was followed here, where the main field is separated by a stationary shock with Mach number \text{Ma}$_s$ and the left- and right-hand initial states satisfy the Rankine-–Hugoniot jump conditions.
For a pre-shock state of $(\rho, p, u_x=\text{Ma}_s c_{s}, u_y=0)_l$ on the left-hand side, where the flow velocity is given from the imposed Ma number via $c_{s,l} = \sqrt{\gamma T_l} = \sqrt{\gamma p_l/\rho_l}$, the post-shock state $(\rho, p, u_x, u_y=0)_r$ is found as
\begin{align}
    \rho_r  &= \rho_l \frac{(\gamma+1)\text{Ma}_s^2}{(\gamma-1)\text{Ma}_s^2+2} \text{,} \\
    p_r     &= p_l \frac{2\gamma \text{Ma}_s^2-(\gamma-1)}{(\gamma+1)} \text{,}  \\ 
    u_{x,r} &= u_{x,l} \frac{(\gamma-1)\text{Ma}_s^2+2}{(\gamma+1)\text{Ma}_s^2} 
    \text{.} 
\end{align}
The resulting initial field $(\rho, p, u_x, u_y)_\infty$ is perturbed by an isentropic vortex which is advected through the shock. 
The maximum tangential velocity of the vortex defines the vortex Mach number as Ma$_v = u_{\varphi}^{max} / c_{s,l}$.
In Cartesian coordinates, the initial conditions for the vortex read
\begin{align}
    u_x &= u_{x,\infty} + c_{s,l} \text{Ma}_v \frac{y-y_v}{r_v} \cdot e^{(1-r^2)/2}  \text{,} \\
    u_y &= u_{y,\infty} - c_{s,l} \text{Ma}_v \frac{x-x_v}{r_v} \cdot e^{(1-r^2)/2} \text{,} \\
    \rho &= \rho_{\infty} \left[ 1 - \frac{\gamma -1}{2} \text{Ma}_v^2 \cdot e^{(1-r^2)} \right]^{1/(\gamma -1)} \text{,} \\
    p &= p_{\infty} \left[ 1 - \frac{\gamma -1}{2} \text{Ma}_v^2 \cdot e^{(1-r^2)} \right]^{\gamma/(\gamma -1)} \text{.}
\end{align}
Note that the field is perturbed on both sides of the shock to match the reference solution in the DNS setup \cite{inoue1999sound}, where the influenced region of the vortex in the IC overlaps the shock slightly.
The reduced radius $r$ is defined with the vortex center position ($x_v$, $y_v$) and the vortex radius $r_v$ as $r = \sqrt{(x-x_v)^2+(y-y_v)^2}/r_v$, where the vortex radius is connected to the dynamic viscosity of the fluid via the Reynolds number defined as Re$_v = \rho_l c_{s,l} r_v / \mu$.

\begin{figure}[b]
\centering
\includegraphics[width=.99\linewidth]{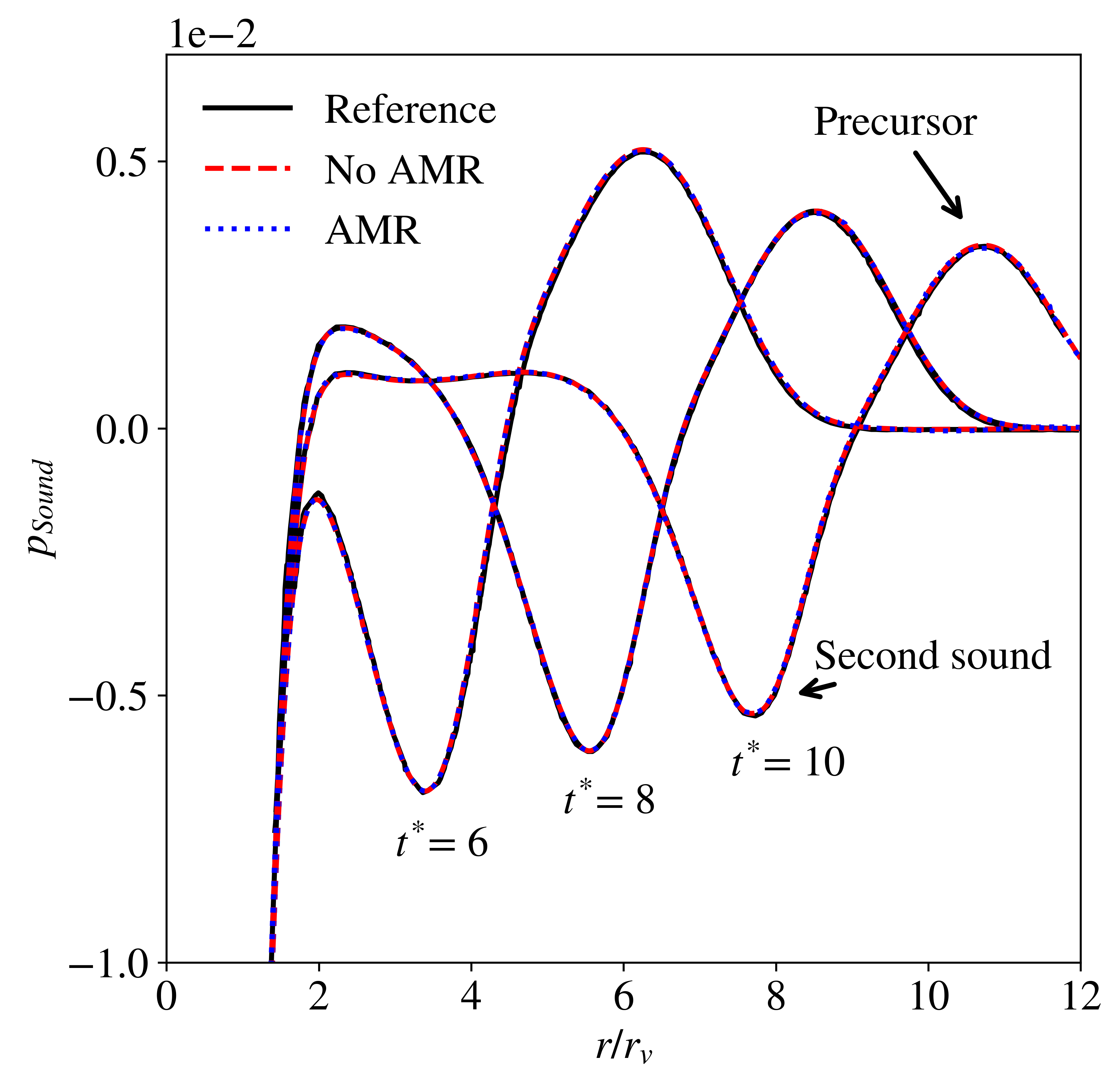}
\caption{Comparison of the radial sound pressure distribution of case C (\text{Ma}$_s = 1.2$ and \text{Ma}$_v = 0.25$) at three different non-dimensional times $t^*=6, 8, 10$, measured in the direction of $\varphi = 45^\circ$ degrees with origin in the vortex center.}
\label{fig:ShockVortexSoundpressureA}
\end{figure}

\begin{figure*}[t]
\centering
    \includegraphics[width=.325\linewidth]{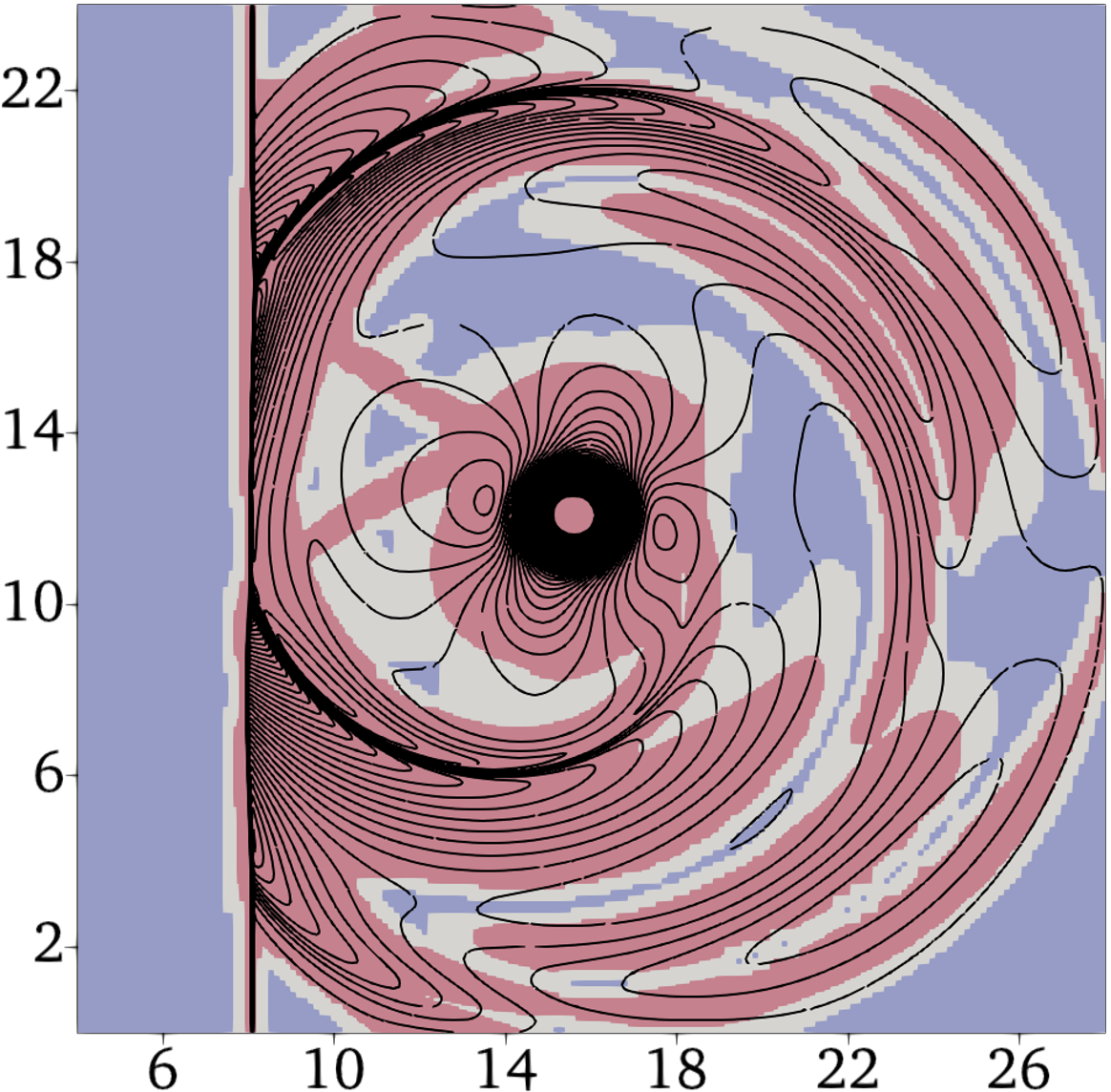}
\hfill
    \includegraphics[width=.325\linewidth]{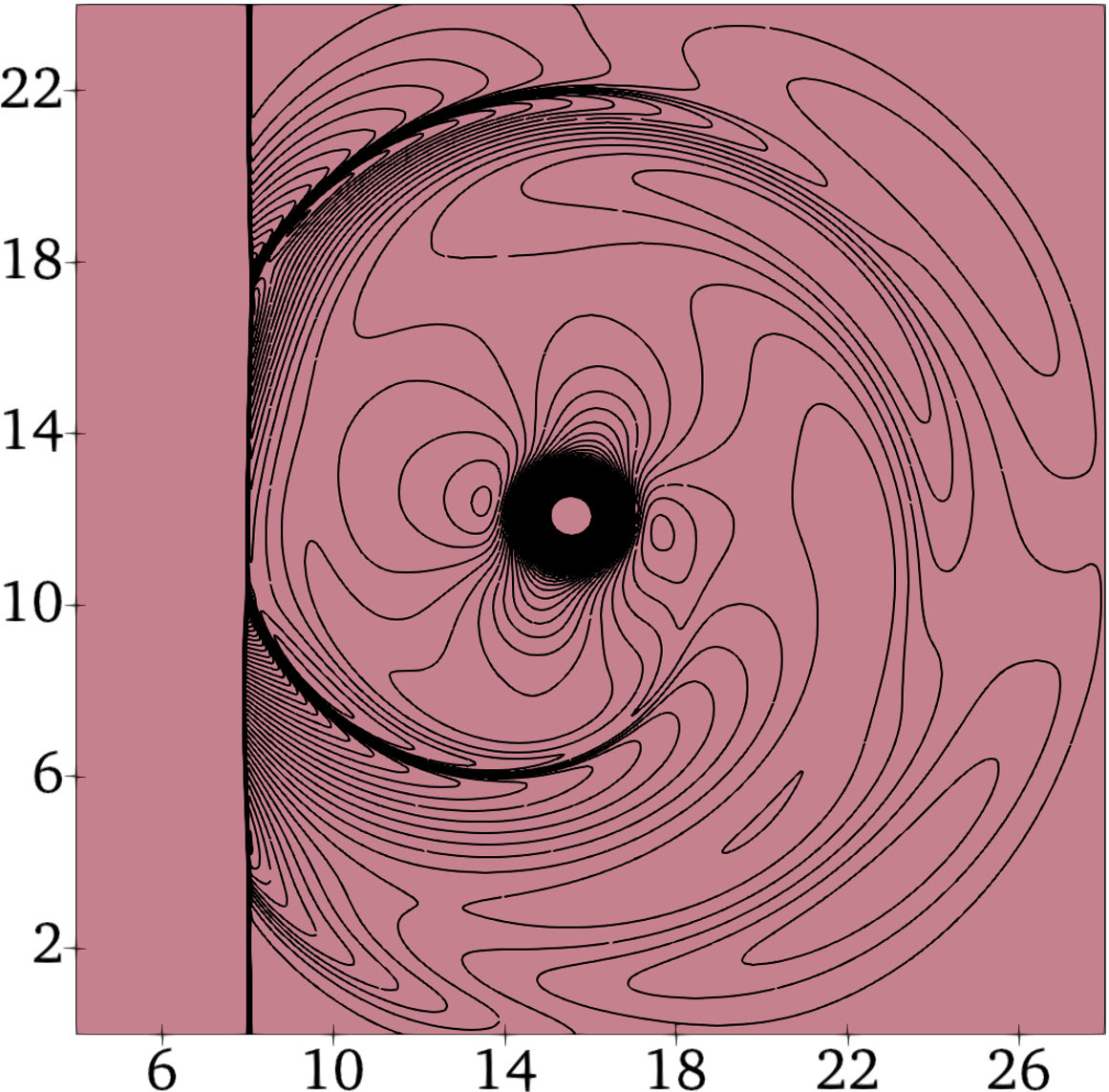}
\hfill
    \includegraphics[width=.325\linewidth]{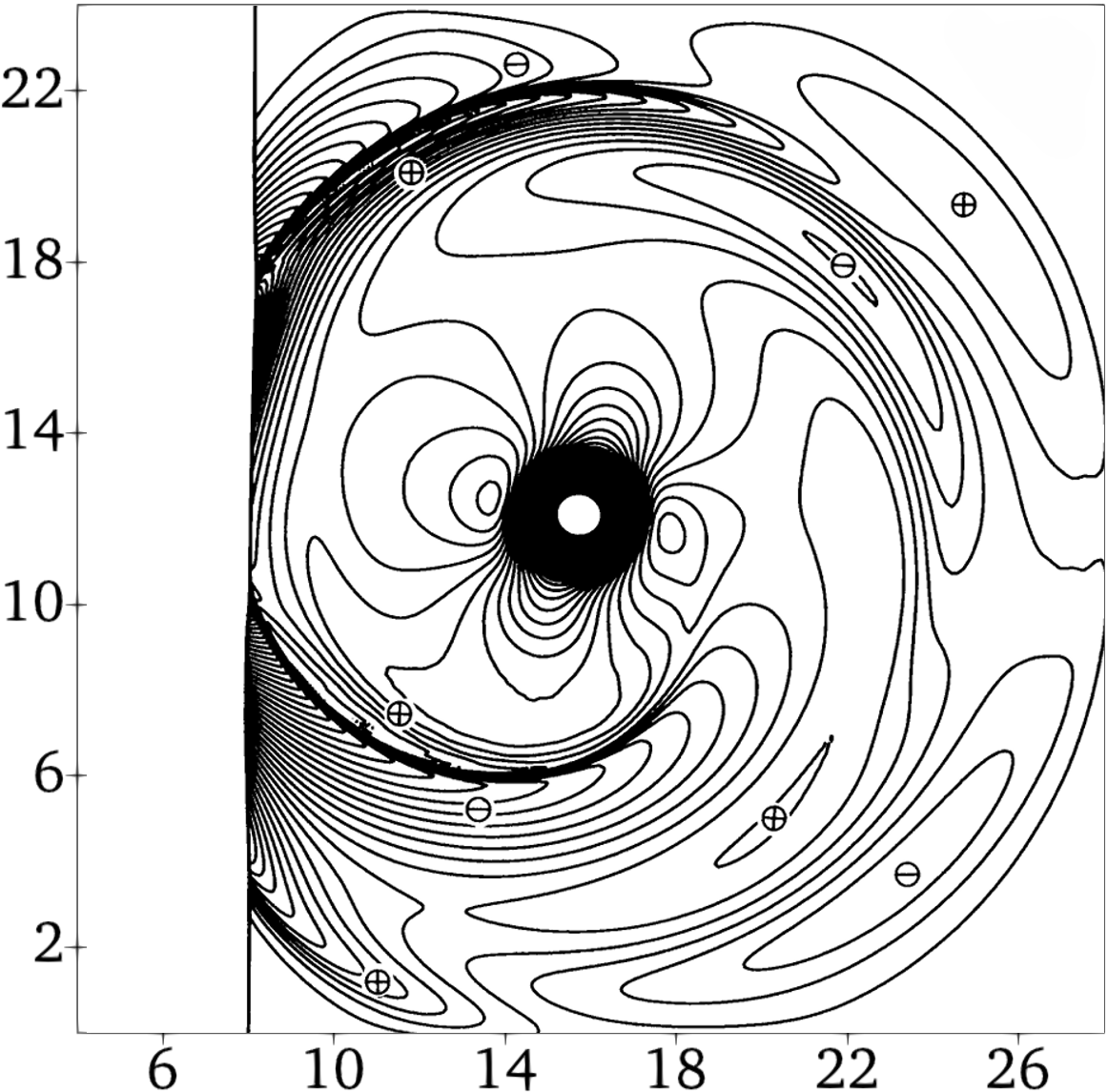}
\caption{
Solutions of the shock-vortex interaction case G (\text{Ma}$_s = 1.29$ and \text{Ma}$_v = 0.39$) at $t^*=10.3$ depicted with $90$ equidistant contours of the sound pressure $p_{Sound}$. 
The solution with AMR is depicted on the left, while the solution on a uniform grid is shown in the middle and the reference solution from Inoue \& Hattori \cite{inoue1999sound} on the right 
(The signs $\oplus$ and $\ominus$ in the reference denote regions of positive and negative sound pressure, respectively).
The spatial resolution of the present solver is indicated as colored background (blue: $\delta x = \delta y = L_x/420 = L_y/360$, gray: $\delta x = \delta y = L_x/820 = L_y/760$, red: $\delta x = \delta y = L_x/1680 = L_y/1440$).
}
\label{fig:ShockVortexLevelsB}
\end{figure*}

The shock position was initialized at $x_s$ = $8$ with $\rho_l = 1$ and $p_l=1$ in a domain $x \in [0, 28]$ and $y \in [0, 24]$.
The vortex with $r_v = 1$, rotating in clock-wise direction, was centered at \mbox{($x_v$, $y_v$)} \mbox{$= (6, 12)$}.
The Reynolds number was set to Re$_v = 800$ and the Prandtl number to $Pr = 0.75$.
As in the original setup, periodic BCs were used for the boundaries in y-direction.
Cases C and G from \cite{inoue1999sound} were run.
For case C, \text{Ma}$_s = 1.2$ and \text{Ma}$_v = 0.25$ was imposed, 
whereas for case G, these numbers were set to \text{Ma}$_s = 1.29$ and \text{Ma}$_v = 0.39$.
Vorticity as well as the gradients of pressure, density and vorticity were used as refinement sensors.
A resolution of $\delta x = \delta y = L_x/1680 = L_y/1440$ was applied for the case without AMR and as the peak resolution in AMR, i.e. $\delta x_0 = \delta y_0 = L_x/420 = L_y/360$ on level $l=0$.

Fig.~\ref{fig:ShockVortexLevelsA} depicts the sound pressure contours at the non-dimensional time $t^* = 6$ for case C with and without AMR, as compared to the reference solution.
Thereby, the sound pressure is defined as $p_{Sound} = p/p_r -1$, with $p_r$ being the initial pressure in the post-shock region, and the non-dimensional time is given by $t^* = t c_{s,l} / r_v$.
Note that the sound pressure usually amounts to a small perturbation in the order of $\lesssim 1\%$ of the hydrodynamic pressure on top of it and is therefore a rather sensitive quantity.
A good agreement of the pressure contours with the reference DNS solution of \cite{inoue1999sound} can be observed. 
The AMR solver perfectly replicates the solution of the uniform high-resolution grid, since the most important features of the flow, where the relevant dissipative physics occurs, are equally resolved.
Importantly, the deformation of the shock, including the two shock reflections, is also well captured.
For further quantitative comparison, the radial distribution of the sound pressure was measured in the direction of $\varphi=45^\circ$ degrees with the origin at the vortex center.
The results with and without AMR as compared to the reference simulations are depicted in Fig.~\ref{fig:ShockVortexSoundpressureA} for three different non-dimensional times $t^*$.
It can clearly be seen that the reference solutions are perfectly matched.
In particular, the temporal development of the position and magnitude of the peak sound pressure of the first sound (precursor) and the second sound emerging from the shock-vortex interaction are well captured.

\begin{figure}[b]
\centering
\includegraphics[width=.99\linewidth]{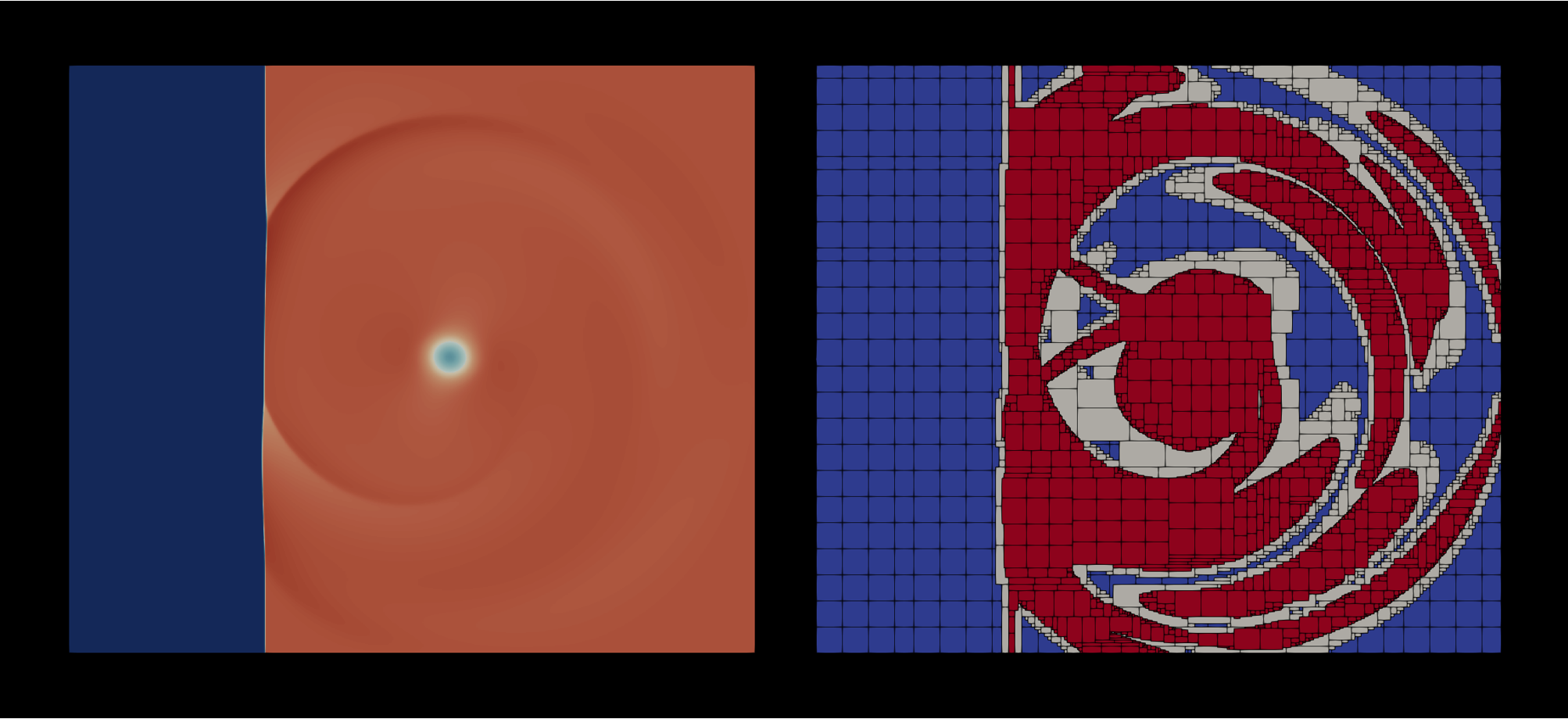}
\caption{
A video (Multimedia available online) of the AMR simulation in Fig.~\ref{fig:ShockVortexLevelsB} displaying the temporal evolution of the sound pressure and AMR levels including individual AMR patches indicated by black lines (color bars are given in the video and the coloring of resolution follows Fig.~\ref{fig:ShockVortexLevelsB}).
}
\label{fig:SVG_video}
\end{figure}

\begin{figure}[h]
\centering
\includegraphics[width=.99\linewidth]{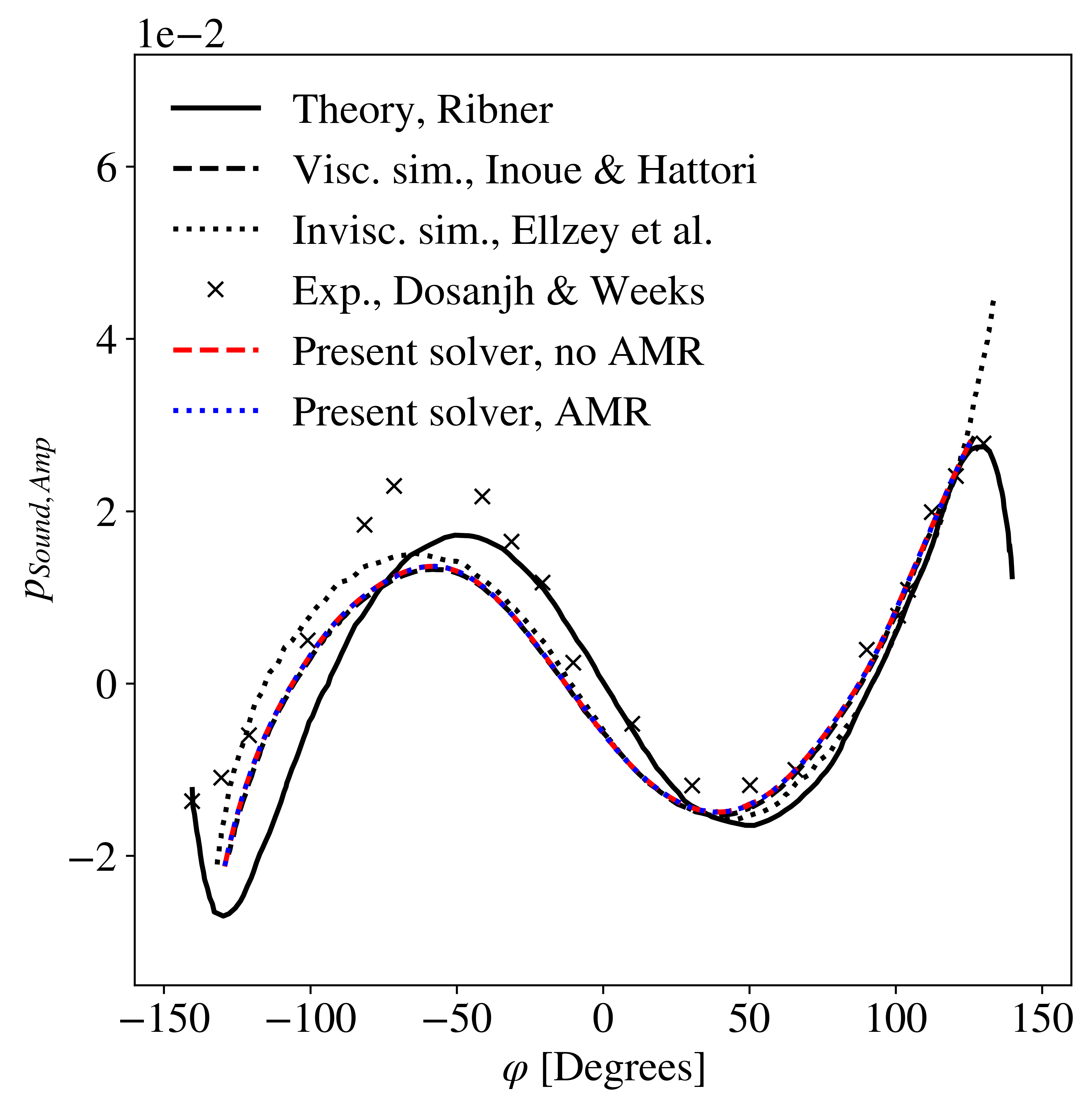}
\caption{
Comparison of the circumferential distribution of the sound pressure amplitude of case G (\text{Ma}$_s = 1.29$ and \text{Ma}$_v = 0.39$) at non-dimensional time $t^*=10.3$ with 
theoretical results of Ribner \cite{Ribner1985}, 
viscous DNS simulations of Inoue \& Hattori \cite{inoue1999sound}, 
inviscid simulations by Ellzey et al. \cite{Ellzey1995}, 
and experimental results by Dosanjh \& Weeks \cite{DOSANJH1965}.
}
\label{fig:ShockVortexSoundpressureB}
\end{figure}

The sound pressure contours of case G at non-dimensional time $t^* = 10.3$ are depicted in Fig.~\ref{fig:ShockVortexLevelsB}.
Again, excellent agreement can be seen between the solver with or without AMR and the reference solution, despite operating in a flow regime in terms of the Ma number very close to the limits of Galilean invariance.
A video of the temporal evolution of the AMR simulation presented in Fig.~\ref{fig:ShockVortexLevelsB} is given as an integral multimedia in Fig.~\ref{fig:SVG_video} (Multimedia available online).
This time, the sound pressure amplitude $p_{sound, amp}$ is quantitatively compared to inviscid and viscous DNS simulations, experimental results as well as theory.
Thereby, the sound pressure amplitude is computed as $p_{Sound, Amp} = (p_2 - p_1)/p_r$, i.e. the difference between the peak sound pressure of the precursor, denoted as $p_{Sound,1} = p_1/p_r - 1$, and the second sound, $p_{Sound,2} = p_2/p_r - 1$, where $p_1$ and $p_2$ are measured equivalently to \cite{inoue1999sound} at $r/r_v = 10.8$ and $r/r_v = 8.8$ from the vortex center, respectively. 
The results for the circumferential distribution are depicted in Fig.~\ref{fig:ShockVortexSoundpressureB}.
It can be seen that the reference solution of the viscous simulation \cite{inoue1999sound} is perfectly matched, both with and without AMR.
The results lie well within the knowledge band acquired by former simulations \cite{inoue1999sound, Ellzey1995}, experimental measurements conducted at a much higher Reynolds number of Re $\approx 160'000$ \cite{DOSANJH1965}, and theoretical results \cite{Ribner1985}, which further validates the presented model and mesh refinement methodology.
Therefore, these results demonstrate that the solver performs excellently in obtaining sensitive quantities for complex setups in compressible and moderately supersonic flows at the NSF level.

\subsection{Computational efficiency}

Lastly, the computational efficiency of AMR compared to simulations with a uniform grid of the same peak resolution is assessed.
It should be emphasized that the computational efficiency of AMR is very problem specific, as it depends, on one hand, on the amount of relevant flow features in the domain.
This is due to the proportion of cells in the domain that need to be refined to adequately resolve a given flow feature.
On the other hand, there is an AMR overhead, which includes for example 
the filling of level-wise BCs, 
the application of conservative refluxing, 
but also 
the regridding procedure 
which strongly depends on the temporal evolution of the flow features.
In summary, this can lead to extreme efficiencies for problems which are only sparsely populated with relevant flow features and if certain AMR parameters, e.g. regridding times, grid efficiency, or patch sizes, can be set to produce minimal overhead, and vice versa.
For this reason, the following assessment of cost reduction is restricted to the two-dimensional cases along with their mentioned AMR parameters as presented in this paper.
These cases represent challenging benchmarks for AMR cost in terms of relevant flow feature densities and evolution patterns.

An estimate of the efficiency gains in AMR can be made in terms of the number of advanced cells over the whole simulation in the non-AMR case normalized by the AMR case.
The results can be found in table~\ref{tab:Cost}.
Over all considered benchmarks, factors of $4$ to $9$ were achieved.
These numbers provide an approximate estimate for the reduction in memory usage, as memory usage is roughly proportional to the number of advanced cells in the simulation when neglecting minor contributions from using flux registers.
These values also provide an estimate for the reduction in computational cost without considerations of the AMR overhead.

\begin{table}[h]
\footnotesize	
\centering
\begin{tabular}{|c|c|}
    \hline
         Case & Normalized number of advanced cells \\
    \hline
        Riemann \#7    &  9.019 \\
        Riemann \#8    &  8.256 \\
        Riemann \#12   &  7.464 \\
        Riemann \#3*   &  6.168 \\
        Shock-vortex C &  4.549 \\
        Shock-vortex G &  4.051 \\
    \hline
\end{tabular}
\caption{
Reduction factor in computational cost and memory usage of AMR without overhead for the presented two-dimensional test cases with the same peak resolution. 
The shock-vortex case C is compared for the full length of $t^*=10$.
}
\label{tab:Cost}
\end{table}

Note that the optimal number of processors generally differs between simulations with and without AMR for the same peak resolution, hence, a fair estimation of AMR overhead has to be conducted using optimal parallelization in both cases.
Moreover, the AMR overhead also depends on the parallelization strategy, e.g. communications, parallel repartitioning and load balancing of the dynamic grid as mentioned in paragraph~\ref{par:parallelAMR}.
For an analysis of the parallelization approach and scaling effects of the employed library, which is not the scope of this work, the interested reader is referred to the original references \cite{amrex1, amrex2, amrex3}.
For an optimal number of processors in both the AMR and non-AMR case of the shock-vortex interaction G, the performance of the AMR solver decreased to $69.5$\% of the single level-solver, indicating that the rest, i.e. $30.5$\%, is spent as AMR overhead.
Importantly, the AMR parameters employed for this consideration were chosen to penalize the AMR overhead; in particular the regridding, which was done on every level at every second level time step with a high grid efficiency of $0.98$ using multiple refinement sensors.
Therefore, the estimated AMR overhead should be considered as an upper bound, as the regridding part of the AMR overhead can be minimized significantly by choosing these parameters in a more application oriented way, for example most notably by elevating the regridding time by an order of magnitude.
All in all, these results mark a significant improvement, which further validates the efficiency and effectiveness of the presented discrete kinetic AMR solver.

\section{Summary and conclusions\label{sec:conslusions}}

In this paper, a discrete velocity Boltzmann solver with adaptive mesh refinement (AMR) is presented for the simulation of compressible Euler and Navier--Stokes--Fourier level flows.
The kinetic model based on the single relaxation time Bhatnagar--Gross--Krook collision kernel makes use of double-distribution functions as well as the quasi-equilibrium approach to capture various specific heat ratios and non-unity Prandtl number flows, respectively.
The model employs a discrete set of designer particle velocities in phase space, where the distribution function is discretized by a Gauss--Hermite quadrature.
Strict conservation is ensured by space-time discretization with the finite-volume method.
Furthermore, the parallel block-structured AMR approach is applied to locally refine the grid.
Throughout the manuscript, emphasis is placed on the correct recovery of hydrodynamics as well as restoring strict conservation in the AMR methodology by employing refluxing operations.

A validation of the AMR methodology has shown that conservation laws are strictly preserved.
Moreover, the correct hydrodynamics was accurately recovered by the solver, including correct and Galilean invariant dispersion rates for a temperature range over $3$ orders of magnitude and dissipation rates of all eigen-modes up to Mach $\approx 1.8$, respectively.
Results for further one- and two-dimensional benchmarks up to Mach numbers of $3.2$ and temperature ratios of $7$, such as the Sod and Lax shock tubes, the Shu--Osher problem, several Riemann problems as well as viscous shock-vortex interactions, demonstrated that the solver accurately captures the reference solutions.
Excellent performance in obtaining sensitive quantities was proven, for example in the test case involving nonlinear acoustics, whilst, for the same accuracy and fidelity of the solution, the AMR methodology significantly reduced computational cost and memory footprints.
Over all demonstrated two-dimensional problems, up to a $4$- to $9$-fold reduction was achieved and an upper limit of the AMR overhead of $30$\% was found in a case with very cost-intensive parameter choice.

The presented discrete kinetic model has known limitations in the temperature and Mach number range dictated by the static reference frame.
Therefore, future work will focus on efficient AMR implementations of locally shifted velocity sets for high Mach number applications and the particles on demand method for compressible flows with strong discontinuities.

\section*{Acknowledgments}
We thank the organization of the 33rd Discrete Simulation of Fluid Dynamics Conference (DSFD) in Zürich, Switzerland held over July 9-12 2024 for creating the platform at which this work was first presented.
This work was supported by European Research Council (ERC) Advanced Grant No.\ 834763-PonD and by the Swiss National Science Foundation (SNSF) Grant 200021-228065. 
Computational resources at the Swiss National Super Computing Center (CSCS) were provided under Grant No. s1286.
R.M.S. acknowledges Nilesh Sawant and Abhimanyu Bhadauria for useful discussions regarding some implementations used in this work.

\section*{Author Declarations}

\noindent\textbf{Conflict of Interest}\\
The authors have no conflicts to disclose.

\noindent\textbf{Ethical approval}\\
The work presented here by the authors did not require ethics approval or consent to participate.

\noindent\textbf{Author contributions}\\
R.M.S.:
Conceptualization of the study, 
derivation and implementation of the solver, 
formal analysis,
evaluation of the solver, 
data analysis, 
writing- initial manuscript and revised versions. 
S.A.H.:
Conceptualization of the study, 
formal analysis,
writing- initial manuscript and revised versions. 
I.V.K.:
Conceptualization of the study, 
writing- initial manuscript and revised versions,
funding acquisition, resources.
All authors approved the final manuscript.

\noindent\textbf{Data availability statement}\\
The data that support the findings of this study are available within the article or from the corresponding author(s) upon reasonable request.

\section*{Appendix}
\appendix
\section{Multi-scale analysis}
\label{app:multiscale}
Let us consider the following system of equations,
\begin{multline}
	\mathcal{D}_t \{f, g\} = \frac{1}{\tau_1}\left(\{f^{\rm eq}, g^{\rm eq}\} - \{f, g\}\right) \\ + \left(\frac{1}{\tau_1}-\frac{1}{\tau_2}\right) \left(\{f^{*}, g^{*}\} - \{f^{\rm eq}, g^{\rm eq}\}\right),
\end{multline}
where, $\mathcal{D}_t=\partial_t + \bm{v}\cdot\bm{\nabla}$. For the multi-scale analysis let us introduce the following parameters: characteristic flow velocity $\mathcal{U}$, characteristic flow scale $\mathcal{L}$, characteristic flow time $\mathcal{T}=\mathcal{L}/\mathcal{U}$, characteristic density $\bar{\rho}$, speed of sound of ideal gas $c_s=\sqrt{\gamma R T}$. With these, the variables are reduced as follows (primes denote non-dimensional variables): time $t=\mathcal{T}t'$, space $\bm{x}=\mathcal{L}\bm{x}'$, flow velocity $\bm{u}=\mathcal{U}\bm{u}'$, particle velocity $\bm{v}=c_s\bm{v}'$, density $\rho$=$\bar{\rho}\rho'$, distribution function $f=\bar{\rho}c_s^{-3}f'$.
Furthermore, the following non-dimensional groups are introduced: Knudsen number ${\rm Kn}={\tau c_s}/{\mathcal{L}}$ and Mach number ${\rm Ma}={\mathcal{U}}/{c_s}$. With this, the equations are rescaled as follows:
\begin{multline}
	{\rm Kn}\,\mathcal{D}_t' \{f', g'\} = \frac{1}{\tau_1'}\left(\{f^{\rm eq'}, g^{\rm eq'}\} - \{f', g'\}\right) \\ + \left(\frac{1}{\tau_1'}-\frac{1}{\tau_2'}\right) \left(\{f^{*'}, g^{*'}\} - \{f^{\rm eq'}, g^{\rm eq'}\}\right),
\end{multline}
Assuming ${\rm Kn}\sim\epsilon$ and dropping the primes for the sake of readability,
	\begin{multline}
	    \epsilon \mathcal{D}_t \{f, g\} = \frac{1}{\tau_1}\left(\{f^{\rm eq}, g^{\rm eq}\} - \{f, g\}\right) \\ + \left(\frac{1}{\tau_1}-\frac{1}{\tau_2}\right) \left(\{f^{*}, g^{*}\} - \{f^{\rm eq}, g^{\rm eq}\}\right).
	\end{multline}
Then introducing multi-scale expansions:
	\begin{multline}
	    \{f, g\} = \{f^{(0)}, g^{(0)}\} + \epsilon \{f^{(1)}, g^{(1)}\} + \epsilon^2 \{f^{(2)}, g^{(2)}\} 
     \\+ O(\epsilon^3),
	\end{multline}
and,
	\begin{multline}
	    \{f^*, g^*\} = \{f^{*(0)}, g^{*(0)}\} + \epsilon \{f^{*(1)}, g^{*(1)}\} + \epsilon^2 \{f^{*(2)}, g^{*(2)}\} 
     \\+ O(\epsilon^3),
	\end{multline}
the following equations are recovered at scales $\epsilon$ and $\epsilon^2$:
	\begin{subequations}
	\begin{align}
	\epsilon &: \mathcal{D}_{t}^{(1)} \{f^{(0)}, g^{(0)}\} = 
 \nonumber \\&-\frac{1}{\tau_1} \{f^{(1)} , g^{(1)}\} + \left(\frac{1}{\tau_1}-\frac{1}{\tau_2}\right) \{f^{*(1)}, g^{*(1)}\}, \label{Eq:CE_Eq_orders_1}\\
	\epsilon^2 &: \partial_t^{(2)}\{f^{(0)}, g^{(0)}\} + \mathcal{D}_{t}^{(1)}\{f^{(1)}, g^{(1)}\} = 
 \nonumber \\&-\frac{1}{\tau} \{f^{(2)} , g^{(2)}\ + \left(\frac{1}{\tau_1}-\frac{1}{\tau_2}\right) \{f^{*(2)}, g^{*(2)}\}, \label{Eq:CE_Eq_orders_2}
	\end{align}
    \label{Eq:CE_Eq_orders}
    \end{subequations}
with $\{f^{(0)}, g^{(0)}\}=\{f^{\rm eq}, g^{\rm eq}\}$. Note that for this system, the solvability conditions are:
	\begin{subequations}
	\begin{align}
    \forall k>0, & \int_{\mathbb{R}^D} f^{(k)} d\bm{v}= \int_{\mathbb{R}^D} f^{*(k)} d\bm{v} = 0,\\
    \forall k>0, & \int_{\mathbb{R}^D} \bm{v} f^{(k)} d\bm{v} = \int_{\mathbb{R}^D} \bm{v} f^{*(k)} d\bm{v} = 0,\\
    \forall k>0, & \int_{\mathbb{R}^D} \left(g^{(k)} + \frac{\bm{v}^2}{2} f^{(k)}\right) d\bm{v} =
    \nonumber \\& \int_{\mathbb{R}^D} \left(g^{*(k)} + \frac{\bm{v}^2}{2} f^{*(k)}\right) d\bm{v} = 0.
	\end{align}
    \end{subequations}
Taking the moments, $\int\{f, \bm{v} f\}d\bm{v}$, of the Chapman--Enskog-expanded equations at order $\epsilon$:
	\begin{eqnarray}
	    \partial_t^{(1)}\rho + \bm{\nabla}\cdot\rho \bm{u} &=& 0,\label{eq:approach2_continuity1}\\
	    \partial_t^{(1)}\rho \bm{u} + \bm{\nabla}\cdot\rho \bm{u}\otimes\bm{u} + \bm{\nabla}\cdot p\bm{I} &=& 0.\label{eq:approach2_NS1}
	\end{eqnarray}
For the energy balance equation, taking the moment $\int_{\mathbb{R}^D} \left(g + \frac{\bm{v}^2}{2}f\right) d\bm{v}$,
\begin{equation}
	\partial_t^{(1)}E + \bm{\nabla}\cdot E \bm{u} + \bm{\nabla}\cdot p\bm{u} = 0.
\end{equation}
Using the Euler level momentum balance and continuity equations, a balance equation for kinetic energy can be derived as,
\begin{equation}
    \partial_t^{(1)}K + \bm{\nabla}\cdot K \bm{u} + \bm{u}\cdot \bm{\nabla}p = 0,
\end{equation}
which in turn can be used to derive a balance equation for internal energy,
\begin{equation}
    \partial_t^{(1)} U + \bm{\nabla}\cdot U \bm{u} + p \bm{\nabla}\cdot\bm{u} = 0,
\end{equation}
and pressure,
\begin{equation}
    \partial_t^{(1)} p + \bm{\nabla}\cdot p \bm{u} + \frac{R}{c_V} p \bm{\nabla}\cdot\bm{u} = 0,\label{eq:pressure_euler_CE}
\end{equation}
where
\begin{equation}
        \frac{\partial U}{\partial T} = \rho c_V.
\end{equation}
Going up one order in $\epsilon$, at order $\epsilon^2$ the continuity equation is:
	\begin{equation}
	    \partial_t^{(2)}\rho = 0,\label{eq:approach2_continuity2}
	\end{equation}
while for the momentum balance equation one has:
    \begin{equation}
        \partial_t^{(2)}\rho \bm{u} + \bm{\nabla}\left(\int_{\mathbb{R}^D}\bm{v}\otimes\bm{v} f^{(1)}\right) = 0.
    \end{equation}
The second term can be further expanded using the first-order-in-$\epsilon$ equation as:
    \begin{multline}
        \int_{\mathbb{R}^D}\bm{v}\otimes\bm{v} f^{(1)}d\bm{v} =\\ -\tau_1\left[\partial_t^{(1)}\left(\int_{\mathbb{R}^D}\bm{v}\otimes\bm{v}f^{(0)}d\bm{v}\right) + \bm{\nabla}\cdot \left(\int_{\mathbb{R}^D}\bm{v}\otimes\bm{v}\otimes\bm{v}f^{(0)}d\bm{v}\right) \right. \\ \left. + -\left(\frac{1}{\tau_1}-\frac{1}{\tau_2}\right)\left(\int_{\mathbb{R}^D}\bm{v}\otimes\bm{v} f^{*(1)}d\bm{v}\right)\right].
    \end{multline}
    The remainder of the expansion, depends on the construction of the quasi-equilibria attractors. In the case of a quasi-equilibrium state constructed using minimization under constraint on the following moments $\{\rho, \rho u_\alpha, E, P_{\alpha\beta}\}$ where the latter are the pressure tensor components,
    \begin{equation}
        \int_{\mathbb{R}^D}\bm{v}\otimes\bm{v} f^{*(1)}d\bm{v} = 0,
    \end{equation}
    one gets,
    \begin{multline}
        \int_{\mathbb{R}^D}\bm{v}\otimes\bm{v} f^{(1)}d\bm{v} =\\ -\tau_1\left[\partial_t^{(1)}\left(\rho\bm{u}\otimes\bm{u} + p\bm{I}\right) + \bm{\nabla}\cdot \rho\bm{u}\otimes\bm{u}\otimes\bm{u} \right. \\ \left. + \bm{\nabla}p\bm{u} + \bm{\nabla}p\bm{u}^{\dagger} + \bm{I}\bm{\nabla}\cdot p\bm{u}\right].
    \end{multline}
    Note that different from a mono-atomic system, in 3-D this leads to constraints on 11 moments. In the case of constraints on the following moments $\{\rho, \rho u_\alpha, E, q_{\alpha}\}$ where the latter are contracted third-order moments,
    \begin{equation}
        \int_{\mathbb{R}^D}\bm{v}\otimes\bm{v} f^{*}d\bm{v} = \int_{\mathbb{R}^D}\bm{v}\otimes\bm{v} fd\bm{v},
    \end{equation}
    one gets,
    \begin{multline}
        \int_{\mathbb{R}^D}\bm{v}\otimes\bm{v} f^{(1)}d\bm{v} =\\ -\tau_2\left[\partial_t^{(1)}\left(\rho\bm{u}\otimes\bm{u} + p\bm{I}\right) + \bm{\nabla}\cdot \rho\bm{u}\otimes\bm{u}\otimes\bm{u} \right. \\ \left. + \bm{\nabla}p\bm{u} + \bm{\nabla}p\bm{u}^{\dagger} + \bm{I}\bm{\nabla}\cdot p\bm{u}\right].
    \end{multline}
Next expanding,
    \begin{multline}
        \partial_t^{(1)}\left(\rho\bm{u}\otimes\bm{u} + p\bm{I}\right) = \bm{u}\otimes\partial_t^{(1)}\rho \bm{u} + \bm{u}\otimes\partial_t^{(1)}\rho \bm{u}^{\dagger} \\ - \bm{u}\otimes\bm{u}\partial_t^{(1)}\rho + \partial_t^{(1)}p\bm{I},
    \end{multline}
where we use,
    \begin{equation}
        \bm{u}\otimes\partial_t^{(1)}\rho \bm{u} = -\bm{u}\otimes\left[\bm{\nabla}\cdot\rho\bm{u}\otimes\bm{u} + \bm{\nabla}p\right],
    \end{equation}
and,
    \begin{equation}
        \bm{u}\otimes\bm{u}\partial_t^{(1)}\rho = - \bm{u}\otimes\bm{u} \bm{\nabla}\cdot\rho \bm{u},
    \end{equation}
to arrive at
    \begin{multline}
        \partial_t^{(1)}\left(\rho\bm{u}\otimes\bm{u} + p\bm{I}\right) = -\bm{\nabla}\cdot\rho\bm{u}\otimes\bm{u}\otimes\bm{u} -\bm{u}\otimes\bm{\nabla}p  \\ - {\left(\bm{u}\otimes\bm{\nabla}p\right)}^\dagger +  \partial_t^{(1)}p\bm{I},
    \end{multline}
where one can use Eq.~\eqref{eq:pressure_euler_CE} to get to,
    \begin{multline}
        \int \bm{v}\otimes\bm{v} f^{(1)}d\bm{v} =\\ -p\{\tau_1, \tau_2\}\left[ \left( \bm{\nabla}\bm{u} + \bm{\nabla}\bm{u}^{\dagger}\right) - \frac{R}{c_V}\bm{\nabla}\cdot\bm{u}\bm{I}\right].
    \end{multline}
Plugging this final expression into the momentum balance equation at order $\epsilon^2$,
    \begin{equation}
        \partial_t^{(2)}\rho \bm{u} + \bm{\nabla}\cdot\bm{T}_{\rm NS}  = 0,
    \end{equation}
where
    \begin{multline}
        \bm{T}_{\rm NS} = \{\tau_1,\tau_2\} p\left[\bm{\nabla}\bm{u} + \bm{\nabla}\bm{u}^\dagger - \frac{2}{D}\bm{\nabla}\cdot\bm{u}\bm{I}\right] \\ + \{\tau_1,\tau_2\} p \left(\frac{2}{D}-\frac{R}{c_V}\right)\bm{\nabla}\cdot\bm{u}\bm{I}.
    \end{multline}
Note that here,
\begin{equation}
    \mu = p\{\tau_1, \tau_2\}.
\end{equation}
Moving on to the energy balance equation at order $\epsilon^2$,
\begin{equation}
    \partial_t^{(2)} E + \bm{\nabla}\cdot\left(\int_{\mathbb{R}^D}\bm{v}g^{(1)}d\bm{v}\right) = 0,
\end{equation}    
where
\begin{multline}\label{eq:CE_noneq_q}
    \int_{\mathbb{R}^D}\bm{v}g^{(1)}d\bm{v} = -\{\tau_2, \tau_1\}\left[\partial_t^{(1)}\left(p+E\right)\bm{u} \right.
    \\\left. + \bm{\nabla}\cdot\left(\frac{p}{\rho}\left(E + p\right) + p\bm{u}\otimes\bm{u} + \left(p + E\right)\bm{u}\otimes\bm{u} \right) \right].
\end{multline}
To expand this equation, we multiply the pressure balance equation at Euler level by $\bm{u}$ and obtain the following balance equation using momentum and continuity balance,
\begin{equation}
    \partial_t^{(1)}p\bm{u} + \bm{\nabla}\cdot p \bm{u}\otimes\bm{u} + \frac{p}{\rho}\bm{\nabla}p + \frac{R}{c_V}p\bm{u}\bm{\nabla}\cdot\bm{u} = 0.
\end{equation}
Using the same approach we can also derive the following additional balance equation,
\begin{equation}
    \partial_t^{(1)}E \bm{u} + \bm{\nabla}\cdot\left(E + p\right)\bm{u}\otimes\bm{u} + \frac{E}{\rho}\bm{\nabla}p - p\bm{u}\cdot\bm{\nabla}\bm{u} = 0.
\end{equation}
Plugging these equations into Eq.~\eqref{eq:CE_noneq_q} and after some algebra,
\begin{equation}\label{eq:vg1_eq}
    \int_{\mathbb{R}^D}\bm{v}g^{(1)}d\bm{v} = -\{\tau_2,\tau_1\} p \left(c_V+R\right)\bm{\nabla}T - \bm{u}\cdot\bm{T}_{\rm NS},
\end{equation}
which leads to:
\begin{equation}\label{eq:g_eps_2}
    \partial_t^{(2)}E - \bm{\nabla}\cdot\left[\{\tau_2,\tau_1\} p \left(c_V+R\right)\bm{\nabla}T\right] + \bm{\nabla}\cdot\left(\bm{u}\cdot\bm{T}_{\rm NS}\right) = 0.
\end{equation}
 Here,
 \begin{equation}
     \lambda = \{\tau_2, \tau_1\}p\left(c_V+R\right).
 \end{equation}
The Prandtl number is readily shown to be,
\begin{equation}
    {\rm Pr} = \frac{\{\tau_1, \tau_2\}}{\{\tau_2, \tau_1\}}.
\end{equation}
In the limit of a BGK collision operator, $\tau_1=\tau_2$,
\begin{equation}
    {\rm Pr} = 1.
\end{equation}

\section{Moments of the distribution functions}
\label{app:moments}
The moments of the distribution functions $f_i$ and $g_i$ are
\begin{align}
    &\sum_i f_i^{\rm eq} = \sum_i f_i = \rho 
    \\
    &\sum_i f_i^{\rm eq} v_{i\alpha} = \sum_i f_i v_{i\alpha} = \rho u_\alpha
    \\
    &\sum_i f_i^{\rm eq} v_{i\alpha} v_{i\beta}= \rho u_\alpha u_\beta + \rho T \delta_{\alpha \beta} 
    \\
    &\sum_i f_i^{\rm eq} v_{i\alpha} v_{i\beta} v_{i\gamma}= \rho  u_\alpha u_\beta u_\gamma + \rho T 
    [u_\alpha \delta_{\beta \gamma}]_{cyc} 
    \\
    &\begin{aligned}
    \sum_i f_i^{\rm eq}& v_{i\alpha} v_{i\beta} v_{i\gamma} v_{i\delta} 
    = \rho  u_\alpha u_\beta u_\gamma u_\delta + \rho T [u_\alpha u_\beta \delta_{\gamma \delta}]_{cyc} 
    \\
    &+ \rho T^2 [\delta_{\alpha\beta}\delta_{\gamma\delta}]_{cyc}
    \end{aligned}
    \\
    &\begin{aligned}\sum_i g_i^{\rm eq}& + \sum_i f_i^{\rm eq} v_{i\gamma} v_{i\gamma} = \sum_i g_i + \sum_i f_i v_{i\gamma} v_{i\gamma} 
    \\ 
    &= 2 \rho \left(C_v T + \frac{u_\gamma u_\gamma}{2}\right)=2 \rho E 
    \end{aligned}
    \\
    &\sum_i g_i^{\rm eq} v_{i\alpha} + \sum_i f_i^{\rm eq} v_{i\gamma} v_{i\gamma} v_{i\alpha} = 2 \rho u_\alpha (E+T) 
    \\
    &\begin{aligned}
    \sum_i g_i^{\rm eq}& v_{i\alpha} v_{i\beta} + \sum_i f_i^{\rm eq} v_{i\gamma} v_{i\gamma} v_{i\alpha} v_{i\beta} 
    \\
    &\hfill = 2 \rho T u_\alpha u_\beta + 2 \rho (u_\alpha u_\beta + T \delta_{\alpha \beta}) (E+T)
    ,
    \end{aligned}
\end{align}
where $"[]_{cyc}"$ denotes cyclic non-repetitive permutation over the indices.

\section{Hermite polynomials and coefficients of the Grad expansion}
\label{app:gradexpansion}

Populations can be approximated with a Grad expansion,
\begin{equation}
     f_i = w_i \sum_{n=0}^{N} \frac{1}{n!} \bm{a}^{(n)}(\bm{M}, T_{ref}) : \bm{\mathcal{H}}^{(n)} (\bm{v}_i)
     \text{,}
\end{equation}
here shown for a set of moments $\bm{M}$ up to the fourth-order and a truncation at $N=4$,
using the Hermite polynomials
\begin{align}
    \mathcal{H}^{(0)}(\bm{v}_i) 
        =& 1 
        \text{,} 
        \\
    \mathcal{H}^{(1)}_{\alpha}(\bm{v}_i) 
        =& v_{i\alpha}
        \text{,} 
        \\
    \mathcal{H}^{(2)}_{\alpha \beta}(\bm{v}_i) 
        =& v_{i\alpha} v_{i\beta} - T_{ref}\delta_{\alpha \beta} 
        \text{,} 
        \\
    \mathcal{H}^{(3)}_{\alpha \beta \gamma}(\bm{v}_i) 
        =& v_{i\alpha} v_{i\beta} v_{i\gamma} 
        - T_{ref} [v_{i\alpha} \delta_{\beta \gamma}]_{cyc} 
       \text{,} 
        \\  
    \mathcal{H}^{(4)}_{\alpha \beta \gamma \delta}(\bm{v}_i) 
        =& v_{i\alpha} v_{i\beta} v_{i\gamma} v_{i\delta} 
        - T_{ref} [v_{i\alpha} v_{i\beta} \delta_{\gamma\delta}]_{cyc} 
        \nonumber \\
        &+ T_{ref}^2 [\delta_{\alpha\beta} \delta_{\gamma\delta}]_{cyc} 
        \text{.}
\end{align}
For the computation of equilibrium populations $f_i^{eq}$, the corresponding coefficients $a^{eq,(n)}$ are found as a function of the set of equilibrium moments $\bm{M}^{eq}$, as
\begin{align}
    a^{eq,(0)} 
    =& M^{eq,(0)} = \rho 
    \text{,} \\
    a^{eq,(1)}_\alpha 
    =& \frac{1}{T_{ref}} \Bigl[ M^{eq,(1)}_\alpha \Bigr]  = \frac{1}{T_{ref}} \Bigl[ \rho u_{\alpha} \Bigr] 
   \text{,} \\
    a^{eq,(2)}_{\alpha\beta} 
    =& \frac{1}{ T_{ref}^2} \Bigl[ M_{\alpha \beta}^{eq,(2)} - T_{ref} M^{eq,(0)} \delta_{\alpha \beta} \Bigr]
    \nonumber \\
    =& \frac{1}{ T_{ref}^2} \Bigl[ \rho u_\alpha u_\beta + T_{ref} (\theta - 1) \rho \delta_{\alpha\beta} \Bigr]
    \text{,} \\
    a^{eq,(3)}_{\alpha\beta\gamma} 
    =& \frac{1}{T_{ref}^3} \Bigl[ M^{eq,(3)}_{\alpha\beta\gamma} -  T_{ref} [ M_\alpha^{eq,(1)} \delta_{\beta\gamma} ]_{cyc} \Bigr]
    \nonumber \\
    =& \frac{1}{T_{ref}^3}  \Bigl[  
    \rho u_\alpha u_\beta u_\gamma +  T_{ref} (\theta -1) [\rho  u_\alpha \delta_{\beta\gamma}]_{cyc} 
    \Bigr] 
    \text{,} \\
    a^{eq,(4)}_{\alpha\beta\gamma\delta} 
    =& \frac{1}{T_{ref}^4}  \Bigl[  
    M^{eq,(4)}_{\alpha\beta\gamma\delta} 
    - T_{ref}[M^{eq,(2)}_{\alpha\beta}\delta_{\alpha\beta}]_{cyc}  
    \nonumber \\
    &+ T_{ref}^2 [ M^{eq,(0)} \delta_{\alpha\beta} \delta_{\gamma\delta}]_{cyc} 
    \Bigr] 
    \nonumber \\
    =& \frac{1}{T_{ref}^4}  \Bigl[  
    \rho u_\alpha u_\beta u_\gamma u_\delta + T_{ref} (\theta -1) [ \rho u_\alpha u_\beta \delta_{\gamma\delta}]_{cyc}  
    \nonumber \\ 
    &+ T_{ref}^2 (\theta -1)^2   [\rho \delta_{\alpha\beta} \delta_{\gamma\delta}]_{cyc} 
    \Bigr] 
    \text{,} 
\end{align}
where $"[]_{cyc}"$ denotes cyclic non-repetitive permutation over the indices.
$D$ denotes the physical dimension and $\theta$ is the normalized temperature defined as $\theta = T / T_{ref}$, with $T_{ref}$ given as a consequence of the velocity set.
Note that the contracted Hermite polynomials and equilibrium coefficients of order three and four, i.e.
\begin{align}
    \mathcal{H}^{(3)}_{\alpha}(\bm{v}_i) 
        =& v_{i\alpha} \left(  \bm{v}_i^2 - T_{ref}\left( D + 2 \right) \right)
       \text{,} 
        \\  
    \mathcal{H}^{(4)}_{\alpha \beta}(\bm{v}_i) 
        =& v_{i\alpha} v_{i\beta}  \left(  \bm{v}_i^2 - T_{ref}\left( D + 4 \right) \right) 
        \nonumber \\
        &- T_{ref} \delta_{\alpha \beta} \left(  \bm{v}_i^2 - T_{ref}\left( D + 2 \right) \right)
       \text{,}  \\
a^{eq,(3)}_{\alpha} 
    =& \frac{1}{T_{ref}^3}  \Bigl[  
    \rho u_{\alpha} \left( \bm{u}^2 + T_{ref}\left(\theta-1\right)\left(D+2\right)\right)
    \Bigr] 
    \text{,} \\
 a^{eq,(4)}_{\alpha\beta} 
    =& \frac{1}{T_{ref}^4}  \Bigl[  
   \rho u_{\alpha} u_{\beta} \left( \bm{u}^2+T_{ref}\left(\theta-1\right)\left(D+4\right)\right)
    \nonumber \\
    &+T_{ref}\left(\theta-1\right) \delta_{\alpha\beta} \left(\bm{u}^2+T_{ref}\left(\theta-1\right)\left(D+2\right)\right)
    \Bigr] 
    \text{,} 
\end{align}
are sufficient for Euler and NSF level flows compared to the full third- and fourth-order expansion, respectively.

\section{Specifications of the velocity sets} 
\label{app:velocityset}

The D2Q16 and D2Q25 velocity sets, weights and reference temperatures are listed in table~\ref{tab:theD2Q16Lattice} and ~\ref{tab:theD2Q25Lattice}, respectively.

\begin{table}[htb]
\footnotesize
\centering
\begin{tabular}{ |p{1.6cm}|p{3.4cm}|p{3.2cm}|  }
 \hline
 \multicolumn{3}{|c|}{ D2Q16, $T_{ref}=1$ } \\
 \hline
 $i$ & $(v_{ix},v_{iy}) $ & $w_i$ \\
 \hline
$0-3$  & $(\pm \sqrt{3-\sqrt{6}},\pm \sqrt{3-\sqrt{6}})$   &   $\frac{3+\sqrt{6}}{12} \times \frac{3+\sqrt{6}}{12}$  \\
$4-7$   &$(\pm \sqrt{3-\sqrt{6}} ,\pm \sqrt{3+\sqrt{6}})$  &   $\frac{3+\sqrt{6}}{12} \times \frac{3-\sqrt{6}}{12}$  \\
$8-11$ &   $(\pm \sqrt{3+\sqrt{6}},\pm \sqrt{3-\sqrt{6}})$   &   $\frac{3-\sqrt{6}}{12} \times \frac{3+\sqrt{6}}{12}$ \\
$12-15$  & $(\pm \sqrt{3+\sqrt{6}},\pm \sqrt{3+\sqrt{6}})$   &   $\frac{3-\sqrt{6}}{12} \times \frac{3-\sqrt{6}}{12}$  \\
 \hline
\end{tabular}
\caption{Forth-order quadrature in two dimensions (D2Q16).}
\label{tab:theD2Q16Lattice}
\end{table}

\begin{table}[htb]
\footnotesize
\centering
\begin{tabular}{ |p{1.6cm}|p{3.4cm}|p{3.2cm}|  }
 \hline
 \multicolumn{3}{|c|}{ D2Q25, $T_{ref}=1$ } \\
 \hline
 $i$ & $(v_{ix},v_{iy}) $ & $w_i$  \\
 \hline
$0$ & ($0$, $0$) &  $\frac{8}{15}$ \\
$1,3$ & ($\pm \sqrt{5-\sqrt{10}}$, $0$) &  $\frac{8}{15} \times (7+2\sqrt{10})$ \\
$2,4$ & ($0$, $\pm \sqrt{5-\sqrt{10}}$) &  $(7+2\sqrt{10}) \times \frac{8}{15} $ \\
 $5,6,7,8$ & ($\pm \sqrt{5-\sqrt{10}}$, $\pm \sqrt{5-\sqrt{10}}$) &  $(7+2\sqrt{10}) \times (7+2\sqrt{10})$ \\
 $9,11$ & ($\pm \sqrt{5+\sqrt{10}}$, $0$) &  $(7-2\sqrt{10}) \times \frac{8}{15}$ \\
 $10,12$ & ($0$, $\pm \sqrt{5+\sqrt{10}}$) &  $\frac{8}{15} \times (7-2\sqrt{10})$ \\
$13,16,17,20$ & ($\pm \sqrt{5+\sqrt{10}}, \pm \sqrt{5-\sqrt{10}})$ &  $(7-2\sqrt{10}) \times (7+2\sqrt{10})$ \\
$14,15,18,19$ & $(\pm \sqrt{5-\sqrt{10}}, \pm \sqrt{5+\sqrt{10}})$ &  $(7+2\sqrt{10}) \times (7-2\sqrt{10})$ \\
$21,22,23,24$ & $(\pm \sqrt{5+\sqrt{10}}, \pm \sqrt{5+\sqrt{10}})$ &  $(7-2\sqrt{10}) \times (7-2\sqrt{10})$ \\
 \hline
\end{tabular}
\caption{Fifth-order quadrature in two dimensions (D2Q25).}
\label{tab:theD2Q25Lattice}
\end{table}

\section*{References}
\bibliography{_aipsamp}

\begin{thebibliography}{100}%
\makeatletter
\providecommand \@ifxundefined [1]{%
 \@ifx{#1\undefined}
}%
\providecommand \@ifnum [1]{%
 \ifnum #1\expandafter \@firstoftwo
 \else \expandafter \@secondoftwo
 \fi
}%
\providecommand \@ifx [1]{%
 \ifx #1\expandafter \@firstoftwo
 \else \expandafter \@secondoftwo
 \fi
}%
\providecommand \natexlab [1]{#1}%
\providecommand \enquote  [1]{``#1''}%
\providecommand \bibnamefont  [1]{#1}%
\providecommand \bibfnamefont [1]{#1}%
\providecommand \citenamefont [1]{#1}%
\providecommand \href@noop [0]{\@secondoftwo}%
\providecommand \href [0]{\begingroup \@sanitize@url \@href}%
\providecommand \@href[1]{\@@startlink{#1}\@@href}%
\providecommand \@@href[1]{\endgroup#1\@@endlink}%
\providecommand \@sanitize@url [0]{\catcode `\\12\catcode `\$12\catcode `\&12\catcode `\#12\catcode `\^12\catcode `\_12\catcode `\%12\relax}%
\providecommand \@@startlink[1]{}%
\providecommand \@@endlink[0]{}%
\providecommand \url  [0]{\begingroup\@sanitize@url \@url }%
\providecommand \@url [1]{\endgroup\@href {#1}{\urlprefix }}%
\providecommand \urlprefix  [0]{URL }%
\providecommand \Eprint [0]{\href }%
\providecommand \doibase [0]{http://dx.doi.org/}%
\providecommand \selectlanguage [0]{\@gobble}%
\providecommand \bibinfo  [0]{\@secondoftwo}%
\providecommand \bibfield  [0]{\@secondoftwo}%
\providecommand \translation [1]{[#1]}%
\providecommand \BibitemOpen [0]{}%
\providecommand \bibitemStop [0]{}%
\providecommand \bibitemNoStop [0]{.\EOS\space}%
\providecommand \EOS [0]{\spacefactor3000\relax}%
\providecommand \BibitemShut  [1]{\csname bibitem#1\endcsname}%
\let\auto@bib@innerbib\@empty
\bibitem [{\citenamefont {Pirozzoli}(2011)}]{pirozzoli2011numerical}%
  \BibitemOpen
  \bibfield  {author} {\bibinfo {author} {\bibfnamefont {S.}~\bibnamefont {Pirozzoli}},\ }\href@noop {} {\bibfield  {journal} {\bibinfo  {journal} {Annual review of fluid mechanics}\ }\textbf {\bibinfo {volume} {43}},\ \bibinfo {pages} {163} (\bibinfo {year} {2011})}\BibitemShut {NoStop}%
\bibitem [{\citenamefont {Yee}\ and\ \citenamefont {Sj{\"o}green}(2018)}]{yee2018recent}%
  \BibitemOpen
  \bibfield  {author} {\bibinfo {author} {\bibfnamefont {H.~C.}\ \bibnamefont {Yee}}\ and\ \bibinfo {author} {\bibfnamefont {B.}~\bibnamefont {Sj{\"o}green}},\ }\href@noop {} {\bibfield  {journal} {\bibinfo  {journal} {computers \& Fluids}\ }\textbf {\bibinfo {volume} {169}},\ \bibinfo {pages} {331} (\bibinfo {year} {2018})}\BibitemShut {NoStop}%
\bibitem [{\citenamefont {Krüger}\ \emph {et~al.}(2017)\citenamefont {Krüger}, \citenamefont {Kusumaatmaja}, \citenamefont {Kuzmin}, \citenamefont {Shardt}, \citenamefont {Silva},\ and\ \citenamefont {Viggen}}]{LBMBookKrueger}%
  \BibitemOpen
  \bibfield  {author} {\bibinfo {author} {\bibfnamefont {T.}~\bibnamefont {Krüger}}, \bibinfo {author} {\bibfnamefont {H.}~\bibnamefont {Kusumaatmaja}}, \bibinfo {author} {\bibfnamefont {A.}~\bibnamefont {Kuzmin}}, \bibinfo {author} {\bibfnamefont {O.}~\bibnamefont {Shardt}}, \bibinfo {author} {\bibfnamefont {G.}~\bibnamefont {Silva}}, \ and\ \bibinfo {author} {\bibfnamefont {E.~M.}\ \bibnamefont {Viggen}},\ }\href@noop {} {\bibfield  {journal} {\bibinfo  {journal} {Springer International Publishing}\ }\textbf {\bibinfo {volume} {10}},\ \bibinfo {pages} {4} (\bibinfo {year} {2017})}\BibitemShut {NoStop}%
\bibitem [{\citenamefont {Succi}(2001)}]{succiBook}%
  \BibitemOpen
  \bibfield  {author} {\bibinfo {author} {\bibfnamefont {S.}~\bibnamefont {Succi}},\ }\href@noop {} {\emph {\bibinfo {title} {The lattice Boltzmann equation: for fluid dynamics and beyond}}}\ (\bibinfo  {publisher} {Oxford university press},\ \bibinfo {year} {2001})\BibitemShut {NoStop}%
\bibitem [{\citenamefont {Hosseini}\ and\ \citenamefont {Karlin}(2023)}]{hosseini2023lattice}%
  \BibitemOpen
  \bibfield  {author} {\bibinfo {author} {\bibfnamefont {S.~A.}\ \bibnamefont {Hosseini}}\ and\ \bibinfo {author} {\bibfnamefont {I.~V.}\ \bibnamefont {Karlin}},\ }\href@noop {} {\bibfield  {journal} {\bibinfo  {journal} {Physics {R}eports}\ }\textbf {\bibinfo {volume} {1030}},\ \bibinfo {pages} {1} (\bibinfo {year} {2023})}\BibitemShut {NoStop}%
\bibitem [{\citenamefont {Chapman}\ and\ \citenamefont {Cowling}(1970)}]{Chapman}%
  \BibitemOpen
  \bibfield  {author} {\bibinfo {author} {\bibfnamefont {S.}~\bibnamefont {Chapman}}\ and\ \bibinfo {author} {\bibfnamefont {T.~G.}\ \bibnamefont {Cowling}},\ }\href@noop {} {\emph {\bibinfo {title} {The mathematical theory of non-uniform gases. an account of the kinetic theory of viscosity, thermal conduction and diffusion in gases}}}\ (\bibinfo {year} {1970})\BibitemShut {NoStop}%
\bibitem [{\citenamefont {Bhatnagar}\ \emph {et~al.}(1954)\citenamefont {Bhatnagar}, \citenamefont {Gross},\ and\ \citenamefont {Krook}}]{bhatnagar1954model}%
  \BibitemOpen
  \bibfield  {author} {\bibinfo {author} {\bibfnamefont {P.~L.}\ \bibnamefont {Bhatnagar}}, \bibinfo {author} {\bibfnamefont {E.~P.}\ \bibnamefont {Gross}}, \ and\ \bibinfo {author} {\bibfnamefont {M.}~\bibnamefont {Krook}},\ }\href@noop {} {\bibfield  {journal} {\bibinfo  {journal} {Physical review}\ }\textbf {\bibinfo {volume} {94}},\ \bibinfo {pages} {511} (\bibinfo {year} {1954})}\BibitemShut {NoStop}%
\bibitem [{\citenamefont {Qian}\ \emph {et~al.}(1992)\citenamefont {Qian}, \citenamefont {D'Humières},\ and\ \citenamefont {Lallemand}}]{QianLBM}%
  \BibitemOpen
  \bibfield  {author} {\bibinfo {author} {\bibfnamefont {Y.~H.}\ \bibnamefont {Qian}}, \bibinfo {author} {\bibfnamefont {D.}~\bibnamefont {D'Humières}}, \ and\ \bibinfo {author} {\bibfnamefont {P.}~\bibnamefont {Lallemand}},\ }\href {https://dx.doi.org/10.1209/0295-5075/17/6/001} {\bibfield  {journal} {\bibinfo  {journal} {Europhysics Letters}\ }\textbf {\bibinfo {volume} {17}},\ \bibinfo {pages} {479} (\bibinfo {year} {1992})}\BibitemShut {NoStop}%
\bibitem [{\citenamefont {Chen}\ \emph {et~al.}(1992)\citenamefont {Chen}, \citenamefont {Chen},\ and\ \citenamefont {Matthaeus}}]{ChenLBM}%
  \BibitemOpen
  \bibfield  {author} {\bibinfo {author} {\bibfnamefont {H.}~\bibnamefont {Chen}}, \bibinfo {author} {\bibfnamefont {S.}~\bibnamefont {Chen}}, \ and\ \bibinfo {author} {\bibfnamefont {W.~H.}\ \bibnamefont {Matthaeus}},\ }\href {https://link.aps.org/doi/10.1103/PhysRevA.45.R5339} {\bibfield  {journal} {\bibinfo  {journal} {Phys. Rev. A}\ }\textbf {\bibinfo {volume} {45}},\ \bibinfo {pages} {R5339} (\bibinfo {year} {1992})}\BibitemShut {NoStop}%
\bibitem [{\citenamefont {Guo}\ and\ \citenamefont {Xu}(2021)}]{guo2021progress}%
  \BibitemOpen
  \bibfield  {author} {\bibinfo {author} {\bibfnamefont {Z.}~\bibnamefont {Guo}}\ and\ \bibinfo {author} {\bibfnamefont {K.}~\bibnamefont {Xu}},\ }\href@noop {} {\bibfield  {journal} {\bibinfo  {journal} {Advances in Aerodynamics}\ }\textbf {\bibinfo {volume} {3}},\ \bibinfo {pages} {1} (\bibinfo {year} {2021})}\BibitemShut {NoStop}%
\bibitem [{\citenamefont {Xu}\ \emph {et~al.}(2018)\citenamefont {Xu}, \citenamefont {Zhang},\ and\ \citenamefont {Zhang}}]{xu2018discrete}%
  \BibitemOpen
  \bibfield  {author} {\bibinfo {author} {\bibfnamefont {A.}~\bibnamefont {Xu}}, \bibinfo {author} {\bibfnamefont {G.}~\bibnamefont {Zhang}}, \ and\ \bibinfo {author} {\bibfnamefont {Y.}~\bibnamefont {Zhang}},\ }\href@noop {} {\bibfield  {journal} {\bibinfo  {journal} {Kinetic theory}\ ,\ \bibinfo {pages} {450}} (\bibinfo {year} {2018})}\BibitemShut {NoStop}%
\bibitem [{\citenamefont {Mieussens}(2000)}]{mieussens2000discrete}%
  \BibitemOpen
  \bibfield  {author} {\bibinfo {author} {\bibfnamefont {L.}~\bibnamefont {Mieussens}},\ }\href@noop {} {\bibfield  {journal} {\bibinfo  {journal} {Journal of Computational Physics}\ }\textbf {\bibinfo {volume} {162}},\ \bibinfo {pages} {429} (\bibinfo {year} {2000})}\BibitemShut {NoStop}%
\bibitem [{\citenamefont {Hosseini}\ \emph {et~al.}(2024{\natexlab{a}})\citenamefont {Hosseini}, \citenamefont {Boivin}, \citenamefont {Th{\'e}venin},\ and\ \citenamefont {Karlin}}]{hosseini2024lattice}%
  \BibitemOpen
  \bibfield  {author} {\bibinfo {author} {\bibfnamefont {S.~A.}\ \bibnamefont {Hosseini}}, \bibinfo {author} {\bibfnamefont {P.}~\bibnamefont {Boivin}}, \bibinfo {author} {\bibfnamefont {D.}~\bibnamefont {Th{\'e}venin}}, \ and\ \bibinfo {author} {\bibfnamefont {I.}~\bibnamefont {Karlin}},\ }\href@noop {} {\bibfield  {journal} {\bibinfo  {journal} {Progress in Energy and Combustion Science}\ }\textbf {\bibinfo {volume} {102}},\ \bibinfo {pages} {101140} (\bibinfo {year} {2024}{\natexlab{a}})}\BibitemShut {NoStop}%
\bibitem [{\citenamefont {Coveney}\ \emph {et~al.}(2002)\citenamefont {Coveney}, \citenamefont {Succi}, \citenamefont {d'Humières}, \citenamefont {Ginzburg}, \citenamefont {Krafczyk}, \citenamefont {Lallemand},\ and\ \citenamefont {Luo}}]{Coveney_MRT}%
  \BibitemOpen
  \bibfield  {author} {\bibinfo {author} {\bibfnamefont {P.~V.}\ \bibnamefont {Coveney}}, \bibinfo {author} {\bibfnamefont {S.}~\bibnamefont {Succi}}, \bibinfo {author} {\bibfnamefont {D.}~\bibnamefont {d'Humières}}, \bibinfo {author} {\bibfnamefont {I.}~\bibnamefont {Ginzburg}}, \bibinfo {author} {\bibfnamefont {M.}~\bibnamefont {Krafczyk}}, \bibinfo {author} {\bibfnamefont {P.}~\bibnamefont {Lallemand}}, \ and\ \bibinfo {author} {\bibfnamefont {L.-S.}\ \bibnamefont {Luo}},\ }\href {https://royalsocietypublishing.org/doi/abs/10.1098/rsta.2001.0955} {\bibfield  {journal} {\bibinfo  {journal} {Philosophical Transactions of the Royal Society of London. Series A: Mathematical, Physical and Engineering Sciences}\ }\textbf {\bibinfo {volume} {360}},\ \bibinfo {pages} {437} (\bibinfo {year} {2002})}\BibitemShut {NoStop}%
\bibitem [{\citenamefont {Geier}\ \emph {et~al.}(2006)\citenamefont {Geier}, \citenamefont {Greiner},\ and\ \citenamefont {Korvink}}]{Geier_Cascaded}%
  \BibitemOpen
  \bibfield  {author} {\bibinfo {author} {\bibfnamefont {M.}~\bibnamefont {Geier}}, \bibinfo {author} {\bibfnamefont {A.}~\bibnamefont {Greiner}}, \ and\ \bibinfo {author} {\bibfnamefont {J.~G.}\ \bibnamefont {Korvink}},\ }\href {https://link.aps.org/doi/10.1103/PhysRevE.73.066705} {\bibfield  {journal} {\bibinfo  {journal} {Phys. Rev. E}\ }\textbf {\bibinfo {volume} {73}},\ \bibinfo {pages} {066705} (\bibinfo {year} {2006})}\BibitemShut {NoStop}%
\bibitem [{\citenamefont {Geier}\ \emph {et~al.}(2015)\citenamefont {Geier}, \citenamefont {Schönherr}, \citenamefont {Pasquali},\ and\ \citenamefont {Krafczyk}}]{Geier_Cumulant}%
  \BibitemOpen
  \bibfield  {author} {\bibinfo {author} {\bibfnamefont {M.}~\bibnamefont {Geier}}, \bibinfo {author} {\bibfnamefont {M.}~\bibnamefont {Schönherr}}, \bibinfo {author} {\bibfnamefont {A.}~\bibnamefont {Pasquali}}, \ and\ \bibinfo {author} {\bibfnamefont {M.}~\bibnamefont {Krafczyk}},\ }\href {https://www.sciencedirect.com/science/article/pii/S0898122115002126} {\bibfield  {journal} {\bibinfo  {journal} {computers \& Mathematics with Applications}\ }\textbf {\bibinfo {volume} {70}},\ \bibinfo {pages} {507} (\bibinfo {year} {2015})}\BibitemShut {NoStop}%
\bibitem [{\citenamefont {Malaspinas}(2015)}]{malaspinas_recursivereg}%
  \BibitemOpen
  \bibfield  {author} {\bibinfo {author} {\bibfnamefont {O.}~\bibnamefont {Malaspinas}},\ }\href {https://arxiv.org/abs/1505.06900} {\enquote {\bibinfo {title} {Increasing stability and accuracy of the lattice boltzmann scheme: recursivity and regularization},}\ } (\bibinfo {year} {2015})\BibitemShut {NoStop}%
\bibitem [{\citenamefont {Karlin}\ and\ \citenamefont {Succi}(1998)}]{Karlin_entropicEQ}%
  \BibitemOpen
  \bibfield  {author} {\bibinfo {author} {\bibfnamefont {I.~V.}\ \bibnamefont {Karlin}}\ and\ \bibinfo {author} {\bibfnamefont {S.}~\bibnamefont {Succi}},\ }\href {https://link.aps.org/doi/10.1103/PhysRevE.58.R4053} {\bibfield  {journal} {\bibinfo  {journal} {Phys. Rev. E}\ }\textbf {\bibinfo {volume} {58}},\ \bibinfo {pages} {R4053} (\bibinfo {year} {1998})}\BibitemShut {NoStop}%
\bibitem [{\citenamefont {Karlin}\ \emph {et~al.}(2014)\citenamefont {Karlin}, \citenamefont {B\"osch},\ and\ \citenamefont {Chikatamarla}}]{Karlin_Gibbs}%
  \BibitemOpen
  \bibfield  {author} {\bibinfo {author} {\bibfnamefont {I.~V.}\ \bibnamefont {Karlin}}, \bibinfo {author} {\bibfnamefont {F.}~\bibnamefont {B\"osch}}, \ and\ \bibinfo {author} {\bibfnamefont {S.~S.}\ \bibnamefont {Chikatamarla}},\ }\href {https://link.aps.org/doi/10.1103/PhysRevE.90.031302} {\bibfield  {journal} {\bibinfo  {journal} {Phys. Rev. E}\ }\textbf {\bibinfo {volume} {90}},\ \bibinfo {pages} {031302} (\bibinfo {year} {2014})}\BibitemShut {NoStop}%
\bibitem [{\citenamefont {Hosseini}\ \emph {et~al.}(2023)\citenamefont {Hosseini}, \citenamefont {Atif}, \citenamefont {Ansumali},\ and\ \citenamefont {Karlin}}]{Ali_reviewEntropic}%
  \BibitemOpen
  \bibfield  {author} {\bibinfo {author} {\bibfnamefont {S.~A.}\ \bibnamefont {Hosseini}}, \bibinfo {author} {\bibfnamefont {M.}~\bibnamefont {Atif}}, \bibinfo {author} {\bibfnamefont {S.}~\bibnamefont {Ansumali}}, \ and\ \bibinfo {author} {\bibfnamefont {I.~V.}\ \bibnamefont {Karlin}},\ }\href {https://www.sciencedirect.com/science/article/pii/S0045793023001093} {\bibfield  {journal} {\bibinfo  {journal} {computers \& Fluids}\ }\textbf {\bibinfo {volume} {259}},\ \bibinfo {pages} {105884} (\bibinfo {year} {2023})}\BibitemShut {NoStop}%
\bibitem [{\citenamefont {Shakhov}(1968)}]{Shakhov}%
  \BibitemOpen
  \bibfield  {author} {\bibinfo {author} {\bibfnamefont {E.~M.}\ \bibnamefont {Shakhov}},\ }\href {https://doi.org/10.1007/BF01029546} {\bibfield  {journal} {\bibinfo  {journal} {Fluid Dynamics}\ }\textbf {\bibinfo {volume} {3}},\ \bibinfo {pages} {95} (\bibinfo {year} {1968})}\BibitemShut {NoStop}%
\bibitem [{\citenamefont {Holway Lowell~H.}(1966)}]{HolwayES}%
  \BibitemOpen
  \bibfield  {author} {\bibinfo {author} {\bibfnamefont {J.}~\bibnamefont {Holway Lowell~H.}},\ }\href {https://doi.org/10.1063/1.1761920} {\bibfield  {journal} {\bibinfo  {journal} {The Physics of Fluids}\ }\textbf {\bibinfo {volume} {9}},\ \bibinfo {pages} {1658} (\bibinfo {year} {1966})}\BibitemShut {NoStop}%
\bibitem [{\citenamefont {Gorban}\ and\ \citenamefont {Karlin}(1994)}]{gorban1994}%
  \BibitemOpen
  \bibfield  {author} {\bibinfo {author} {\bibfnamefont {A.~N.}\ \bibnamefont {Gorban}}\ and\ \bibinfo {author} {\bibfnamefont {I.~V.}\ \bibnamefont {Karlin}},\ }\href@noop {} {\bibfield  {journal} {\bibinfo  {journal} {Physica A: Statistical Mechanics and its Applications}\ }\textbf {\bibinfo {volume} {206}},\ \bibinfo {pages} {401} (\bibinfo {year} {1994})}\BibitemShut {NoStop}%
\bibitem [{\citenamefont {Ansumali}\ \emph {et~al.}(2007)\citenamefont {Ansumali}, \citenamefont {Arcidiacono}, \citenamefont {Chikatamarla}, \citenamefont {Prasianakis}, \citenamefont {Gorban},\ and\ \citenamefont {Karlin}}]{QE_ansumali-2007}%
  \BibitemOpen
  \bibfield  {author} {\bibinfo {author} {\bibfnamefont {S.}~\bibnamefont {Ansumali}}, \bibinfo {author} {\bibfnamefont {S.}~\bibnamefont {Arcidiacono}}, \bibinfo {author} {\bibfnamefont {S.~S.}\ \bibnamefont {Chikatamarla}}, \bibinfo {author} {\bibfnamefont {N.~I.}\ \bibnamefont {Prasianakis}}, \bibinfo {author} {\bibfnamefont {A.~N.}\ \bibnamefont {Gorban}}, \ and\ \bibinfo {author} {\bibfnamefont {I.~V.}\ \bibnamefont {Karlin}},\ }\href {https://doi.org/10.1140/epjb/e2007-00100-1} {\bibfield  {journal} {\bibinfo  {journal} {The European Physical Journal B}\ }\textbf {\bibinfo {volume} {56}},\ \bibinfo {pages} {135} (\bibinfo {year} {2007})}\BibitemShut {NoStop}%
\bibitem [{\citenamefont {Rykov}(1976)}]{rykov_model_1976}%
  \BibitemOpen
  \bibfield  {author} {\bibinfo {author} {\bibfnamefont {V.~A.}\ \bibnamefont {Rykov}},\ }\href@noop {} {\bibfield  {journal} {\bibinfo  {journal} {Fluid {D}ynamics}\ }\textbf {\bibinfo {volume} {10}},\ \bibinfo {pages} {959} (\bibinfo {year} {1976})}\BibitemShut {NoStop}%
\bibitem [{\citenamefont {Karlin}\ \emph {et~al.}(2013)\citenamefont {Karlin}, \citenamefont {Sichau},\ and\ \citenamefont {Chikatamarla}}]{KarlinTwoPop}%
  \BibitemOpen
  \bibfield  {author} {\bibinfo {author} {\bibfnamefont {I.~V.}\ \bibnamefont {Karlin}}, \bibinfo {author} {\bibfnamefont {D.}~\bibnamefont {Sichau}}, \ and\ \bibinfo {author} {\bibfnamefont {S.~S.}\ \bibnamefont {Chikatamarla}},\ }\href {https://link.aps.org/doi/10.1103/PhysRevE.88.063310} {\bibfield  {journal} {\bibinfo  {journal} {Phys. Rev. E}\ }\textbf {\bibinfo {volume} {88}},\ \bibinfo {pages} {063310} (\bibinfo {year} {2013})}\BibitemShut {NoStop}%
\bibitem [{\citenamefont {Hosseini}\ \emph {et~al.}(2024{\natexlab{b}})\citenamefont {Hosseini}, \citenamefont {Bhadauria},\ and\ \citenamefont {Karlin}}]{ProbingDoubleDist2024}%
  \BibitemOpen
  \bibfield  {author} {\bibinfo {author} {\bibfnamefont {S.~A.}\ \bibnamefont {Hosseini}}, \bibinfo {author} {\bibfnamefont {A.}~\bibnamefont {Bhadauria}}, \ and\ \bibinfo {author} {\bibfnamefont {I.~V.}\ \bibnamefont {Karlin}},\ }\href {https://link.aps.org/doi/10.1103/PhysRevE.110.045313} {\bibfield  {journal} {\bibinfo  {journal} {Phys. Rev. E}\ }\textbf {\bibinfo {volume} {110}},\ \bibinfo {pages} {045313} (\bibinfo {year} {2024}{\natexlab{b}})}\BibitemShut {NoStop}%
\bibitem [{\citenamefont {Prasianakis}\ and\ \citenamefont {Karlin}(2007)}]{Prasianakis2007}%
  \BibitemOpen
  \bibfield  {author} {\bibinfo {author} {\bibfnamefont {N.~I.}\ \bibnamefont {Prasianakis}}\ and\ \bibinfo {author} {\bibfnamefont {I.~V.}\ \bibnamefont {Karlin}},\ }\href {https://link.aps.org/doi/10.1103/PhysRevE.76.016702} {\bibfield  {journal} {\bibinfo  {journal} {Phys. Rev. E}\ }\textbf {\bibinfo {volume} {76}},\ \bibinfo {pages} {016702} (\bibinfo {year} {2007})}\BibitemShut {NoStop}%
\bibitem [{\citenamefont {Saadat}\ \emph {et~al.}(2019)\citenamefont {Saadat}, \citenamefont {B\"osch},\ and\ \citenamefont {Karlin}}]{Saadat2019}%
  \BibitemOpen
  \bibfield  {author} {\bibinfo {author} {\bibfnamefont {M.~H.}\ \bibnamefont {Saadat}}, \bibinfo {author} {\bibfnamefont {F.}~\bibnamefont {B\"osch}}, \ and\ \bibinfo {author} {\bibfnamefont {I.~V.}\ \bibnamefont {Karlin}},\ }\href {https://link.aps.org/doi/10.1103/PhysRevE.99.013306} {\bibfield  {journal} {\bibinfo  {journal} {Phys. Rev. E}\ }\textbf {\bibinfo {volume} {99}},\ \bibinfo {pages} {013306} (\bibinfo {year} {2019})}\BibitemShut {NoStop}%
\bibitem [{\citenamefont {Hosseini}\ \emph {et~al.}(2020)\citenamefont {Hosseini}, \citenamefont {Darabiha},\ and\ \citenamefont {Th{\'e}venin}}]{hosseini2020compressibility}%
  \BibitemOpen
  \bibfield  {author} {\bibinfo {author} {\bibfnamefont {S.~A.}\ \bibnamefont {Hosseini}}, \bibinfo {author} {\bibfnamefont {N.}~\bibnamefont {Darabiha}}, \ and\ \bibinfo {author} {\bibfnamefont {D.}~\bibnamefont {Th{\'e}venin}},\ }\href@noop {} {\bibfield  {journal} {\bibinfo  {journal} {Philosophical Transactions of the Royal Society A}\ }\textbf {\bibinfo {volume} {378}},\ \bibinfo {pages} {20190399} (\bibinfo {year} {2020})}\BibitemShut {NoStop}%
\bibitem [{\citenamefont {Feng}\ \emph {et~al.}(2015)\citenamefont {Feng}, \citenamefont {Sagaut},\ and\ \citenamefont {Tao}}]{FengCorrection}%
  \BibitemOpen
  \bibfield  {author} {\bibinfo {author} {\bibfnamefont {Y.}~\bibnamefont {Feng}}, \bibinfo {author} {\bibfnamefont {P.}~\bibnamefont {Sagaut}}, \ and\ \bibinfo {author} {\bibfnamefont {W.}~\bibnamefont {Tao}},\ }\href {https://www.sciencedirect.com/science/article/pii/S0021999115005926} {\bibfield  {journal} {\bibinfo  {journal} {Journal of Computational Physics}\ }\textbf {\bibinfo {volume} {303}},\ \bibinfo {pages} {514} (\bibinfo {year} {2015})}\BibitemShut {NoStop}%
\bibitem [{\citenamefont {Li}\ \emph {et~al.}(2012)\citenamefont {Li}, \citenamefont {Luo}, \citenamefont {He}, \citenamefont {Gao},\ and\ \citenamefont {Tao}}]{li2012coupling}%
  \BibitemOpen
  \bibfield  {author} {\bibinfo {author} {\bibfnamefont {Q.}~\bibnamefont {Li}}, \bibinfo {author} {\bibfnamefont {K.~H.}\ \bibnamefont {Luo}}, \bibinfo {author} {\bibfnamefont {Y.~L.}\ \bibnamefont {He}}, \bibinfo {author} {\bibfnamefont {Y.~J.}\ \bibnamefont {Gao}}, \ and\ \bibinfo {author} {\bibfnamefont {W.~Q.}\ \bibnamefont {Tao}},\ }\href@noop {} {\bibfield  {journal} {\bibinfo  {journal} {Physical Review E—Statistical, Nonlinear, and Soft Matter Physics}\ }\textbf {\bibinfo {volume} {85}},\ \bibinfo {pages} {016710} (\bibinfo {year} {2012})}\BibitemShut {NoStop}%
\bibitem [{\citenamefont {Feng}\ \emph {et~al.}(2016)\citenamefont {Feng}, \citenamefont {Sagaut},\ and\ \citenamefont {Tao}}]{FENG_FV_DBM}%
  \BibitemOpen
  \bibfield  {author} {\bibinfo {author} {\bibfnamefont {Y.}~\bibnamefont {Feng}}, \bibinfo {author} {\bibfnamefont {P.}~\bibnamefont {Sagaut}}, \ and\ \bibinfo {author} {\bibfnamefont {W.-Q.}\ \bibnamefont {Tao}},\ }\href {https://www.sciencedirect.com/science/article/pii/S0045793016300652} {\bibfield  {journal} {\bibinfo  {journal} {computers \& Fluids}\ }\textbf {\bibinfo {volume} {131}},\ \bibinfo {pages} {45} (\bibinfo {year} {2016})}\BibitemShut {NoStop}%
\bibitem [{\citenamefont {Chikatamarla}\ and\ \citenamefont {Karlin}(2006)}]{ChikatamarlaMultispeed}%
  \BibitemOpen
  \bibfield  {author} {\bibinfo {author} {\bibfnamefont {S.~S.}\ \bibnamefont {Chikatamarla}}\ and\ \bibinfo {author} {\bibfnamefont {I.~V.}\ \bibnamefont {Karlin}},\ }\href {https://link.aps.org/doi/10.1103/PhysRevLett.97.190601} {\bibfield  {journal} {\bibinfo  {journal} {Phys. Rev. Lett.}\ }\textbf {\bibinfo {volume} {97}},\ \bibinfo {pages} {190601} (\bibinfo {year} {2006})}\BibitemShut {NoStop}%
\bibitem [{\citenamefont {Frapolli}\ \emph {et~al.}(2014)\citenamefont {Frapolli}, \citenamefont {Chikatamarla},\ and\ \citenamefont {Karlin}}]{FrapolliMultispeed}%
  \BibitemOpen
  \bibfield  {author} {\bibinfo {author} {\bibfnamefont {N.}~\bibnamefont {Frapolli}}, \bibinfo {author} {\bibfnamefont {S.}~\bibnamefont {Chikatamarla}}, \ and\ \bibinfo {author} {\bibfnamefont {I.}~\bibnamefont {Karlin}},\ }\href@noop {} {\bibfield  {journal} {\bibinfo  {journal} {Physical review. E, Statistical physics, plasmas, fluids, and related interdisciplinary topics}\ }\textbf {\bibinfo {volume} {90}} (\bibinfo {year} {2014})}\BibitemShut {NoStop}%
\bibitem [{\citenamefont {Frapolli}\ \emph {et~al.}(2016{\natexlab{a}})\citenamefont {Frapolli}, \citenamefont {Chikatamarla},\ and\ \citenamefont {Karlin}}]{Frapolli2016b}%
  \BibitemOpen
  \bibfield  {author} {\bibinfo {author} {\bibfnamefont {N.}~\bibnamefont {Frapolli}}, \bibinfo {author} {\bibfnamefont {S.~S.}\ \bibnamefont {Chikatamarla}}, \ and\ \bibinfo {author} {\bibfnamefont {I.~V.}\ \bibnamefont {Karlin}},\ }\href {https://link.aps.org/doi/10.1103/PhysRevLett.117.010604} {\bibfield  {journal} {\bibinfo  {journal} {Phys. Rev. Lett.}\ }\textbf {\bibinfo {volume} {117}},\ \bibinfo {pages} {010604} (\bibinfo {year} {2016}{\natexlab{a}})}\BibitemShut {NoStop}%
\bibitem [{\citenamefont {Hosseini}\ \emph {et~al.}(2019)\citenamefont {Hosseini}, \citenamefont {Coreixas}, \citenamefont {Darabiha},\ and\ \citenamefont {Th\'evenin}}]{Ali_shiftedStencils}%
  \BibitemOpen
  \bibfield  {author} {\bibinfo {author} {\bibfnamefont {S.~A.}\ \bibnamefont {Hosseini}}, \bibinfo {author} {\bibfnamefont {C.}~\bibnamefont {Coreixas}}, \bibinfo {author} {\bibfnamefont {N.}~\bibnamefont {Darabiha}}, \ and\ \bibinfo {author} {\bibfnamefont {D.}~\bibnamefont {Th\'evenin}},\ }\href {https://link.aps.org/doi/10.1103/PhysRevE.100.063301} {\bibfield  {journal} {\bibinfo  {journal} {Phys. Rev. E}\ }\textbf {\bibinfo {volume} {100}},\ \bibinfo {pages} {063301} (\bibinfo {year} {2019})}\BibitemShut {NoStop}%
\bibitem [{\citenamefont {Coreixas}\ and\ \citenamefont {Latt}(2020)}]{Coreixas2020}%
  \BibitemOpen
  \bibfield  {author} {\bibinfo {author} {\bibfnamefont {C.}~\bibnamefont {Coreixas}}\ and\ \bibinfo {author} {\bibfnamefont {J.}~\bibnamefont {Latt}},\ }\href {http://dx.doi.org/10.1063/5.0027986} {\bibfield  {journal} {\bibinfo  {journal} {Physics of Fluids}\ }\textbf {\bibinfo {volume} {32}} (\bibinfo {year} {2020})}\BibitemShut {NoStop}%
\bibitem [{\citenamefont {Dorschner}\ \emph {et~al.}(2018)\citenamefont {Dorschner}, \citenamefont {Bösch},\ and\ \citenamefont {Karlin}}]{PonD18}%
  \BibitemOpen
  \bibfield  {author} {\bibinfo {author} {\bibfnamefont {B.}~\bibnamefont {Dorschner}}, \bibinfo {author} {\bibfnamefont {F.}~\bibnamefont {Bösch}}, \ and\ \bibinfo {author} {\bibfnamefont {I.~V.}\ \bibnamefont {Karlin}},\ }\href {https://link.aps.org/doi/10.1103/PhysRevLett.121.130602} {\bibfield  {journal} {\bibinfo  {journal} {Phys. Rev. Lett.}\ }\textbf {\bibinfo {volume} {121}},\ \bibinfo {pages} {130602} (\bibinfo {year} {2018})}\BibitemShut {NoStop}%
\bibitem [{\citenamefont {Reyhanian}\ \emph {et~al.}(2020)\citenamefont {Reyhanian}, \citenamefont {Dorschner},\ and\ \citenamefont {Karlin}}]{Reyhanian20}%
  \BibitemOpen
  \bibfield  {author} {\bibinfo {author} {\bibfnamefont {E.}~\bibnamefont {Reyhanian}}, \bibinfo {author} {\bibfnamefont {B.}~\bibnamefont {Dorschner}}, \ and\ \bibinfo {author} {\bibfnamefont {I.~V.}\ \bibnamefont {Karlin}},\ }\href {\doibase 10.1103/PhysRevE.102.020103} {\bibfield  {journal} {\bibinfo  {journal} {Phys. Rev. E}\ }\textbf {\bibinfo {volume} {102}},\ \bibinfo {pages} {020103} (\bibinfo {year} {2020})}\BibitemShut {NoStop}%
\bibitem [{\citenamefont {Reyhanian}\ \emph {et~al.}(2021)\citenamefont {Reyhanian}, \citenamefont {Dorschner},\ and\ \citenamefont {Karlin}}]{Reyhanian21}%
  \BibitemOpen
  \bibfield  {author} {\bibinfo {author} {\bibfnamefont {E.}~\bibnamefont {Reyhanian}}, \bibinfo {author} {\bibfnamefont {B.}~\bibnamefont {Dorschner}}, \ and\ \bibinfo {author} {\bibfnamefont {I.}~\bibnamefont {Karlin}},\ }\href {\doibase 10.3390/computation9020013} {\bibfield  {journal} {\bibinfo  {journal} {Computation}\ }\textbf {\bibinfo {volume} {9}},\ \bibinfo {pages} {13} (\bibinfo {year} {2021})}\BibitemShut {NoStop}%
\bibitem [{\citenamefont {Sawant}\ \emph {et~al.}(2022)\citenamefont {Sawant}, \citenamefont {Dorschner},\ and\ \citenamefont {Karlin}}]{sawant2022detonation}%
  \BibitemOpen
  \bibfield  {author} {\bibinfo {author} {\bibfnamefont {N.}~\bibnamefont {Sawant}}, \bibinfo {author} {\bibfnamefont {B.}~\bibnamefont {Dorschner}}, \ and\ \bibinfo {author} {\bibfnamefont {I.~V.}\ \bibnamefont {Karlin}},\ }\href@noop {} {\bibfield  {journal} {\bibinfo  {journal} {{AIP} advances}\ }\textbf {\bibinfo {volume} {12}},\ \bibinfo {pages} {075107} (\bibinfo {year} {2022})}\BibitemShut {NoStop}%
\bibitem [{\citenamefont {Bhadauria}\ and\ \citenamefont {Karlin}(2024)}]{Bhaduria23}%
  \BibitemOpen
  \bibfield  {author} {\bibinfo {author} {\bibfnamefont {A.}~\bibnamefont {Bhadauria}}\ and\ \bibinfo {author} {\bibfnamefont {I.}~\bibnamefont {Karlin}},\ }\href {https://www.sciencedirect.com/science/article/pii/S0045793023003778} {\bibfield  {journal} {\bibinfo  {journal} {computers \& Fluids}\ }\textbf {\bibinfo {volume} {271}},\ \bibinfo {pages} {106152} (\bibinfo {year} {2024})}\BibitemShut {NoStop}%
\bibitem [{\citenamefont {Bardow}\ \emph {et~al.}(2008)\citenamefont {Bardow}, \citenamefont {Karlin},\ and\ \citenamefont {Gusev}}]{BardowMultispeedSSLBM}%
  \BibitemOpen
  \bibfield  {author} {\bibinfo {author} {\bibfnamefont {A.}~\bibnamefont {Bardow}}, \bibinfo {author} {\bibfnamefont {I.~V.}\ \bibnamefont {Karlin}}, \ and\ \bibinfo {author} {\bibfnamefont {A.~A.}\ \bibnamefont {Gusev}},\ }\href {https://link.aps.org/doi/10.1103/PhysRevE.77.025701} {\bibfield  {journal} {\bibinfo  {journal} {Phys. Rev. E}\ }\textbf {\bibinfo {volume} {77}},\ \bibinfo {pages} {025701} (\bibinfo {year} {2008})}\BibitemShut {NoStop}%
\bibitem [{\citenamefont {Wilde}\ \emph {et~al.}(2020)\citenamefont {Wilde}, \citenamefont {Kr\"amer}, \citenamefont {Reith},\ and\ \citenamefont {Foysi}}]{Wilde_SSLBM_CF}%
  \BibitemOpen
  \bibfield  {author} {\bibinfo {author} {\bibfnamefont {D.}~\bibnamefont {Wilde}}, \bibinfo {author} {\bibfnamefont {A.}~\bibnamefont {Kr\"amer}}, \bibinfo {author} {\bibfnamefont {D.}~\bibnamefont {Reith}}, \ and\ \bibinfo {author} {\bibfnamefont {H.}~\bibnamefont {Foysi}},\ }\href {https://link.aps.org/doi/10.1103/PhysRevE.101.053306} {\bibfield  {journal} {\bibinfo  {journal} {Phys. Rev. E}\ }\textbf {\bibinfo {volume} {101}},\ \bibinfo {pages} {053306} (\bibinfo {year} {2020})}\BibitemShut {NoStop}%
\bibitem [{\citenamefont {Wilde}\ \emph {et~al.}(2021)\citenamefont {Wilde}, \citenamefont {Krämer}, \citenamefont {Bedrunka}, \citenamefont {Reith},\ and\ \citenamefont {Foysi}}]{Wilde_CubatureRules}%
  \BibitemOpen
  \bibfield  {author} {\bibinfo {author} {\bibfnamefont {D.}~\bibnamefont {Wilde}}, \bibinfo {author} {\bibfnamefont {A.}~\bibnamefont {Krämer}}, \bibinfo {author} {\bibfnamefont {M.}~\bibnamefont {Bedrunka}}, \bibinfo {author} {\bibfnamefont {D.}~\bibnamefont {Reith}}, \ and\ \bibinfo {author} {\bibfnamefont {H.}~\bibnamefont {Foysi}},\ }\href {https://www.sciencedirect.com/science/article/pii/S1877750321000508} {\bibfield  {journal} {\bibinfo  {journal} {Journal of Computational Science}\ }\textbf {\bibinfo {volume} {51}},\ \bibinfo {pages} {101355} (\bibinfo {year} {2021})}\BibitemShut {NoStop}%
\bibitem [{\citenamefont {Xu}\ \emph {et~al.}(2021)\citenamefont {Xu}, \citenamefont {Chen},\ and\ \citenamefont {Cai}}]{XuFV}%
  \BibitemOpen
  \bibfield  {author} {\bibinfo {author} {\bibfnamefont {L.}~\bibnamefont {Xu}}, \bibinfo {author} {\bibfnamefont {R.}~\bibnamefont {Chen}}, \ and\ \bibinfo {author} {\bibfnamefont {X.-C.}\ \bibnamefont {Cai}},\ }\href {https://link.aps.org/doi/10.1103/PhysRevE.103.023306} {\bibfield  {journal} {\bibinfo  {journal} {Phys. Rev. E}\ }\textbf {\bibinfo {volume} {103}},\ \bibinfo {pages} {023306} (\bibinfo {year} {2021})}\BibitemShut {NoStop}%
\bibitem [{\citenamefont {Ji}\ \emph {et~al.}(2024)\citenamefont {Ji}, \citenamefont {Hosseini}, \citenamefont {Dorschner}, \citenamefont {Luo},\ and\ \citenamefont {Karlin}}]{Ji24}%
  \BibitemOpen
  \bibfield  {author} {\bibinfo {author} {\bibfnamefont {Y.}~\bibnamefont {Ji}}, \bibinfo {author} {\bibfnamefont {S.~A.}\ \bibnamefont {Hosseini}}, \bibinfo {author} {\bibfnamefont {B.}~\bibnamefont {Dorschner}}, \bibinfo {author} {\bibfnamefont {K.~H.}\ \bibnamefont {Luo}}, \ and\ \bibinfo {author} {\bibfnamefont {I.~V.}\ \bibnamefont {Karlin}},\ }\href {https://doi.org/10.1017/jfm.2024.94} {\bibfield  {journal} {\bibinfo  {journal} {Journal of Fluid Mechanics}\ }\textbf {\bibinfo {volume} {983}},\ \bibinfo {pages} {A11} (\bibinfo {year} {2024})}\BibitemShut {NoStop}%
\bibitem [{\citenamefont {Liu}\ \emph {et~al.}(2014)\citenamefont {Liu}, \citenamefont {Yu}, \citenamefont {Xu},\ and\ \citenamefont {Zhong}}]{UGKS_14}%
  \BibitemOpen
  \bibfield  {author} {\bibinfo {author} {\bibfnamefont {S.}~\bibnamefont {Liu}}, \bibinfo {author} {\bibfnamefont {P.}~\bibnamefont {Yu}}, \bibinfo {author} {\bibfnamefont {K.}~\bibnamefont {Xu}}, \ and\ \bibinfo {author} {\bibfnamefont {C.}~\bibnamefont {Zhong}},\ }\href@noop {} {\bibfield  {journal} {\bibinfo  {journal} {Journal of Computational Physics}\ }\textbf {\bibinfo {volume} {259}},\ \bibinfo {pages} {96} (\bibinfo {year} {2014})}\BibitemShut {NoStop}%
\bibitem [{\citenamefont {Guo}\ \emph {et~al.}(2015)\citenamefont {Guo}, \citenamefont {Wang},\ and\ \citenamefont {Xu}}]{DUGKS_compressiblecase}%
  \BibitemOpen
  \bibfield  {author} {\bibinfo {author} {\bibfnamefont {Z.}~\bibnamefont {Guo}}, \bibinfo {author} {\bibfnamefont {R.}~\bibnamefont {Wang}}, \ and\ \bibinfo {author} {\bibfnamefont {K.}~\bibnamefont {Xu}},\ }\href {https://link.aps.org/doi/10.1103/PhysRevE.91.033313} {\bibfield  {journal} {\bibinfo  {journal} {Phys. Rev. E}\ }\textbf {\bibinfo {volume} {91}},\ \bibinfo {pages} {033313} (\bibinfo {year} {2015})}\BibitemShut {NoStop}%
\bibitem [{\citenamefont {Guo}\ \emph {et~al.}(2023)\citenamefont {Guo}, \citenamefont {Wang},\ and\ \citenamefont {Qi}}]{GuoDUGKS23}%
  \BibitemOpen
  \bibfield  {author} {\bibinfo {author} {\bibfnamefont {Z.}~\bibnamefont {Guo}}, \bibinfo {author} {\bibfnamefont {L.-P.}\ \bibnamefont {Wang}}, \ and\ \bibinfo {author} {\bibfnamefont {Y.}~\bibnamefont {Qi}},\ }\href {https://link.aps.org/doi/10.1103/PhysRevE.107.025304} {\bibfield  {journal} {\bibinfo  {journal} {Phys. Rev. E}\ }\textbf {\bibinfo {volume} {107}},\ \bibinfo {pages} {025304} (\bibinfo {year} {2023})}\BibitemShut {NoStop}%
\bibitem [{\citenamefont {Kallikounis}\ \emph {et~al.}(2022)\citenamefont {Kallikounis}, \citenamefont {Dorschner},\ and\ \citenamefont {Karlin}}]{Kallikounis22}%
  \BibitemOpen
  \bibfield  {author} {\bibinfo {author} {\bibfnamefont {N.~G.}\ \bibnamefont {Kallikounis}}, \bibinfo {author} {\bibfnamefont {B.}~\bibnamefont {Dorschner}}, \ and\ \bibinfo {author} {\bibfnamefont {I.~V.}\ \bibnamefont {Karlin}},\ }\href {\doibase 10.1103/PhysRevE.106.015301} {\bibfield  {journal} {\bibinfo  {journal} {Phys. Rev. E}\ }\textbf {\bibinfo {volume} {106}},\ \bibinfo {pages} {015301} (\bibinfo {year} {2022})}\BibitemShut {NoStop}%
\bibitem [{\citenamefont {Kallikounis}\ and\ \citenamefont {Karlin}(2024)}]{Kallikounis23}%
  \BibitemOpen
  \bibfield  {author} {\bibinfo {author} {\bibfnamefont {N.~G.}\ \bibnamefont {Kallikounis}}\ and\ \bibinfo {author} {\bibfnamefont {I.~V.}\ \bibnamefont {Karlin}},\ }\href {https://link.aps.org/doi/10.1103/PhysRevE.109.015304} {\bibfield  {journal} {\bibinfo  {journal} {Phys. Rev. E}\ }\textbf {\bibinfo {volume} {109}},\ \bibinfo {pages} {015304} (\bibinfo {year} {2024})}\BibitemShut {NoStop}%
\bibitem [{\citenamefont {Kallikounis}\ \emph {et~al.}(2021)\citenamefont {Kallikounis}, \citenamefont {Dorschner},\ and\ \citenamefont {Karlin}}]{Kallikounis_multiscale}%
  \BibitemOpen
  \bibfield  {author} {\bibinfo {author} {\bibfnamefont {N.~G.}\ \bibnamefont {Kallikounis}}, \bibinfo {author} {\bibfnamefont {B.}~\bibnamefont {Dorschner}}, \ and\ \bibinfo {author} {\bibfnamefont {I.~V.}\ \bibnamefont {Karlin}},\ }\href {https://link.aps.org/doi/10.1103/PhysRevE.103.063305} {\bibfield  {journal} {\bibinfo  {journal} {Phys. Rev. E}\ }\textbf {\bibinfo {volume} {103}},\ \bibinfo {pages} {063305} (\bibinfo {year} {2021})}\BibitemShut {NoStop}%
\bibitem [{\citenamefont {Kallikounis}(2023)}]{PhD_Kallikounis}%
  \BibitemOpen
  \bibfield  {author} {\bibinfo {author} {\bibfnamefont {N.}~\bibnamefont {Kallikounis}},\ }\emph {\bibinfo {title} {Multiscale Kineitc Modeling of High Speed Gas Dynamics}},\ \href@noop {} {Ph.D. thesis},\ \bibinfo  {school} {ETH Zürich} (\bibinfo {year} {2023})\BibitemShut {NoStop}%
\bibitem [{\citenamefont {Dorschner}\ \emph {et~al.}(2016)\citenamefont {Dorschner}, \citenamefont {Frapolli}, \citenamefont {Chikatamarla},\ and\ \citenamefont {Karlin}}]{DorschnerRefinementELBM}%
  \BibitemOpen
  \bibfield  {author} {\bibinfo {author} {\bibfnamefont {B.}~\bibnamefont {Dorschner}}, \bibinfo {author} {\bibfnamefont {N.}~\bibnamefont {Frapolli}}, \bibinfo {author} {\bibfnamefont {S.~S.}\ \bibnamefont {Chikatamarla}}, \ and\ \bibinfo {author} {\bibfnamefont {I.~V.}\ \bibnamefont {Karlin}},\ }\href {https://link.aps.org/doi/10.1103/PhysRevE.94.053311} {\bibfield  {journal} {\bibinfo  {journal} {Phys. Rev. E}\ }\textbf {\bibinfo {volume} {94}},\ \bibinfo {pages} {053311} (\bibinfo {year} {2016})}\BibitemShut {NoStop}%
\bibitem [{\citenamefont {Guzik}\ \emph {et~al.}(2013)\citenamefont {Guzik}, \citenamefont {Gao}, \citenamefont {Weisgraber}, \citenamefont {Alder},\ and\ \citenamefont {Colella}}]{Guzik}%
  \BibitemOpen
  \bibfield  {author} {\bibinfo {author} {\bibfnamefont {S.}~\bibnamefont {Guzik}}, \bibinfo {author} {\bibfnamefont {X.}~\bibnamefont {Gao}}, \bibinfo {author} {\bibfnamefont {T.}~\bibnamefont {Weisgraber}}, \bibinfo {author} {\bibfnamefont {B.}~\bibnamefont {Alder}}, \ and\ \bibinfo {author} {\bibfnamefont {P.}~\bibnamefont {Colella}}\ }(\bibinfo {year} {2013})\BibitemShut {NoStop}%
\bibitem [{\citenamefont {Eitel-Amor}\ \emph {et~al.}(2013)\citenamefont {Eitel-Amor}, \citenamefont {Meinke},\ and\ \citenamefont {Schröder}}]{EitelAmor_AMR_LBM}%
  \BibitemOpen
  \bibfield  {author} {\bibinfo {author} {\bibfnamefont {G.}~\bibnamefont {Eitel-Amor}}, \bibinfo {author} {\bibfnamefont {M.}~\bibnamefont {Meinke}}, \ and\ \bibinfo {author} {\bibfnamefont {W.}~\bibnamefont {Schröder}},\ }\href {https://www.sciencedirect.com/science/article/pii/S0045793013000315} {\bibfield  {journal} {\bibinfo  {journal} {computers \& Fluids}\ }\textbf {\bibinfo {volume} {75}},\ \bibinfo {pages} {127} (\bibinfo {year} {2013})}\BibitemShut {NoStop}%
\bibitem [{\citenamefont {He}\ \emph {et~al.}(2022)\citenamefont {He}, \citenamefont {Huang}, \citenamefont {Yin}, \citenamefont {Hu},\ and\ \citenamefont {Li}}]{He_AMR_LBM}%
  \BibitemOpen
  \bibfield  {author} {\bibinfo {author} {\bibfnamefont {Q.}~\bibnamefont {He}}, \bibinfo {author} {\bibfnamefont {W.}~\bibnamefont {Huang}}, \bibinfo {author} {\bibfnamefont {Y.}~\bibnamefont {Yin}}, \bibinfo {author} {\bibfnamefont {Y.}~\bibnamefont {Hu}}, \ and\ \bibinfo {author} {\bibfnamefont {D.}~\bibnamefont {Li}},\ }\href {https://doi.org/10.1063/5.0104362} {\bibfield  {journal} {\bibinfo  {journal} {Physics of Fluids}\ }\textbf {\bibinfo {volume} {34}},\ \bibinfo {pages} {093321} (\bibinfo {year} {2022})}\BibitemShut {NoStop}%
\bibitem [{\citenamefont {Schukmann}\ \emph {et~al.}(2023)\citenamefont {Schukmann}, \citenamefont {Schneider}, \citenamefont {Haas},\ and\ \citenamefont {Böhle}}]{Schukmann_AMR_LBM}%
  \BibitemOpen
  \bibfield  {author} {\bibinfo {author} {\bibfnamefont {A.}~\bibnamefont {Schukmann}}, \bibinfo {author} {\bibfnamefont {A.}~\bibnamefont {Schneider}}, \bibinfo {author} {\bibfnamefont {V.}~\bibnamefont {Haas}}, \ and\ \bibinfo {author} {\bibfnamefont {M.}~\bibnamefont {Böhle}},\ }\href {https://www.mdpi.com/2311-5521/8/3/103} {\bibfield  {journal} {\bibinfo  {journal} {Fluids}\ }\textbf {\bibinfo {volume} {8}} (\bibinfo {year} {2023})}\BibitemShut {NoStop}%
\bibitem [{\citenamefont {Fakhari}\ and\ \citenamefont {Lee}(2014)}]{Fakhari_FD_LBM}%
  \BibitemOpen
  \bibfield  {author} {\bibinfo {author} {\bibfnamefont {A.}~\bibnamefont {Fakhari}}\ and\ \bibinfo {author} {\bibfnamefont {T.}~\bibnamefont {Lee}},\ }\href {https://link.aps.org/doi/10.1103/PhysRevE.89.033310} {\bibfield  {journal} {\bibinfo  {journal} {Phys. Rev. E}\ }\textbf {\bibinfo {volume} {89}},\ \bibinfo {pages} {033310} (\bibinfo {year} {2014})}\BibitemShut {NoStop}%
\bibitem [{\citenamefont {Huang}\ \emph {et~al.}(2023)\citenamefont {Huang}, \citenamefont {Chen}, \citenamefont {Zhang}, \citenamefont {Wang},\ and\ \citenamefont {Wang}}]{Huang_AMR_RLBFS}%
  \BibitemOpen
  \bibfield  {author} {\bibinfo {author} {\bibfnamefont {X.}~\bibnamefont {Huang}}, \bibinfo {author} {\bibfnamefont {J.}~\bibnamefont {Chen}}, \bibinfo {author} {\bibfnamefont {J.}~\bibnamefont {Zhang}}, \bibinfo {author} {\bibfnamefont {L.}~\bibnamefont {Wang}}, \ and\ \bibinfo {author} {\bibfnamefont {Y.}~\bibnamefont {Wang}},\ }\href@noop {} {\bibfield  {journal} {\bibinfo  {journal} {Symmetry}\ }\textbf {\bibinfo {volume} {15}},\ \bibinfo {pages} {1909} (\bibinfo {year} {2023})}\BibitemShut {NoStop}%
\bibitem [{\citenamefont {Frapolli}\ \emph {et~al.}(2016{\natexlab{b}})\citenamefont {Frapolli}, \citenamefont {Chikatamarla},\ and\ \citenamefont {Karlin}}]{2016Entropic}%
  \BibitemOpen
  \bibfield  {author} {\bibinfo {author} {\bibfnamefont {N.}~\bibnamefont {Frapolli}}, \bibinfo {author} {\bibfnamefont {S.~S.}\ \bibnamefont {Chikatamarla}}, \ and\ \bibinfo {author} {\bibfnamefont {I.~V.}\ \bibnamefont {Karlin}},\ }\href@noop {} {\bibfield  {journal} {\bibinfo  {journal} {Physical Review E}\ }\textbf {\bibinfo {volume} {93}},\ \bibinfo {pages} {063302} (\bibinfo {year} {2016}{\natexlab{b}})}\BibitemShut {NoStop}%
\bibitem [{\citenamefont {Grad}(1949)}]{grad1949kinetic}%
  \BibitemOpen
  \bibfield  {author} {\bibinfo {author} {\bibfnamefont {H.}~\bibnamefont {Grad}},\ }\href@noop {} {\bibfield  {journal} {\bibinfo  {journal} {Communications on pure and applied mathematics}\ }\textbf {\bibinfo {volume} {2}},\ \bibinfo {pages} {331} (\bibinfo {year} {1949})}\BibitemShut {NoStop}%
\bibitem [{\citenamefont {Zhang}(1988)}]{NND1}%
  \BibitemOpen
  \bibfield  {author} {\bibinfo {author} {\bibfnamefont {H.~X.}\ \bibnamefont {Zhang}},\ }\href@noop {} {\bibfield  {journal} {\bibinfo  {journal} {Acta Aerodynamica Sinica}\ }\textbf {\bibinfo {volume} {6}},\ \bibinfo {pages} {143} (\bibinfo {year} {1988})}\BibitemShut {NoStop}%
\bibitem [{\citenamefont {Zhang}\ and\ \citenamefont {Zhuang}(2020)}]{NND2}%
  \BibitemOpen
  \bibfield  {author} {\bibinfo {author} {\bibfnamefont {H.}~\bibnamefont {Zhang}}\ and\ \bibinfo {author} {\bibfnamefont {F.}~\bibnamefont {Zhuang}},\ }\href@noop {} {\bibfield  {journal} {\bibinfo  {journal} {Journal of Computational Physics}\ }\textbf {\bibinfo {volume} {415}},\ \bibinfo {pages} {109491} (\bibinfo {year} {2020})}\BibitemShut {NoStop}%
\bibitem [{\citenamefont {Roe}(1986)}]{limiterRoe1986}%
  \BibitemOpen
  \bibfield  {author} {\bibinfo {author} {\bibfnamefont {P.~L.}\ \bibnamefont {Roe}},\ }\href {https://www.annualreviews.org/doi/10.1146/annurev.fl.18.010186.002005} {\bibfield  {journal} {\bibinfo  {journal} {Annual Review of Fluid Mechanics}\ }\textbf {\bibinfo {volume} {18}},\ \bibinfo {pages} {337} (\bibinfo {year} {1986})}\BibitemShut {NoStop}%
\bibitem [{\citenamefont {Zheng}\ and\ \citenamefont {Groth}(2012)}]{OverviewAMRStrategies}%
  \BibitemOpen
  \bibfield  {author} {\bibinfo {author} {\bibfnamefont {J.~Z.~X.}\ \bibnamefont {Zheng}}\ and\ \bibinfo {author} {\bibfnamefont {C.~P.~T.}\ \bibnamefont {Groth}},\ }\href@noop {} {\  (\bibinfo {year} {2012})}\BibitemShut {NoStop}%
\bibitem [{\citenamefont {Berger}\ and\ \citenamefont {Oliger}(1984)}]{Berger_Patchbased1}%
  \BibitemOpen
  \bibfield  {author} {\bibinfo {author} {\bibfnamefont {M.~J.}\ \bibnamefont {Berger}}\ and\ \bibinfo {author} {\bibfnamefont {J.}~\bibnamefont {Oliger}},\ }\href {https://www.sciencedirect.com/science/article/pii/0021999184900731} {\bibfield  {journal} {\bibinfo  {journal} {Journal of Computational Physics}\ }\textbf {\bibinfo {volume} {53}},\ \bibinfo {pages} {484} (\bibinfo {year} {1984})}\BibitemShut {NoStop}%
\bibitem [{\citenamefont {Beckingsale}(2015)}]{BeckingsalePhD}%
  \BibitemOpen
  \bibfield  {author} {\bibinfo {author} {\bibfnamefont {D.}~\bibnamefont {Beckingsale}},\ }\emph {\bibinfo {title} {Towards Scalable Adaptive Mesh Refinement on Future Parallel Architectures}},\ \href@noop {} {Ph.D. thesis} (\bibinfo {year} {2015})\BibitemShut {NoStop}%
\bibitem [{\citenamefont {Khokhlov}(1998)}]{KHOKHLOV1998}%
  \BibitemOpen
  \bibfield  {author} {\bibinfo {author} {\bibfnamefont {A.~M.}\ \bibnamefont {Khokhlov}},\ }\href {https://www.sciencedirect.com/science/article/pii/S0021999198999983} {\bibfield  {journal} {\bibinfo  {journal} {Journal of Computational Physics}\ }\textbf {\bibinfo {volume} {143}},\ \bibinfo {pages} {519} (\bibinfo {year} {1998})}\BibitemShut {NoStop}%
\bibitem [{\citenamefont {Gargantini}(1982)}]{octtrees}%
  \BibitemOpen
  \bibfield  {author} {\bibinfo {author} {\bibfnamefont {I.}~\bibnamefont {Gargantini}},\ }\href {https://www.sciencedirect.com/science/article/pii/0146664X82900582} {\bibfield  {journal} {\bibinfo  {journal} {Computer Graphics and Image Processing}\ }\textbf {\bibinfo {volume} {20}},\ \bibinfo {pages} {365} (\bibinfo {year} {1982})}\BibitemShut {NoStop}%
\bibitem [{\citenamefont {MacNeice}\ \emph {et~al.}(2000)\citenamefont {MacNeice}, \citenamefont {Olson}, \citenamefont {Mobarry}, \citenamefont {de~Fainchtein},\ and\ \citenamefont {Packer}}]{PARAMESH}%
  \BibitemOpen
  \bibfield  {author} {\bibinfo {author} {\bibfnamefont {P.}~\bibnamefont {MacNeice}}, \bibinfo {author} {\bibfnamefont {K.~M.}\ \bibnamefont {Olson}}, \bibinfo {author} {\bibfnamefont {C.}~\bibnamefont {Mobarry}}, \bibinfo {author} {\bibfnamefont {R.}~\bibnamefont {de~Fainchtein}}, \ and\ \bibinfo {author} {\bibfnamefont {C.}~\bibnamefont {Packer}},\ }\href {https://www.sciencedirect.com/science/article/pii/S0010465599005019} {\bibfield  {journal} {\bibinfo  {journal} {Computer Physics Communications}\ }\textbf {\bibinfo {volume} {126}},\ \bibinfo {pages} {330} (\bibinfo {year} {2000})}\BibitemShut {NoStop}%
\bibitem [{\citenamefont {Zhang}\ \emph {et~al.}(2019)\citenamefont {Zhang}, \citenamefont {Almgren}, \citenamefont {Beckner}, \citenamefont {Bell}, \citenamefont {Blaschke}, \citenamefont {Chan}, \citenamefont {Day}, \citenamefont {Friesen}, \citenamefont {Gott}, \citenamefont {Graves} \emph {et~al.}}]{amrex1}%
  \BibitemOpen
  \bibfield  {author} {\bibinfo {author} {\bibfnamefont {W.}~\bibnamefont {Zhang}}, \bibinfo {author} {\bibfnamefont {A.}~\bibnamefont {Almgren}}, \bibinfo {author} {\bibfnamefont {V.}~\bibnamefont {Beckner}}, \bibinfo {author} {\bibfnamefont {J.}~\bibnamefont {Bell}}, \bibinfo {author} {\bibfnamefont {J.}~\bibnamefont {Blaschke}}, \bibinfo {author} {\bibfnamefont {C.}~\bibnamefont {Chan}}, \bibinfo {author} {\bibfnamefont {M.}~\bibnamefont {Day}}, \bibinfo {author} {\bibfnamefont {B.}~\bibnamefont {Friesen}}, \bibinfo {author} {\bibfnamefont {K.}~\bibnamefont {Gott}}, \bibinfo {author} {\bibfnamefont {D.}~\bibnamefont {Graves}},  \emph {et~al.},\ }\href@noop {} {\bibfield  {journal} {\bibinfo  {journal} {The Journal of Open Source Software}\ }\textbf {\bibinfo {volume} {4}},\ \bibinfo {pages} {1370} (\bibinfo {year} {2019})}\BibitemShut {NoStop}%
\bibitem [{\citenamefont {Zhang}\ \emph {et~al.}(2021)\citenamefont {Zhang}, \citenamefont {Myers}, \citenamefont {Gott}, \citenamefont {Almgren},\ and\ \citenamefont {Bell}}]{amrex2}%
  \BibitemOpen
  \bibfield  {author} {\bibinfo {author} {\bibfnamefont {W.}~\bibnamefont {Zhang}}, \bibinfo {author} {\bibfnamefont {A.}~\bibnamefont {Myers}}, \bibinfo {author} {\bibfnamefont {K.}~\bibnamefont {Gott}}, \bibinfo {author} {\bibfnamefont {A.}~\bibnamefont {Almgren}}, \ and\ \bibinfo {author} {\bibfnamefont {J.}~\bibnamefont {Bell}},\ }\href@noop {} {\bibfield  {journal} {\bibinfo  {journal} {The International Journal of High Performance Computing Applications}\ }\textbf {\bibinfo {volume} {35}},\ \bibinfo {pages} {508} (\bibinfo {year} {2021})}\BibitemShut {NoStop}%
\bibitem [{\citenamefont {Myers}\ \emph {et~al.}(2024)\citenamefont {Myers}, \citenamefont {Zhang}, \citenamefont {Almgren}, \citenamefont {Antoun}, \citenamefont {Bell}, \citenamefont {Huebl},\ and\ \citenamefont {Sinn}}]{amrex3}%
  \BibitemOpen
  \bibfield  {author} {\bibinfo {author} {\bibfnamefont {A.}~\bibnamefont {Myers}}, \bibinfo {author} {\bibfnamefont {W.}~\bibnamefont {Zhang}}, \bibinfo {author} {\bibfnamefont {A.}~\bibnamefont {Almgren}}, \bibinfo {author} {\bibfnamefont {T.}~\bibnamefont {Antoun}}, \bibinfo {author} {\bibfnamefont {J.}~\bibnamefont {Bell}}, \bibinfo {author} {\bibfnamefont {A.}~\bibnamefont {Huebl}}, \ and\ \bibinfo {author} {\bibfnamefont {A.}~\bibnamefont {Sinn}},\ }\href {https://api.semanticscholar.org/CorpusID:268532087} {\bibfield  {journal} {\bibinfo  {journal} {ArXiv}\ }\textbf {\bibinfo {volume} {abs/2403.12179}} (\bibinfo {year} {2024})}\BibitemShut {NoStop}%
\bibitem [{\citenamefont {Berger}(1986)}]{Berger_Patchbased2}%
  \BibitemOpen
  \bibfield  {author} {\bibinfo {author} {\bibfnamefont {M.~J.}\ \bibnamefont {Berger}},\ }\href {https://doi.org/10.1137/0907061} {\bibfield  {journal} {\bibinfo  {journal} {SIAM Journal on Scientific and Statistical Computing}\ }\textbf {\bibinfo {volume} {7}},\ \bibinfo {pages} {904} (\bibinfo {year} {1986})}\BibitemShut {NoStop}%
\bibitem [{\citenamefont {Berger}\ and\ \citenamefont {Colella}(1989)}]{Berger_Patchbased3}%
  \BibitemOpen
  \bibfield  {author} {\bibinfo {author} {\bibfnamefont {M.~J.}\ \bibnamefont {Berger}}\ and\ \bibinfo {author} {\bibfnamefont {P.}~\bibnamefont {Colella}},\ }\href@noop {} {\bibfield  {journal} {\bibinfo  {journal} {Journal of Computational Physics}\ }\textbf {\bibinfo {volume} {82}},\ \bibinfo {pages} {64} (\bibinfo {year} {1989})}\BibitemShut {NoStop}%
\bibitem [{\citenamefont {Bell}\ \emph {et~al.}(1994)\citenamefont {Bell}, \citenamefont {Berger}, \citenamefont {Saltzman},\ and\ \citenamefont {Welcome}}]{Berger_Patchbased4}%
  \BibitemOpen
  \bibfield  {author} {\bibinfo {author} {\bibfnamefont {J.}~\bibnamefont {Bell}}, \bibinfo {author} {\bibfnamefont {M.}~\bibnamefont {Berger}}, \bibinfo {author} {\bibfnamefont {J.}~\bibnamefont {Saltzman}}, \ and\ \bibinfo {author} {\bibfnamefont {M.}~\bibnamefont {Welcome}},\ }\href {https://doi.org/10.1137/0915008} {\bibfield  {journal} {\bibinfo  {journal} {SIAM Journal on Scientific Computing}\ }\textbf {\bibinfo {volume} {15}},\ \bibinfo {pages} {127} (\bibinfo {year} {1994})}\BibitemShut {NoStop}%
\bibitem [{\citenamefont {Berger}\ and\ \citenamefont {Rigoutsos}(1991)}]{BergerRigoutsos}%
  \BibitemOpen
  \bibfield  {author} {\bibinfo {author} {\bibfnamefont {M.}~\bibnamefont {Berger}}\ and\ \bibinfo {author} {\bibfnamefont {I.}~\bibnamefont {Rigoutsos}},\ }\href@noop {} {\bibfield  {journal} {\bibinfo  {journal} {IEEE Transactions on Systems, Man and Cybernetics}\ }\textbf {\bibinfo {volume} {21}},\ \bibinfo {pages} {1278} (\bibinfo {year} {1991})}\BibitemShut {NoStop}%
\bibitem [{\citenamefont {Deiterding}(2011)}]{DeiterdingESAIM}%
  \BibitemOpen
  \bibfield  {author} {\bibinfo {author} {\bibfnamefont {R.}~\bibnamefont {Deiterding}},\ }\href {https://doi.org/10.1051/proc/201134002} {\bibfield  {journal} {\bibinfo  {journal} {ESAIM: Proc.}\ }\textbf {\bibinfo {volume} {34}},\ \bibinfo {pages} {97} (\bibinfo {year} {2011})}\BibitemShut {NoStop}%
\bibitem [{\citenamefont {Deiterding}(2003)}]{DeiterdingPhD}%
  \BibitemOpen
  \bibfield  {author} {\bibinfo {author} {\bibfnamefont {R.}~\bibnamefont {Deiterding}},\ }\emph {\bibinfo {title} {Parallel adaptive simulation of multi-dimensional detonation structures}},\ \href {https://eprints.soton.ac.uk/380602/} {Ph.D. thesis},\ \bibinfo  {school} {Brandenburgische Technische Universitat Cottbus} (\bibinfo {year} {2003})\BibitemShut {NoStop}%
\bibitem [{\citenamefont {Dellar}(2001)}]{dellar2001bulk}%
  \BibitemOpen
  \bibfield  {author} {\bibinfo {author} {\bibfnamefont {P.~J.}\ \bibnamefont {Dellar}},\ }\href@noop {} {\bibfield  {journal} {\bibinfo  {journal} {Physical Review E}\ }\textbf {\bibinfo {volume} {64}},\ \bibinfo {pages} {031203} (\bibinfo {year} {2001})}\BibitemShut {NoStop}%
\bibitem [{\citenamefont {Sod}(1978)}]{Sodtube}%
  \BibitemOpen
  \bibfield  {author} {\bibinfo {author} {\bibfnamefont {G.~A.}\ \bibnamefont {Sod}},\ }\href@noop {} {\bibfield  {journal} {\bibinfo  {journal} {Journal of Computational Physics}\ }\textbf {\bibinfo {volume} {27}},\ \bibinfo {pages} {1} (\bibinfo {year} {1978})}\BibitemShut {NoStop}%
\bibitem [{\citenamefont {Lax}(1954)}]{Laxtube}%
  \BibitemOpen
  \bibfield  {author} {\bibinfo {author} {\bibfnamefont {P.~D.}\ \bibnamefont {Lax}},\ }\href {https://onlinelibrary.wiley.com/doi/abs/10.1002/cpa.3160070112} {\bibfield  {journal} {\bibinfo  {journal} {Communications on Pure and Applied Mathematics}\ }\textbf {\bibinfo {volume} {7}},\ \bibinfo {pages} {159} (\bibinfo {year} {1954})}\BibitemShut {NoStop}%
\bibitem [{\citenamefont {Shu}\ and\ \citenamefont {Osher}(1988)}]{ShuOsher}%
  \BibitemOpen
  \bibfield  {author} {\bibinfo {author} {\bibfnamefont {C.-W.}\ \bibnamefont {Shu}}\ and\ \bibinfo {author} {\bibfnamefont {S.}~\bibnamefont {Osher}},\ }\href {https://www.sciencedirect.com/science/article/pii/0021999188901775} {\bibfield  {journal} {\bibinfo  {journal} {Journal of Computational Physics}\ }\textbf {\bibinfo {volume} {77}},\ \bibinfo {pages} {439} (\bibinfo {year} {1988})}\BibitemShut {NoStop}%
\bibitem [{\citenamefont {Lax}\ and\ \citenamefont {Liu}(1998)}]{LaxLiu1998}%
  \BibitemOpen
  \bibfield  {author} {\bibinfo {author} {\bibfnamefont {P.~D.}\ \bibnamefont {Lax}}\ and\ \bibinfo {author} {\bibfnamefont {X.-D.}\ \bibnamefont {Liu}},\ }\href {https://doi.org/10.1137/S1064827595291819} {\bibfield  {journal} {\bibinfo  {journal} {SIAM Journal on Scientific Computing}\ }\textbf {\bibinfo {volume} {19}},\ \bibinfo {pages} {319} (\bibinfo {year} {1998})}\BibitemShut {NoStop}%
\bibitem [{\citenamefont {Kurganov}\ and\ \citenamefont {Tadmor}(2002)}]{KurganovR2D}%
  \BibitemOpen
  \bibfield  {author} {\bibinfo {author} {\bibfnamefont {A.}~\bibnamefont {Kurganov}}\ and\ \bibinfo {author} {\bibfnamefont {E.}~\bibnamefont {Tadmor}},\ }\href {https://onlinelibrary.wiley.com/doi/abs/10.1002/num.10025} {\bibfield  {journal} {\bibinfo  {journal} {Numerical Methods for Partial Differential Equations}\ }\textbf {\bibinfo {volume} {18}},\ \bibinfo {pages} {584} (\bibinfo {year} {2002})}\BibitemShut {NoStop}%
\bibitem [{\citenamefont {Schulz-Rinne}\ \emph {et~al.}(1993)\citenamefont {Schulz-Rinne}, \citenamefont {Collins},\ and\ \citenamefont {Glaz}}]{Schulz-Rinne}%
  \BibitemOpen
  \bibfield  {author} {\bibinfo {author} {\bibfnamefont {C.~W.}\ \bibnamefont {Schulz-Rinne}}, \bibinfo {author} {\bibfnamefont {J.~P.}\ \bibnamefont {Collins}}, \ and\ \bibinfo {author} {\bibfnamefont {H.~M.}\ \bibnamefont {Glaz}},\ }\href {https://doi.org/10.1137/0914082} {\bibfield  {journal} {\bibinfo  {journal} {SIAM Journal on Scientific Computing}\ }\textbf {\bibinfo {volume} {14}},\ \bibinfo {pages} {1394} (\bibinfo {year} {1993})}\BibitemShut {NoStop}%
\bibitem [{\citenamefont {Zhang}\ and\ \citenamefont {Zheng}(1990)}]{ZhangZhengRiemann}%
  \BibitemOpen
  \bibfield  {author} {\bibinfo {author} {\bibfnamefont {T.}~\bibnamefont {Zhang}}\ and\ \bibinfo {author} {\bibfnamefont {Y.}~\bibnamefont {Zheng}},\ }\href@noop {} {\bibfield  {journal} {\bibinfo  {journal} {SIAM Journal of Mathematical Analysis}\ }\textbf {\bibinfo {volume} {21}},\ \bibinfo {pages} {593} (\bibinfo {year} {1990})}\BibitemShut {NoStop}%
\bibitem [{\citenamefont {Chang}\ \emph {et~al.}(1995)\citenamefont {Chang}, \citenamefont {Chen},\ and\ \citenamefont {Yang}}]{TungChang1}%
  \BibitemOpen
  \bibfield  {author} {\bibinfo {author} {\bibfnamefont {T.}~\bibnamefont {Chang}}, \bibinfo {author} {\bibfnamefont {G.-Q.}\ \bibnamefont {Chen}}, \ and\ \bibinfo {author} {\bibfnamefont {S.}~\bibnamefont {Yang}},\ }\href {https://www.aimsciences.org/article/id/b675257f-ac90-4311-8762-757ac3ae3597} {\enquote {\bibinfo {title} {On the 2-d riemann problem for the compressible euler equations i. interaction of shocks and rarefaction waves},}\ } (\bibinfo {year} {1995})\BibitemShut {NoStop}%
\bibitem [{\citenamefont {Chang}\ \emph {et~al.}(2000)\citenamefont {Chang}, \citenamefont {Chen},\ and\ \citenamefont {Yang}}]{TungChang2}%
  \BibitemOpen
  \bibfield  {author} {\bibinfo {author} {\bibfnamefont {T.}~\bibnamefont {Chang}}, \bibinfo {author} {\bibfnamefont {G.-Q.}\ \bibnamefont {Chen}}, \ and\ \bibinfo {author} {\bibfnamefont {S.}~\bibnamefont {Yang}},\ }\href {https://www.aimsciences.org/article/id/fa1a74af-57b0-415a-9a3d-0be58f887495} {\enquote {\bibinfo {title} {On the 2-d riemann problem for the compressible euler equations ii. interaction of contact discontinuities},}\ } (\bibinfo {year} {2000})\BibitemShut {NoStop}%
\bibitem [{\citenamefont {Zheng}(2012)}]{RiemannBookZheng}%
  \BibitemOpen
  \bibfield  {author} {\bibinfo {author} {\bibfnamefont {Y.}~\bibnamefont {Zheng}},\ }\href@noop {} {\emph {\bibinfo {title} {Systems of conservation laws: two-dimensional Riemann problems}}},\ Vol.~\bibinfo {volume} {38}\ (\bibinfo  {publisher} {Springer Science and Business Media},\ \bibinfo {year} {2012})\BibitemShut {NoStop}%
\bibitem [{\citenamefont {Sheng}\ and\ \citenamefont {Zhang}(1999)}]{RiemannBookSheng}%
  \BibitemOpen
  \bibfield  {author} {\bibinfo {author} {\bibfnamefont {W.}~\bibnamefont {Sheng}}\ and\ \bibinfo {author} {\bibfnamefont {T.}~\bibnamefont {Zhang}},\ }\href@noop {} {\emph {\bibinfo {title} {The Riemann problem for the transportation equations in gas dynamics}}},\ Vol.\ \bibinfo {volume} {654}\ (\bibinfo  {publisher} {American Mathematical Soc.},\ \bibinfo {year} {1999})\BibitemShut {NoStop}%
\bibitem [{\citenamefont {Pan}\ \emph {et~al.}(2016)\citenamefont {Pan}, \citenamefont {Li},\ and\ \citenamefont {Xu}}]{LR3case}%
  \BibitemOpen
  \bibfield  {author} {\bibinfo {author} {\bibfnamefont {L.}~\bibnamefont {Pan}}, \bibinfo {author} {\bibfnamefont {J.}~\bibnamefont {Li}}, \ and\ \bibinfo {author} {\bibfnamefont {K.}~\bibnamefont {Xu}},\ }\href {https://arxiv.org/abs/1609.04491} {\enquote {\bibinfo {title} {A few benchmark test cases for higher-order euler solvers},}\ } (\bibinfo {year} {2016})\BibitemShut {NoStop}%
\bibitem [{\citenamefont {Inoue}\ and\ \citenamefont {Hattori}(1999)}]{inoue1999sound}%
  \BibitemOpen
  \bibfield  {author} {\bibinfo {author} {\bibfnamefont {O.}~\bibnamefont {Inoue}}\ and\ \bibinfo {author} {\bibfnamefont {Y.}~\bibnamefont {Hattori}},\ }\href@noop {} {\bibfield  {journal} {\bibinfo  {journal} {Journal of Fluid Mechanics}\ }\textbf {\bibinfo {volume} {380}},\ \bibinfo {pages} {81} (\bibinfo {year} {1999})}\BibitemShut {NoStop}%
\bibitem [{\citenamefont {Saadat}\ \emph {et~al.}(2021)\citenamefont {Saadat}, \citenamefont {Hosseini}, \citenamefont {Dorschner},\ and\ \citenamefont {Karlin}}]{saadat2021extended}%
  \BibitemOpen
  \bibfield  {author} {\bibinfo {author} {\bibfnamefont {M.~H.}\ \bibnamefont {Saadat}}, \bibinfo {author} {\bibfnamefont {S.~A.}\ \bibnamefont {Hosseini}}, \bibinfo {author} {\bibfnamefont {B.}~\bibnamefont {Dorschner}}, \ and\ \bibinfo {author} {\bibfnamefont {I.~V.}\ \bibnamefont {Karlin}},\ }\href@noop {} {\bibfield  {journal} {\bibinfo  {journal} {Physics of Fluids}\ }\textbf {\bibinfo {volume} {33}},\ \bibinfo {pages} {046104} (\bibinfo {year} {2021})}\BibitemShut {NoStop}%
\bibitem [{\citenamefont {Ellzey}\ \emph {et~al.}(1995)\citenamefont {Ellzey}, \citenamefont {Henneke}, \citenamefont {Picone},\ and\ \citenamefont {Oran}}]{Ellzey1995}%
  \BibitemOpen
  \bibfield  {author} {\bibinfo {author} {\bibfnamefont {J.~L.}\ \bibnamefont {Ellzey}}, \bibinfo {author} {\bibfnamefont {M.~R.}\ \bibnamefont {Henneke}}, \bibinfo {author} {\bibfnamefont {J.~M.}\ \bibnamefont {Picone}}, \ and\ \bibinfo {author} {\bibfnamefont {E.~S.}\ \bibnamefont {Oran}},\ }\href {https://doi.org/10.1063/1.868738} {\bibfield  {journal} {\bibinfo  {journal} {Physics of Fluids}\ }\textbf {\bibinfo {volume} {7}},\ \bibinfo {pages} {172} (\bibinfo {year} {1995})}\BibitemShut {NoStop}%
\bibitem [{\citenamefont {Dosanj}\ and\ \citenamefont {Weeks}(1965)}]{DOSANJH1965}%
  \BibitemOpen
  \bibfield  {author} {\bibinfo {author} {\bibfnamefont {D.~S.}\ \bibnamefont {Dosanj}}\ and\ \bibinfo {author} {\bibfnamefont {T.~M.}\ \bibnamefont {Weeks}},\ }\href {https://doi.org/10.2514/3.2833} {\bibfield  {journal} {\bibinfo  {journal} {AIAA Journal}\ }\textbf {\bibinfo {volume} {3}},\ \bibinfo {pages} {216} (\bibinfo {year} {1965})}\BibitemShut {NoStop}%
\bibitem [{\citenamefont {Ribner}(1985)}]{Ribner1985}%
  \BibitemOpen
  \bibfield  {author} {\bibinfo {author} {\bibfnamefont {H.~S.}\ \bibnamefont {Ribner}},\ }\href {https://doi.org/10.2514/3.9155} {\bibfield  {journal} {\bibinfo  {journal} {AIAA Journal}\ }\textbf {\bibinfo {volume} {23}},\ \bibinfo {pages} {1708} (\bibinfo {year} {1985})}\BibitemShut {NoStop}%
\end{thebibliography}%

\end{document}